\documentclass[12pt,a4paper]{article}
\pdfoutput=1 
\usepackage{jheppub}
\usepackage{iip-arxiv}
\usepackage[utf8]{inputenc}
\usepackage{ifthen}
\usepackage{xcolor}
\usepackage{amsmath}
\usepackage{mathtools}
\usepackage{bm}
\usepackage{breqn}

\graphicspath{figs/}

\makeatletter

\newcommand{\FN}[1]{\ifdefined\@fnotes \@fnotes{#1}\fi}
\newcommand{\MC}[1]{\ifdefined\@mnotes \@mnotes{#1}\fi}
\newcommand{\CM}[1]{\ifdefined\@cnotes \@cnotes{#1}\fi}
\newcommand{\ML}[1]{\ifdefined\@mlnotes \@mlnotes{#1}\fi}

\makeatother

\newcommand{\by}{\bar y}
\newcommand{\bt}{\bar t}
\newcommand{\bphi}{\bar \phi}

\DeclareMathOperator{\Tr}{Tr}
\DeclareMathOperator{\res}{res}
\DeclareMathOperator*{\genhyp}{F}

\newcommand{\del}[1][z]{\partial_{#1}}

\newcommand{\Dfrac}[3][]{\frac{d^{#1}#2}{d{#3}^{#1}}}

\newcommand{\QNMf}{{}_{s}\omega_{\ell m}}

\newcommand{\rlocal}[2]{R_{#2}^{(#1)}}
\newcommand{\uin}[2][]{R^{(#2)}_{\text{\scriptsize in};\, #1 s }}
\newcommand{\uout}[2][]{R^{(#2)}_{\text{\scriptsize out};\, #1 s }}
\newcommand{\Ain}{A^{(\text{\scriptsize in})}_{\ell m}}
\newcommand{\Aout}{A^{(\text{\scriptsize out})}_{\ell m}}
\newcommand{\gauss}{\sideset{_2}{_{1}}\genhyp}

\newcommand{\lmtheta}[2]{\,\theta_0,\theta_1,\theta_t\ifthenelse{\equal{#1}{}}{}{+#1},\theta_\infty\ifthenelse{\equal{#2}{}}{}{+#2}}

\newcommand{\entao}{\quad\Rightarrow\quad}
\newcommand{\fosum}[2][3]{\sum_{i=1}^{#1}\frac{#2}{z-z_i}}

\newcommand{\tpsi}{\tilde\psi}
\newcommand{\tPsi}{\tilde\Psi}
\newcommand{\Dp}{\partial_{z}}
 
 \newcommand{\defeq}{\equiv}

 \newcommand{\lamslm}{{}_s\lambda_{\ell m}}
 \newcommand{\lamslmbar}{{}_s\bar{\lambda}_{\ell m}}

\title{Kerr--de Sitter Quasinormal Modes  via
  Accessory Parameter Expansion}

\author[a,1]{Fábio Novaes,}
\author[b,2]{Cássio I. S. Marinho,}
\author[a,3]{Máté Lencsés,}
\author[b,c,4]{and Marc Casals}
\affiliation[a]{International Institute of Physics, Federal University of Rio Grande do Norte,
Campus Universitário, Lagoa Nova, Natal-RN 59078-970, Brazil}
\affiliation[b]{Centro Brasileiro de Pesquisas Físicas (CBPF), Rio de Janeiro, 
CEP 22290-180, 
Brazil.}
\affiliation[c]{School of Mathematics and Statistics, University College Dublin, Belfield, Dublin 4, Ireland.}
\email{fabionsantos at gmail.com}
\email{marinho at cbpf.br}
\email{matelencses at gmail.com}
\email{mcasals at cbpf.br}

\abstract{Quasinormal modes are characteristic oscillatory modes that
  control the relaxation of a perturbed physical system back to its
  equilibrium state.  In this work, we calculate QNM frequencies and
  angular eigenvalues of Kerr--de Sitter black holes using a novel
  method based on conformal field theory.  The spin-field perturbation
  equations of this background spacetime essentially reduce to two
  Heun's equations, one for the radial part and one for the angular
  part.  We use the accessory parameter expansion of Heun's equation,
  obtained via the isomonodromic $\tau$-function, in order to find
  analytic expansions for the QNM frequencies and angular eigenvalues.
  The expansion for the frequencies is given as a double series in the
  rotation parameter $a$ and the extremality parameter
  $\epsilon=(r_{C}-r_{+})/L$, where $L$ is the de Sitter radius and
  $r_{C}$ and $r_{+}$ are the radii of, respectively, the cosmological
  and event horizons.  Specifically, we give the frequency expansion
  up to order $\epsilon^2$ for general $a$, and up to order
  $\epsilon^{3}$ with the coefficients expanded up to $(a/L)^{3}$.
  Similarly, the expansion for the angular eigenvalues is given as a
  series up to $(a\omega)^{3}$ with coefficients expanded for small
  $a/L$. We verify the new expansion for the frequencies via a
  numerical analysis and that the expansion for the angular
  eigenvalues agrees with results in the literature.}

\preprint{}

\begin{document}

\maketitle


\section{Introduction}\label{sec:resumocientifico}

Quasinormal Modes (QNMs) are free oscillatory modes of an \emph{open}
physical system. Contrary to normal modes, QNMs oscillate but also
decay exponentially in time. The exponential ringdown of these modes
is thus characterized by complex frequencies
$\omega_{QNM} = \omega_{R} +i\, \omega_{I}$, where
$\omega_{R}\in\mathbb{R}$ corresponds to the frequency of the
oscillation and
$\omega_{I} < 0$ sets the timescale of the decay. The study
of QNMs
and their exponentially-growing counterparts ($\omega_I>0$)
is important for, e.g., the linear stability analysis of
background spacetimes (e.g.,~\cite{whiting1989mode}), to characterize the ringdown
stage of a waveform from a black hole binary
inspiral~\cite{Abbott:2016blz} and for setting the thermalization
scale in holographic gauge
theories \cite{Horowitz:2000aa}. For reviews on QNMs, see, e.g., \cite{Kokkotas1999,berti2009quasinormal}.

Black holes are compact objects characterized by the existence of an
event horizon. Among the class of stationary and axisymmetric
solutions of General Relativity, we can find exact black hole
spacetimes. They are classified by their mass $M$, angular momentum
$J$, charge $Q$ and by the cosmological constant $\Lambda$. In the
following, we use the more convenient angular momentum parameter per
unit mass, $a\equiv J/M$, and the de Sitter radius
$L\equiv \sqrt{3/\Lambda}$ ($\Lambda >0$).

In this paper, we obtain QNMs and angular
eigenvalues for the spin-$s$ perturbation equations of four-dimensional
Kerr--de Sitter black holes ($\Lambda > 0$). These black holes contain four horizons and we
focus on the QNMs obtained from solutions which are valid
between the outer event horizon (at radius $r_{+}$) and the cosmological
horizon (at radius $r_{C}\ge r_+$). 

A standard approach for studying massless spin-field perturbations of
black hole spacetimes is to write the spin-$s$ field equations using
the Newman--Penrose formalism.  This formalism enables the decoupling
and simplification of the linear field perturbation equations into one
single (``master") partial differential equation for any Petrov type D
spacetime. In the absence of sources, this master equation is
separable into two ordinary differential equations (ODEs) for modes
with fixed frequency $\omega$ and azimuthal angular number $m$. This
approach was first pursued by Teukolsky for a Kerr black
hole~\cite{Teukolsky:1973ha}. More generally, one can find a master
equation for spin $s=0,\pm1/2,\pm1,\pm3/2,\pm2$ massless field
perturbations of any Petrov type D spacetime
\cite{Dias2012b,Khanal:1983vb,Chambers1994,Batic2007a}, including
Kerr--de Sitter black holes. The associated ODEs are all reducible to
Heun equations \cite{Suzuki1998,Batic2007a}.  In this work, we exploit
an integrable structure behind Heun equations, deeply related to 2D
conformal field theory (CFT), to find QNMs and angular eigenvalues of
Kerr--de Sitter black holes.

Heun equations are ODEs with four regular singular
points~\cite{ronveaux1995heun}. We can fix these points to be at the
values $z={0,1,x,\infty}$ of the independent variable, without loss of
generality. These equations can then be characterized by the moduli
parameter $x$, its local monodromy data and the so-called accessory
parameter. The monodromy data governs the behaviour of the
analytically continued solutions around the singular points, while the
accessory parameter contains global information of the solutions. Heun
equations can be formally extended to have an extra apparent
singularity, with trivial monodromy. The position of the apparent
singularity can then be continuously changed in a precise way to leave
unaffected the monodromy data. This is called the \emph{isomonodromic
  deformation} of the Heun equation
\cite{Garnier:1912,Iwasaki:1991}. Perhaps surprisingly, this
deformation is Hamiltonian and integrable. One can use this structure
to construct an expansion of the accessory parameter in $x$ with
coefficients given in terms of the local and global monodromy data
\cite{Litvinov:2013sxa,Lencses:2017dgf}. We call this the
\emph{Accessory Parameter Expansion} (APE), revised in
section \ref{sec:how-obtain-ape}. This approach was successfully applied
in the context of 2D conformal field theory in relation to the
classical limit of conformal blocks \cite{Lencses:2017dgf} and for 2D
conformal mappings from a simply connected planar region to the
interior of a circular arc quadrilateral \cite{Anselmo2018}. For other
approaches to obtaining the APE, we refer the reader to
\cite{Litvinov:2013sxa,Menotti2014,Hollands2017}.

In this work, we use the APE to calculate Kerr--de Sitter QNMs and
angular eigenvalues for the first time. In particular, we apply the
APE to the case of Kerr--de Sitter black holes near the extremal limit
$r_C\to r_+$, also called the \emph{rotating Nariai limit}, about both
the upper and lower superradiant bounds.\footnote{Superradiance is a
  scattering phenomenon whereby a field wave extracts rotational
  energy from a rotating black hole.} More specifically, we obtain an
analytical expansion for the QNM frequencies of the form:
\begin{equation}
  \label{eq:omega-expansion}
  _{s}\omega_{\ell m} = m \Omega_{+}+
\epsilon\left[\bar\omega_{0}(a)+\bar{\omega}_{1}(a) \epsilon^{1}
+\bar{\omega}_{2}(a) \epsilon^{2}
+\mathcal{O}(\epsilon^{3})\right],
\end{equation}
in the extremality parameter $\epsilon\equiv (r_{C}-r_{+})/L$,
where $\bar\omega_{0,1,2}$ are  coefficients we determine. Here and in what follows $\Omega_{k}$, $k=+,C$ denotes the angular velocity of the horizon $r_k$, $k=+,C$.  We provide this
expansion: (i) up to third order in $\epsilon$ with the coefficients
expanded up to third order in $a/L$;
(ii) up to second order in
$\epsilon$ for general $a$.  The expansion in Eq.~\eqref{eq:omega-expansion}
is about $m \Omega_{+}$ (unexpanded in $\epsilon$). The expansion about $m\Omega_{C}$ can be
obtained by letting
$\Omega_{+} \rightarrow \Omega_{C},\; \epsilon \rightarrow -\epsilon $
and taking the complex conjugate of the right hand side of
 Eq.~\eqref{eq:omega-expansion}. 
We thus show,
 and check numerically, that the lower superradiant bound
 $\omega=m\Omega_{C}$ is  an accumulation point of
 QNMs, similarly to the upper bound $\omega=m\Omega_{+}$.

To the best of our knowledge, only the term
$\bar\omega_0$ in Eq.~\eqref{eq:omega-expansion} for the specific case of $s=0$ and coupling constant with the Ricci scalar $\xi=0$ had so far been obtained in the literature.
It was obtained in~\cite{anninos2010sitter} by calculating QNMs directly on the rotating
Nariai geometry, which is found by 
a near horizon limit of the Kerr--de Sitter spacetime.
Apart from the APE derivation of Eq.~\eqref{eq:omega-expansion}, we also calculate QNMs directly in the rotating
Nariai geometry in the case of $s=0$ and general $\xi$, thus obtaining an alternative
derivation of $\bar\omega_0$ for this case.
Overall, we believe that the expressions we derive for $\bar\omega_0$ for non-zero spin (as well as for $s=0$ and $\xi\neq 0$) and for
$\bar\omega_{1,2}$ for any spin are new. 
We have checked our 
high-order general-spin Eq.~\eqref{eq:omega-expansion}
against a numerical analysis that we have carried
out using Leaver's method to high accuracy.  This method was first used in \cite{Yoshida2010} to obtain Kerr--de Sitter QNMs and angular eigenvalues and our results are consistent with that work.

Using the same APE method, we derive an analytical expansion for the
angular eigenvalue
\begin{align}
  \label{eq:angular-eigenvalue-true-expansion}
\lamslm
   &=
   \sum_{\substack{n,k=0\\ n+k=3}}^{3}\lambda_{n,k}\,\alpha^n(a\omega)^k+\sum_{\substack{n,k=0\\ n+k=4}}^4\mathcal{O}(\alpha^n(a\omega)^k),
\end{align}
where $\lambda_{n,k}$ 
are the coefficients, obtained as a small $ \alpha \equiv a/L$ expansion from the APE.
For convenience, we will reorganize the explicit coefficients of Eq.~\eqref{eq:angular-eigenvalue-true-expansion} as a finite series
\begin{equation}
  \label{eq:angular-eigenvalue-expansion}
  \lamslm =  \sum_{k=0}^{3}\lambda_{\omega,k}\left(a\omega\right)^{k}+ \mathcal{O}((a\omega)^4),
\end{equation}
where the coefficients $\lambda_{\omega,k}\equiv\sum_{n=0}^{3-k}\lambda_{n,k}\alpha^n + \mathcal{O}(\alpha^{4-k})$, $k=0,\ldots,3$, are given in appendix~\ref{sec:eigen coeffs}. In the rest of this paper, we will only refer to the coefficients in Eq.\eqref{eq:angular-eigenvalue-expansion}. 

Kerr--de Sitter angular eigenvalues were first obtained numerically in \cite{Chambers1994}.
Our expansion for the eigenvalues for small $\alpha$ and $a\omega$ agrees with that already obtained in \cite{Suzuki1998} using a completely different method (and with that in~\cite{berti2006eigenvalues} in the Kerr limit $L\to \infty$).
Our eigenvalue derivation here is a proof of principle of the APE, which is extendable to higher orders than what we present here.
 
Summing up, with
 our method, we obtain: (i) an angular eigenvalue expansion
 for small $a\omega$ and $a/L$
consistent with~\cite{Suzuki1998,berti2006eigenvalues}, and (ii) a new expansion for the QNM frequencies
near the rotating Nariai limit.  For the reader
who is only interested in the final result, the 
coefficients in the
QNM expansion \eqref{eq:omega-expansion}
are given in Eqs.~\eqref{eq:spin-s-rotating-nariai-qnm} and \eqref{eq:bar omega_1}
and in appendix~\ref{sec:omegas},
and 
the  coefficients in the eigenvalue expansion \eqref{eq:angular-eigenvalue-expansion}
are given in appendix~\ref{sec:eigen coeffs}.

There exist various other methods for calculating QNM frequencies, from
asymptotic analyses to numerical techniques.
Ref.~\cite{Yoshida2010} provides a numerical
 calculation of QNMs of Kerr--de Sitter black holes, for modes between $r_{C}$ and
 $r_{+}$, by using a method originally developed by
 Leaver~\cite{Leaver:1985,Leaver:1986a}. Their analysis suggests that
 the QNM frequencies accumulate towards $\omega = m\Omega_{+}$ in each
 of the extremal limits $r_{+}\rightarrow r_{-}$ and
 $r_{+}\rightarrow r_{C}$, where $\Omega_{+}$ is the angular velocity
 of the event horizon $r_{+}$.  The frequency $\omega= m\Omega_{+}$ is
 the upper bound of the superradiant regime.   For non-extremal (i.e., $r_+ \neq r_C$)
 Kerr--de Sitter black holes, the superradiant regime is
 $|m|\Omega_{C}< |\omega| < |m|\Omega_{+}$, where $\Omega_{C}$ is the
 angular velocity of the cosmological horizon $r_{C}$.  Other works we found in the literature
 on QNMs in Kerr--de Sitter are the following:
 ~\cite{tattersall2018kerr} provides an analytic expansion of the QNM with the lowest overtone ($N=0$) in the double approximation of  small $a$ (to linear order) and  large-($\ell+1/2$) (up to order $(\ell+1/2)^{-6}$), where $\ell$ is the multipolar number;
 similarly,~\cite{dyatlov2012asymptotic} provides a Bohr--Sommerfeld type of condition which QNMs for small $a$ must satisfy and which could be solved numerically, being better suited for larger $\ell$;
 within the context of the so-called
  strong cosmic censorship conjecture,
 \cite{Dias2018b} obtained an analytic expression for the QNMs for 
 $\ell=|m|\gg 1$ and  complement it with a numerical study.

We finish the introduction by outlining the contents of this paper.
We describe the geometry and configuration space of the Kerr--de Sitter black hole in section~\ref{sec:kerr-de-sitter}.
In section~\ref{sec:scattering-kerr-de}, we review the spin-field master equation in Kerr--de Sitter and the scattering problem. We express  the product of scattering coefficients, the so-called \emph{greybody factor}, in terms of the monodromies of the radial ODE. We then describe the monodromy conditions for angular eigenvalues and QNMs.
We finish section~\ref{sec:scattering-kerr-de} by obtaining QNMs directly in the rotating Nariai geometry
\cite{anninos2010sitter} and obtaining the same result from the full Heun equation in Kerr--de Sitter under the appropriate limits.
In section \ref{sec:access-param-from}, we review how to obtain the
APE via the isomonodromic $\tau$-function, first proposed in
\cite{Novaes2014c}.
In Subsecs.~\ref{sec:angular-eigenvalues} and
\ref{sec:quasinormal-modes}, we apply the APE around $\epsilon=0$ to obtain expansions for the angular eigenvalues and QNMs.
In section~\ref{sec:comp-with-numer}, we compare our analytic results
with a numerical calculation using Leaver's method.
In section~\ref{sec:conclusions} we present our conclusions and potential
future developments. 

We also include complementary sections with more details on our results. In appendix~\ref{sec:from-self-adjoint}, we introduce the gauge transformations of Fuchsian equations used to reduce the master equations to Heun equations.
In appendix~\ref{sec:altern-deriv}, we present an alternative derivation of the leading order QNM equation.
In appendix~\ref{sec:iso-definitions}, we review how to obtain the APE from the isomonodromic $\tau$-function as in \cite{Lencses:2017dgf}, with the crucial modification of expanding the monodromies in $\epsilon$.
Finally, in appendices~\ref{sec:omegas} and \ref{sec:eigen coeffs} we
display the coefficients of the QNM expansion and angular eigenvalues
respectively. 

In this paper, we choose Planck units
$\hbar=c=G=1$.


\section{Kerr--de Sitter Black Holes}
\label{sec:kerr-de-sitter}

The Kerr--de Sitter metric \cite{Carter1973} can be written in Chambers--Moss
coordinates \cite{Chambers1994}  as
\begin{equation}
  \label{eq:kerradsmetric}
  ds^2 = -\frac{\Delta_{r}}{\rho^2\chi^{4}}\left(dt-a\sin^{2}\theta
    d\phi\right)^2 +
  \frac{\Delta_{\theta}\sin^{2}\theta}{\rho^2\chi^{4}}\left(adt-(r^2+a^{2})d\phi\right)^2
  +\rho^2 \left(\frac{ d\theta^2}{\Delta_{\theta}} +
    \frac{dr^2}{\Delta_{r}}\right),
\end{equation}
where
\begin{equation}
\begin{aligned}
  \label{eq:170}
  \chi^{2}&\equiv 1 + \alpha^{2}, &\quad \rho^{2} &\equiv r^{2} + a^{2}\cos^{2}\theta,\\[5pt] 
  \Delta_{\theta} &\equiv 1 + \alpha^{2}\cos^{2}\theta, &\quad \Delta_{r} &\equiv (r^{2}+a^{2})\left(1-\frac{r^{2}}{L^{2}}\right)-2Mr
\end{aligned}
\end{equation}
and $\alpha \equiv a/L$.
The coordinate ranges are $t,r\in \mathbb{R}$, $\theta \in [0,\pi]$,
$\phi\in [0,2\pi)$. 
 The solutions  are labeled by the
parameters $(a,M,L)$, where $M$  is the mass, $J = M a$ is
the angular momentum
and $L \equiv \sqrt{3/\Lambda}$ is the de Sitter radius, where $\Lambda$ is the cosmological constant. 
In particular, the Ricci scalar is equal to
$R = 12/L^2$.  When $L\rightarrow \infty$, we recover the Kerr metric
in Boyer--Lindquist coordinates and when $a=0$ we recover the
Schwarzschild--de Sitter metric. 

\begin{figure}[tp!]
  \centering
  \includegraphics[width=.85\textwidth]{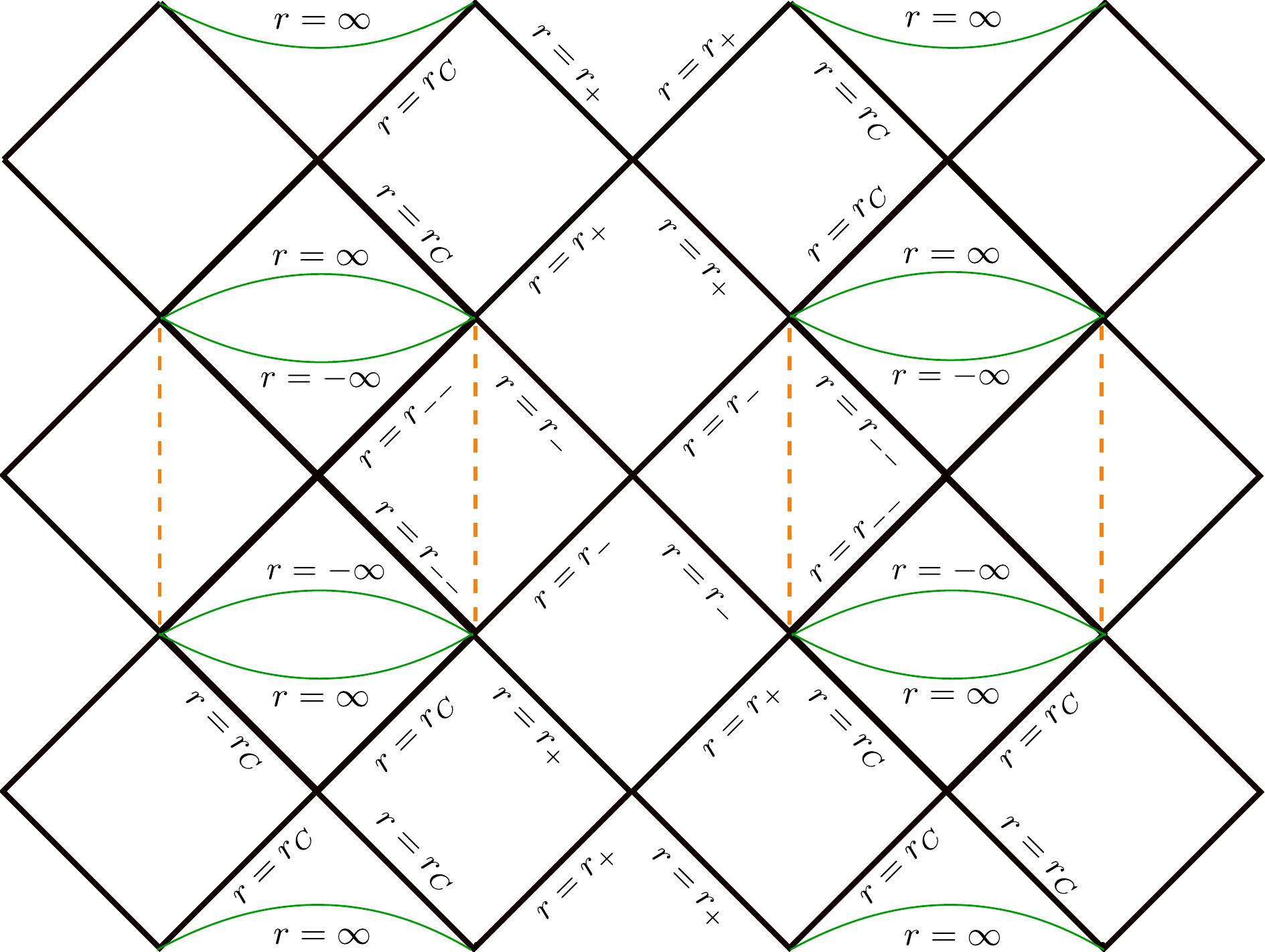}
  \caption{Kerr--de Sitter causal diagram for $\theta=0$. The ring-like
    singularity at $r=0,\, \theta=\pi/2$ is denoted by the orange dashed
    lines.}
  \label{fig:kerr-ds}
\end{figure}

The polynomial $\Delta_{r}$ has four roots: $r_{i},\, i\in \mathcal{I}$, where $\mathcal{I}=\{C,-,+,--\}$,
 with
$r_{--} = -(r_{-}+r_{+}+r_{C})$.  It is easy to check that, in order
to have a black hole in a de Sitter Universe, i.e, for the two largest roots to be
positive, 
$0<r_+\leq r_C$, then the other two roots must be real and
must satisfy $r_{--}<0\leq r_-\leq r_+ < r_C$, and thus $r_{-}$ is the
inner Cauchy horizon.  
This implies that $r_{C}$, $r_{+}$ and $r_{-}$ are the
radii of, respectively, the \emph{cosmological horizon}, the
\emph{(outer) event horizon} and the \emph{(inner) Cauchy horizon}.
The negative horizon $r_{--}$ is reachable if one goes inside the
inner horizon and, avoiding the ring-like singularity (located at $r=0, \theta = \pi/2$), goes to negative values of $r$, thus reaching
another asymptotically de Sitter region.  The global extension of the
Kerr--de Sitter black hole is 
 described
in \cite{Carter1973,Mellor1990,Chambers1994,Akcay2011} and is
shown in figure~\ref{fig:kerr-ds}. 

\begin{figure}
    \centering \includegraphics[width=.7\textwidth]{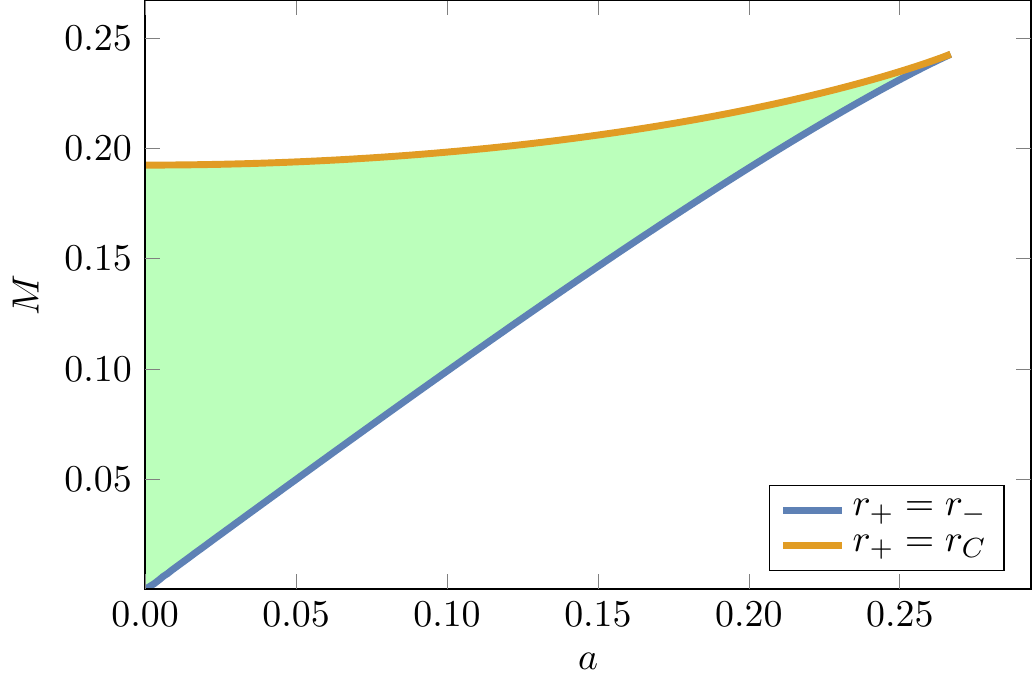}
    \caption{The green region correspond to values of $(M,a)$ (with
    $L=1$)
    representing  black hole solutions.
    The blue line corresponds to $r_{+}=r_{-}$ and the
    orange line to $r_{+}=r_{C}$}
    \label{fig:discriminant}
\end{figure}

\begin{figure}
    \centering
    \includegraphics{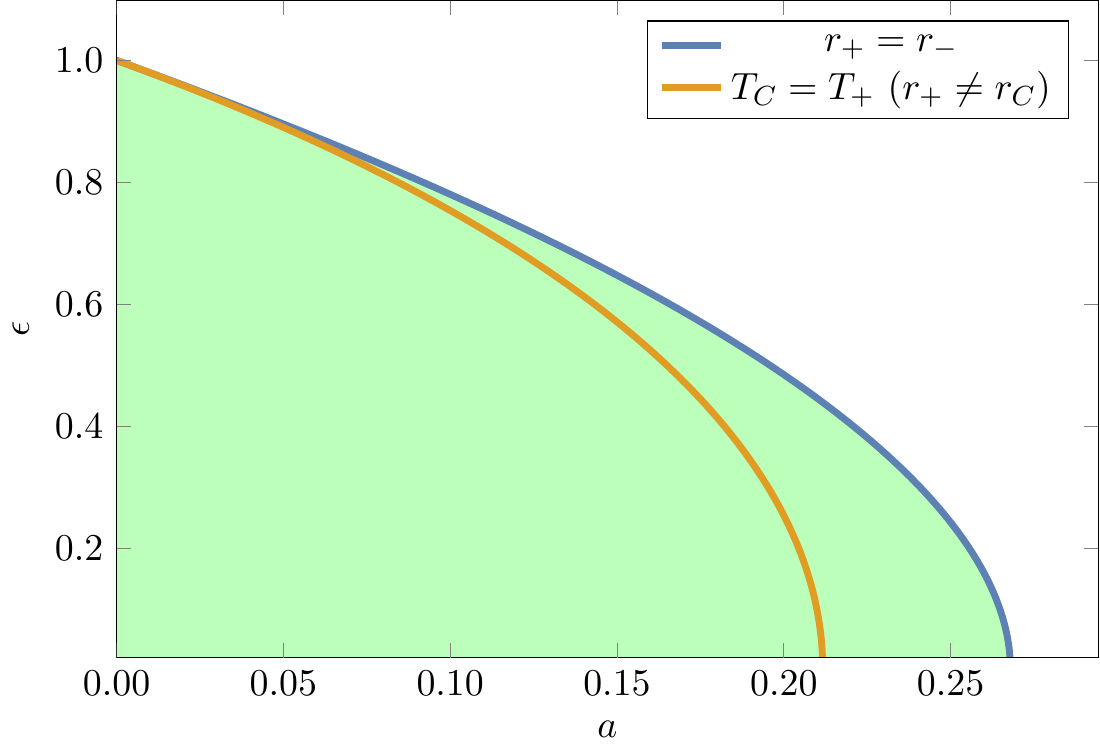}
    \caption{Kerr--de Sitter configuration space in terms of $(\epsilon,a)$
      with $L=1$. The green region corresponds to the same one in
      figure~\ref{fig:discriminant} of Kerr--de Sitter black holes.  The
      $\epsilon=0$ line corresponds to the rotating Nariai limit
      $r_{+}=r_{C}$ and the point $\epsilon=1, a=0$ corresponds to
      empty de Sitter spacetime. The blue line corresponds to the
      extremal limit $r_{+}=r_{-}$, while the orange line corresponds
      to the equal temperature condition $T_{C}=T_{+}$ for
      $r_{+}\neq r_{C}$ (\emph{lukewarm} solutions).}
    \label{fig:phase_eps}
\end{figure}

The limit $r_+\to r_-$ corresponds to an extremal black hole
($r_+=r_-$ is an event horizon), whereas $r_+\to r_C$ corresponds to
an extreme naked singularity ($r_+=r_C$ is a
cosmological horizon)~\cite{Akcay2011} and it is called the \emph{rotating Nariai limit} \cite{Booth1999,anninos2010sitter}.\footnote{We note, however, that
  Ref.~\cite{Akcay2011} mentions that ``there is no clear,
  universally-agreed-upon definition for naked singularities when the
  cosmological constant is non-zero". An observer outside the
  cosmological horizon $r_{C}$ would call the extremal $r_{+}=r_{C}$
  solution an extremal \emph{black hole}, for instance.} The extremal black hole is a maximally-rotating black hole, and it is interesting to note that it can have $a>M$ ($a=M$ being the maximal rotating limit in Kerr). 

Two or more roots of $\Delta_{r}$
coincide if and only if the discriminant
$\mathcal{D} = L^{-12}\prod_{i\neq j \in \mathcal{I}}(r_{i}-r_{j})$ is zero. It is
possible to write the discriminant in terms of only $M/L$ and $a/L$.
One can show that 
$\mathcal{D}\geq 0$ and $M,a\geq0$ if and only if
all the roots of $\Delta_{r}$ are real
and 
$r_{C}\geq r_{+}\geq r_{-}\geq 0>r_{--}$. This
condition thus defines the region in green in figure~\ref{fig:discriminant}, corresponding to the parameter space where
black hole solutions exist \cite{Chambers1994}.

The QNM expansion that we calculate later in this paper is most easily
organized as a series in $a$ and in the extremality parameter
$\epsilon \equiv (r_{C}-r_{+})/L$, instead of $a$ and $M$.  The
horizons are given by
\begin{equation}
  \label{eq:rc&rpOfepsion}
  \begin{aligned}
    r_{+}&=r_{C}-
     \epsilon L,
    \quad r_{--}=-r_{+}-r_{C}-r_{-},\\[5pt]
     r_{C}
    &= \frac{\epsilon L}{2} +\frac{L}{2\sqrt{3}}{\sqrt{2\sqrt{\left(\alpha^2+\epsilon ^2-1\right)^2-12 \alpha^2}-2          \alpha^2+\epsilon ^2+2}},
          \\[10pt]
 r_-
&= \frac{L}{2 \sqrt{3}} \left(
           \sqrt{-4 \sqrt{\left(\alpha^2+\epsilon ^2-1\right)^2-12 \alpha^2}-8 \alpha^2-5 \epsilon ^2+8}     
        \,\,-\right.\\[5pt]
      &\hspace{5cm}\left. 
       \sqrt{2 \sqrt{\left(\alpha^2+\epsilon ^2-1\right)^2-12 \alpha^2}-2\alpha^2+\epsilon ^2+2} \right).
   \end{aligned}
    \end{equation}
Each horizon $r_k,\, k\in \mathcal{I}$,
 has an associated temperature $T_{k}$ and an angular velocity $\Omega_{k}$ given by
\begin{equation}
T_{k} =
\frac{|\Delta'_{r}(r_{k})|}{4\pi\chi^{2}(r_{k}^{2}+a^{2})},\quad
\Omega_{k} = \frac{a}{r_{k}^{2}+a^{2}},
\label{eq:heun_temps} 
\end{equation}
where the prime denotes derivative with respect to the argument of $\Delta_r(r)$. The configuration space in terms of $(\epsilon,a)$ is given in figure~\ref{fig:phase_eps}.
The extremal limit corresponds to the blue line,
while the rotating Nariai limit is clearly the $\epsilon=0$
line. Given the temperatures of the event and cosmological horizons,
respectively, $T_{+}$ and $T_{C}$, the orange line in
figure~\ref{fig:phase_eps} corresponds to the condition $T_{+}=T_{C}$
for $r_{+}\neq r_{C}$, called \emph{lukewarm} solutions in
\cite{anninos2010sitter}. As noted in \cite{anninos2010sitter}, these
solutions are not in thermal equilibrium, as the two horizons can
still exchange angular momentum while keeping the temperature
fixed. 
Finally, it is worth mentioning that, recently,
it was shown that the region between the event and cosmological horizons of Kerr--de Sitter is \emph{non-linearly} stable
 for small angular momentum~\cite{hintz2016global}.


\section{Scattering on Kerr--de Sitter Black Holes}
\label{sec:scattering-kerr-de}

\subsection{Linear Field Perturbations}

The analysis of linear field perturbations is important in order to
address, for example, the linear stability of the spacetime and wave
scattering around black holes. The standard formalism describing
spin-field perturbations of the Kerr black hole was set up by
Teukolsky \cite{Teukolsky:1973ha}.
Teukolsky used the so-called Newman--Penrose formalism, whereby the
various field components are projected onto a null tetrad.  Teukolsky
decoupled the equations for the spin-field perturbations in Kerr,
obtaining one single \emph{master} equation which is obeyed by all
possible spins
 of the field.
Furthermore, in Boyer--Lindquist coordinates, the Teukolsky master
equation essentially separates into two ODEs, one for the radial
coordinate and the other one for the polar angular coordinate. This
approach was later generalized to a Kerr--de Sitter background spacetime in
\cite{Khanal:1983vb,Chambers1994,Suzuki1998}. The two ODEs are
actually of Heun type~\cite{ronveaux1995heun} for any Petrov type D
spacetime with a cosmological constant \cite{Batic2007a}.

We note that there are different tetrad
normalization choices in the literature,
but the resulting equations are easily obtained by simple
transformations
that do not change the linear analysis. Here we follow
the conventions of \cite{Suzuki1998,Yoshida2010}, which are different from those in \cite{Chambers1994,Dias2012b}.
We next review the
radial and angular equations for linear spin-field perturbations of Kerr--de Sitter.

Let
$\psi_{s,\omega,m}(t,\phi,r,\theta) = e^{-i\omega
  t}e^{im\phi}R_{s,\omega,m}(r)S_{s,\omega,m}(\theta)$ be a mode
solution of the
master 
equation for the Kerr--de Sitter metric in Chambers--Moss coordinates
\eqref{eq:kerradsmetric}.\footnote{The mode solutions, as well as various other quantities
  throughout the paper, depend on the value of the angular eigenvalue, which is
  usually labeled by the multipolar index $\ell$. In this paper, we
  omit this label unless strictly necessary, in order to avoid index
  cluttering.}  Here, $\omega\in \mathbb{C}$ is the mode
frequency and $m\in\mathbb{Z}$ its azimuthal number.  
The spin-$s$ master equation separates
into the following angular and radial
ODEs~\cite{Khanal:1983vb,Chambers1994,Suzuki1998}:
\begin{subequations}
\label{eq:master-eqs-2}
\begin{align}
 \label{eq:angulareq}\left[
     \del[u]\Delta_{u}\del[u] -  \frac{1}{\Delta_{u}}\left(
        H+\frac{s}{2}\Delta_{u}'
      \right)^{2} + 2 s H' - X_{s}
  \right]S_{s,\omega,m}(u) &=0,\\[10pt]
  \label{eq:radialeq}
  \left[
  \Delta_{r}^{-s}\del[r]\Delta_{r}^{s+1}\del[r]+
  \frac{1}{\Delta_{r}}
  \left(
  W^{2}-isW\Delta_{r}'
  \right)+ 2isW' -Y_{s}
  \right]R_{s,\omega,m}(r)&=0,
\end{align}
\end{subequations}
where 
\begin{align}
  \label{eq:2HW}
  W(r)&\equiv \chi^{2}[\omega(r^{2}+a^{2})-am],\quad &H(u) &\equiv \chi^{2}[a\omega(1-u^{2})-m],
\end{align}
and
\begin{align}
  Y_{s}(r) &\equiv \frac{2}{L^{2}}(s+1)(2s+1)r^{2} + \lamslm - s\left(1-\alpha^{2}\right),\,\,  &X_{s}(u)& \equiv 2(2s^{2}+1)\alpha^{2}u^{2} -\lamslm, \label{eq:2}\\
  u&\equiv \cos\theta,\quad & \Delta_u(u)& \equiv (1-u^{2})(1+\alpha^{2}u^{2}), \label{eq:angular-poly}
\end{align}
with $\lamslm=\lamslm(a\omega,\alpha)$ being the angular eigenvalue.\footnote{
Our angular eigenvalue $\lamslm$, when evaluated in Kerr (i.e, for $L\to \infty$), is 
equal to $\lambda+s$, where $\lambda$ is in Teukolsky's~\cite{Teukolsky:1973ha}.
} The prime in $\Delta_{r}'$ denotes derivative with respect to $r$
and in $\Delta'_{u}$ with respect to $u$. Notice that the $s=0$ radial
equation above corresponds to the Klein--Gordon equation with
conformal coupling \cite{Yoshida2010}, i.e., with Ricci scalar coupling parameter $\xi=1/6$.

The solutions of the angular equation \eqref{eq:angulareq}, assuming
regularity at $u= -1$ and  $+1$, are usually called (Anti-)de Sitter spheroidal
harmonics \cite{Dias2012b} and its eigenvalues
${}_{s}\lambda_{ \ell m}= {}_{s}\lambda_{\ell
  m}(a\omega,\alpha)$ are only known numerically \cite{Yoshida2010} or
as an expansion to order $\mathcal{O}(\alpha^{2},(a\omega)^2)$, as derived
in~\cite{Suzuki1998}. 

The radial and angular ODEs have certain symmetries.  In particular,
Eq.~\eqref{eq:angulareq} is symmetric under
$(s,\omega,m)\rightarrow (-s,-\omega,-m)$.
This implies that ${}_{-s}\lambda_{\ell, -m}(a\omega,\alpha)=\lamslm (-a\omega,\alpha)$ and a similar symmetry for the angular eigenfunction. Moreover, one can also show that
$\Delta_{r}^{-s}R_{-s,-\omega,-m}$ is a solution of
\eqref{eq:radialeq} if $R_{s,\omega,m}$ is  a solution.

\subsection{Reduction of Radial and Angular Equations to Heun Equations}
\label{sec:reduct-heun-equat}

In the following, we want to analyze the connection problem for the
angular and radial equations \eqref{eq:master-eqs-2}. By the
\emph{connection problem} we mean rewriting the solution near a
certain singular point of the differential equation in terms of a
linear combination of solutions near another singular point.  The
tasks of finding the eigenvalues of the angular equation and the QNMs
of the radial equation can both be described in terms of the
connection problem \cite{Castro2013b,Novaes2014c,daCunha:2015fna}, as
we show in the next subsection. It is thus helpful to reduce the
radial and angular equations to some known special form, so that we can
understand the structure of the solutions.  In this section, we are
going to apply M\"obius transformations to the radial and angular
variables $r$ and $u$. This will map the roots of the polynomials
$\Delta_{r}(r)$ and $\Delta_{u}(u)$ to
$(z_{k})_{k\in \mathcal{J}}\equiv(0,1,x,\infty),$ with
$\mathcal{J} \equiv \{0,1,x,\infty\}$.

Equations \eqref{eq:angulareq} and \eqref{eq:radialeq} naively possess
five singular points each.  In the radial case, these points are the
four roots of $\Delta_{r}(r)$, i.e.,
$r_{k},\, k\in \mathcal{I}=\{C,-,+,--\}$, together with the point
$r=\infty$. In the angular case, the singular points are the four
roots of $\Delta_{u}(u)$, i.e. $u=\pm1,\pm \frac{i}{\alpha}$, together
with the point $u=\infty$.  For $\Lambda \neq 0$, all singular points
of the radial and angular equations are regular and thus both
equations are \emph{Fuchsian}. Surprisingly, as firstly shown in
\cite{Suzuki1998}, both equations can be reduced to \emph{Heun
  equations}~\cite{ronveaux1995heun}, i.e., Fuchsian equations with
\emph{four} regular singular points.  After a series of
transformations, the points $r=\infty$ and $u=\infty$ cease to be
singular points of the corresponding ODEs, thus being characterized as
\emph{removable singularities} \cite{slavyanov2000}. We rederive this
result in a systematic way in appendix~\ref{sec:from-self-adjoint}. On
the other hand, for $\Lambda=0$ there is a confluence of singular
points, producing irregular singular points at $r=\infty$ and at
$u=\infty$. Both radial and angular equations then become
\emph{confluent} Heun equations~\cite{Castro2013b,daCunha:2015ana}. In
the following, we derive the Heun equations from the angular
\eqref{eq:angulareq} and radial \eqref{eq:radialeq} equations.
\\

\noindent\textbf{Generic Heun Equation.} 
The basis of our method is to classify the Heun equation in terms of
its monodromy data.  For a function $y=y(z)$, the Heun equation can be
written in the \emph{canonical form} (see
appendix~\ref{sec:from-self-adjoint} or
\cite{ronveaux1995heun,slavyanov2000})
\begin{equation}
  \label{eq:heuneq}
 \Dfrac[2]{y}{z} +
  \left(\frac{1-2\theta_0}{z} + \frac{1-2\theta_1}{z-1} +
    \frac{1-2\theta_x}{z-x} \right) \Dfrac{y}{z} +
  \left(\frac{b_1b_2}{z(z-1)}-\frac{x(x-1)K_x}{z(z-1)(z-x)}
  \right) y(z) = 0,
\end{equation}
where the singular points are
$(z_{k})_{k\in \mathcal{J}}\equiv (0,1,x,\infty)$, with
$x \in \mathbb{C}$. The
coefficients $\bm{\theta}\equiv (\theta_{0}, \theta_{1}, \theta_{x},\theta_{\infty})$, also
known as the local \emph{monodromy coefficients} (to which we shall simply refer as \emph{monodromies}), correspond to half
of the difference of Frobenius exponents of Frobenius series solutions
$y_{k}^{(j)},\, j=1,2$ close to the singular points. That is,
\begin{equation}
  \label{eq:fronbenius-zk}
  y_{k}^{(1)}(z) =(z-z_{k})^{2\theta_{k}}(1+\mathcal{O}(z-z_{k})),\quad
  y_{k}^{(2)}(z)=(1+\mathcal{O}(z-z_{k})),\quad k=0,1,x.
\end{equation}
In their turn, $b_1$ and $b_2$ are the Frobenius exponents at
$z=\infty$:
\begin{equation}
\label{eq:frobenius-infinity}
  y_{\infty}^{(1)}(z) =z^{-b_1}(1+\mathcal{O}(z^{-1})),\quad
  y_{\infty}^{(2)}(z)=z^{-b_2}(1+\mathcal{O}(z^{-1})),\quad 
\end{equation}
with
\begin{equation}
  \label{eq:thetainf-definition}
  2\theta_{\infty}\equiv b_1-b_2.
\end{equation}
The regularity of $z=\infty$ in \eqref{eq:heuneq} implies the Fuchs
condition 
\begin{equation}
\label{eq:fuchs-condition}
  b_1+b_2+\sum_{k\in \mathcal{J}}2\theta_{k} = 2.
\end{equation}
The two relations in Eqs.~\eqref{eq:thetainf-definition} and \eqref{eq:fuchs-condition} can be used to write the Frobenius exponents
$b_{1,2}$ in terms of the $\theta$'s as
\begin{equation}
  \label{eq:37}
  b_1 = -\sum_{k\in \mathcal{J}}\theta_{k}+1 +\theta_{\infty},\quad b_2 = -\sum_{k\in \mathcal{J}}\theta_{k}+1 -\theta_{\infty}.
\end{equation}
The \emph{accessory parameter} $K_{x}$ is the only parameter which,
for generic monodromy data, cannot be written only in terms of
$\theta$'s, as it also depends on global information of the
solutions. Therefore, the Heun equation is completely determined by
five parameters for fixed moduli parameter $x$:
$\mathcal{M}_{\text{Heun}}\equiv
(\theta_{0},\theta_{1},\theta_{x},\theta_{\infty}; K_x)$. 
\\

\noindent\textbf{Angular Equation.}
The Frobenius exponents of local solutions of \eqref{eq:angulareq}
around
$(u_{k})_{k\in \mathcal{J}}\equiv \left(-1,
  \frac{i}{\alpha},1,-\frac{i}{\alpha}\right)$ are given by
\begin{equation}
  \label{eq:35}
  \theta_{k} =-
   \res_{u=u_{k}}\frac{H(u)}{\Delta_{u}(u)}-\frac{s}{2},\, k\in \mathcal{J}.
\end{equation}
Explicitly, the difference of Frobenius exponents is given by
\begin{equation}
  \label{eq:12}
  \begin{aligned}
    &&  2\theta_{0} &=m-s,  &  2\theta_{1} &= \frac{i \left(\alpha ^2+1\right) a\omega }{\alpha }-i \alpha  m-s, \\[5pt]
    && 2\theta_{x} &= -m-s, & 2\theta_{\infty} &= -\frac{i
      \left(\alpha ^2+1\right) a\omega }{\alpha }+i \alpha m-s.
  \end{aligned}
\end{equation}
Starting with Eq.~\eqref{eq:angulareq}, we make
the coordinate transformation
\begin{equation}\label{eq:u to z}
  z(u) = \zeta_{\infty}\frac{u+1}{u +i/\alpha},\quad \zeta_\infty= \frac{2i}{(i+\alpha)},
\end{equation}
whereby
  \begin{equation}\label{eq:u to z pts}
   (u_{k})_{k\in \mathcal{J}}=\left(-1,
  \frac{i}{\alpha},1,-\frac{i}{\alpha}\right)
    \mapsto (z_{k})_{k\in \mathcal{J}} = (0,1,x,\infty),
  \end{equation}
  and
  \begin{equation}
    \label{eq:8}
  x= \frac{4i\alpha}{(i+\alpha)^2}.
  \end{equation}
  We also make the s-homotopic transformation \cite{slavyanov2000}
 \begin{equation}
   \label{eq:S vs f}
   S_{s,\omega,m}(u)
   =z^{-\theta_{0}}(z-1)^{-\theta_{1}}(z-x)^{-\theta_{x}}(z-\zeta_\infty)f(z),
 \end{equation}
 which, as proven in appendix~\ref{sec:reduct-angul-mast}, reduces
 \eqref{eq:angulareq} to the Heun equation
\begin{multline}
\label{eq:heuncanonicalangular}
  \frac{d^2f}{dz^2}+\left(\frac{1-2\theta_{0}}{z}+\frac{1-2\theta_{1}}{z-1}+
\frac{1-2\theta_{x}}{z-x}\right)\frac{df}{dz}+\\[5pt]
+\left(\frac{(1+2s)(1+2s+2\theta_{\infty})}{z(z-1)}-\frac{x(x-1)K_x}{z(z-1)(z-x)}\right)f=0.
\end{multline}
The   accessory parameter is
\begin{multline}
\label{eq:angularaccessory}
K_x= \frac{(2s+1)(\theta_{0}+\theta_{x}+s)-s}{x}+\frac{(2s+1)(\theta_{1}+\theta_{x}+s)-s}{x-1}+\\[5pt]
\frac{2s^{2}+1}{\zeta_{\infty}-x}-\frac{1}{4\zeta_{\infty}^{2}}\frac{2(2s^{2}+1)x^{2}+\lamslm \zeta_{\infty}^{4}}{x(x-1)},
\end{multline}
where we used the fact that $\alpha^{2}=-x^{2}/\zeta_{\infty}^{4}$. Notice that
\begin{equation}
  \label{eq:13}
  \sum_{k\in \mathcal{J}}\theta_{k} = - 2s -\sum_{k\in \mathcal{J}}H_{k} = -2s,
\end{equation}
as the sum of residues
$\sum_{k}H_{k} = \sum_{k}\res_{u=u_{k}}(H(u)/\Delta_{u}(u))$ is zero
from the Cauchy theorem. It should be clear that the extra property
\eqref{eq:13} is not generally true for an arbitrary Heun equation; it
is a special condition for the black hole case and reduces the number
of parameters from five  to four, as one of
the monodromies
 is determined in terms of the others.

The $x\to0$ ($\alpha=a/L \rightarrow 0$) limit of
\eqref{eq:heuncanonicalangular} is regular for $L$ fixed, as neither
the $\theta_k$'s nor ``$x(x-1)K_{x}$" diverge, whereas the
coefficients diverge when taking the same limit but with $a$ fixed, instead of $L$. 
The first limit is the Schwarzschild--de Sitter limit, whereas the second one corresponds to the Kerr limit. In the limit $a\rightarrow 0$, we have ${}_s\lambda_{\ell m} \rightarrow \ell(\ell+1)-s^{2}$.

\vspace{0.5cm}

\noindent\textbf{Radial Equation.} 
The Frobenius exponents of the radial solutions are
$\theta_{k} = iW_{k}+s/2$, where
$W_{k} \equiv\res_{r=r_{k}}\frac{W(r)}{\Delta_{r}(r)}$, $k\in\mathcal{J}$, are residues (see appendix~\ref{sec:schw-de-sitt}). Explicitly,
\begin{gather}
\label{eq:heun_thetas} 
\theta_{k} = i\chi^{2} \left( \frac{\omega (r_k^{2} +a^{2})
    -am}{\Delta_{r}'(r_k)}\right)+\frac{s}{2} =
\pm\frac{i}{4\pi}\left(\frac{\omega -\Omega_{k}m}{T_{k}}\right)+\frac{s}{2},
\quad k\in \mathcal{J}.
\end{gather}
The plus or minus sign in \eqref{eq:heun_thetas} arises from the sign
of $\Delta'_r(r_k)$, chosen in such a way that the temperatures
$T_{k}$ are positive \cite{Chambers1994,ghezelbash_entropy_2004}. 
From the residue theorem, we have that
\begin{equation}
\label{eq:thetasum}
  \sum_{k\in \mathcal{J}}\theta_{k} = 2s.
\end{equation}
which implies, from Eq.~\eqref{eq:37}, that $b_2=1-2s$ and
$b_1 = 1-2s+\theta_{\infty}$.
In order to obtain the Heun equation from Eq.~\eqref{eq:radialeq}, we apply the
coordinate transformation
\begin{gather}
  \label{eq:radial-mobius}
z(r)= \zeta_{\infty}\frac{r-r_{C}}{r-r_{--}},\quad  
\zeta_{\infty}
  = \frac{r_{-}-r_{--}}{r_{-}-r_{C}},\quad x= \zeta_{\infty}\frac{r_{+}-r_{C}}{r_{+}-r_{--}},
\end{gather}
corresponding to the mapping
\begin{equation}
\label{eq:radial-mobius-choice}
  r=
    (r_{--},r_{-},r_{+},r_{C},\infty)
  \quad \mapsto\quad
 z= (\infty,1,x,0,\zeta_{\infty}), 
 \quad 0< x < 1,
\end{equation}
and the s-homotopic transformation
\begin{equation}
f(z)=z^{\frac{s}{2}+\theta_0}(z-1)^{\frac{s}{2}+\theta_1}(z-x)^{\frac{s}{2}+\theta_x}(z-\zeta_\infty)^{-1-2s}R_{s,\omega,m}(r).
\label{eq:yintermsofr}
\end{equation}
The
transformations \eqref{eq:radial-mobius} and \eqref{eq:yintermsofr}
result in the Heun equation
\begin{multline}
  \label{eq:heuncanonicalradial}
  \frac{d^2f}{dz^2}+\left(\frac{1-2\theta_{0}}{z}+\frac{1-2\theta_{1}}{z-1}+
    \frac{1-2\theta_{x}}{z-x}\right)\frac{df}{dz}+\\[5pt]
  +\left(\frac{(1-2s)(1-2s+2\theta_{\infty})}{z(z-1)}-\frac{x(x-1)K_x}{z(z-1)(z-x)}\right)f=0.
\end{multline}
According to \eqref{eq:canonical-accessory-parameter-2}, the accessory
parameter in \eqref{eq:heuncanonicalradial} is given by
\begin{multline}
  \label{eq:heun_K0}  
  K_{x} =
  \frac{(1-2s)(\theta_{x}+\theta_{0}-s)+2s}{x}+\frac{(1-2s)(\theta_{x}+\theta_{1}-s)+2s}{x-1}-\\[5pt]
  -\frac{(1+s)(1+2s)}{x-\zeta_{\infty}}
  +\frac{L^{2}}{x(x-1)}\frac{ 
2(2s+1)(s+1)r_{+}^{2}/L^2+ \lamslm -s(1-a^2/L^2)}{(r_{C}-r_{-})(r_{+}-r_{--})}.
\end{multline}
Notice that we use some of the same symbols in
  the radial and in the angular equations, but it should be clear from the
  context whether such symbol corresponds to the radial or to the
  angular case.

The limits $x\rightarrow 0$ and $x\rightarrow 1$ correspond to the two
possible extremal limits: $r_+\to r_C$ and $r_+\to r_-$,
respectively. If we do not make any assumptions on $\omega$, two of
the $\theta_k$'s diverge in each of these limits and so the resulting
confluence yields an irregular singular point (confluent Heun
equation). However, if we expand
$\omega = m\Omega_{+} +x \tilde{\omega}+\mathcal{O}(x^{2})$ (or
starting the series with $m\Omega_{C}$), with some coefficient
$\tilde{\omega}$, and take the limit $x\rightarrow 0$ in
Eq.~\eqref{eq:heuncanonicalradial}, then the $\theta_k$'s are all
finite and we obtain a regular singularity from the coalescence of
singularities.  We can also expand
$\omega = m\Omega_{+} +(1-x) \tilde{\omega}+\mathcal{O}((x-1)^{2})$
(or starting with $m\Omega_{-}$), take the limit $x \rightarrow 1$ to
reach the same conclusion.\footnote{These observations are compatible
  with the definition of confluence in \cite{slavyanov2000}: a
  coalescence of singularities only increases the rank of the
  resulting singularity if some of the parameters of the equation
  diverge as $x \rightarrow 0$ (or, similarly, as $x \rightarrow
  1$ or $x \rightarrow \infty$).} As we show below, this procedure for $x\rightarrow 0$ gives
the near-horizon, near-extremal limit associated to the rotating
Nariai limit.

\subsection{Kerr--de Sitter Connection Coefficients and Greybody Factor}
\label{sec:conn-coeff-quas}

A standard approach for analytically calculating (approximations to)
scattering coefficients is to match series expansions near each
singular point of the radial ODE in an intermediate region. In many
cases, the radial ODE can be solved exactly in the asymptotic regions
in terms of known special functions (such as hypergeometric functions)
- see, e.g., \cite{detweiler1980black,Harmark2010,yang2013quasinormal}
for QNMs in near-extremal Kerr. However, when trying that approach for
the Kerr--(Anti-)de Sitter family of solutions, one obtains a Heun or
confluent Heun equation in the simplest of the cases. Therefore, in
order to reduce the ODEs in these cases to the hypergeometric
equation, one needs to make some extra assumption, such as low
frequency \cite{Castro2010} or large (Anti)-de Sitter radius
\cite{Crispino:2013pya}. Another approach for calculating scattering
coefficients in the case of Heun equations is the so-called
Mano--Suzuki--Takasugi (MST)
method~\cite{Mano:1996vt,Sasaki:2003xr,casals2018perturbations,ronveaux1995heun}.
This method consists of expressing the solutions as infinite series of
hypergeometric functions, and it was used in
\cite{Suzuki1998,Suzuki1999,Suzuki2000} in Kerr--de Sitter.  Given a
certain augmented convergence condition, these series solutions
converge in an ellipsis with two singular points as foci.

Here we take a different path and use the monodromy approach to
scattering amplitudes, first discussed in \cite{Castro2013b} and later
generalized in \cite{Novaes2014c,daCunha:2015ana,daCunha:2015fna}. An
alternative monodromy approach for calculating QNMs with large
imaginary part has also been developed in, e.g.,
\cite{Motl2003,PhysRevD.97.024048}.
\\

\noindent\textbf{Ingoing and Outgoing Solutions.}  Let us first focus on the
radial equation \eqref{eq:radialeq} supposing that we know the angular eigenvalue
$\lamslm$. For definiteness, we consider the problem of connecting
a local solution near the black hole horizon $
r_{+}$
 to a local solution near the
cosmological horizon $ 
r_{C}$. 
Let us define  two linearly
independent solutions  
$\rlocal{j}{\pm}(r)$
by their local behaviour 
 near $r=r_{j}$, for $j=+,C,$
 as
\begin{equation} 
  \label{eq:R+/- monodromies} 
  \begin{aligned}
    && \rlocal{+}{\pm}(r) &= (r-r_{+})^{-\frac{s}{2} \pm \theta_{+}
    }(1+\mathcal{O}(r-r_{+})),& \theta_{+}&=
    i\frac{\tilde{\omega}_{+}}{2\kappa_{+}}
    +\frac{s}{2},  &&\\[5pt]
    && \rlocal{C}{\pm}( r) &= (r-r_{C})^{-\frac{s}{2} \pm \theta_{C}
    }(1+\mathcal{O}(r-r_{C})),& \theta_{C} &=
    -i\frac{\tilde{\omega}_{C}}{2\kappa_{C}} +\frac{s}{2}, &&
  \end{aligned}
\end{equation}
where 
\begin{equation} \label{eq:tildeomega,kappa}
\tilde{\omega}_{j} \equiv \omega -m \Omega_{j}, \quad
\kappa_{j}\equiv 2\pi T_{j}
\end{equation}
and $\kappa_{j}$ is the surface gravity (see Eq.~\eqref{eq:heun_temps}
for the angular velocity and temperature definitions and
Eq. \eqref{eq:heun_thetas} for the $\theta$'s).

The notion of ingoing and outgoing solutions is essential to define a
scattering problem. Here these notions should match the ingoing and
outgoing null coordinates in Kerr--de Sitter, $(v,u)$
respectively:
\begin{equation}
  \label{eq:in-out-coordinates}
  dt = dv - \frac{\chi^{2}(r^{2}+a^{2})}{\Delta_{r}}dr,\quad   dt = du + \frac{\chi^{2}(r^{2}+a^{2})}{\Delta_{r}}dr,
\end{equation}
which in turn motivate the definition of the \emph{radial tortoise
  coordinate} $r_{*}$ as
\begin{equation}
  \label{eq:tortoise-coordinate}
  dr_{*} =  \frac{\chi^{2}(r^{2}+a^{2})}{\Delta_{r}}dr.
\end{equation}
This implies that $r_{*}\to +\infty$ as $r\to r_C$ and $r_{*}\to -\infty$ as $r\to r_+$.
Partly in terms of $r_{*}$, the above local solutions may be written as
\begin{equation} 
  \label{eq:R+/- tortoise} 
    \rlocal{j}{\pm}(r) \propto (r-r_{j})^{-\frac12(s\mp s)}e^{\pm i
      \tilde{\omega}_{j}r_{*}}(1+\mathcal{O}(r-r_{j})),\quad j=C,+. 
\end{equation}
Therefore, travelling waves have the form $\exp(-i(\omega t \pm 
\tilde{\omega}_{j} r_{*}))$ and the phase velocity depends on the
relative sign of the real part of the frequencies, i.e., 
\begin{equation}
  \label{eq:in-out-conditions}
 \begin{aligned}
   \Re(\omega)\Re(\tilde{\omega}_{j}) >0 &\entao \rlocal{j}{+}\,
   \text{is
     outgoing and}\, \rlocal{j}{-}\,\text{is ingoing at $r=r_{j}$},\\[5pt]
   \Re(\omega)\Re(\tilde{\omega}_{j}) < 0 &\entao \rlocal{j}{+}\,
   \text{is
     ingoing and}\, \rlocal{j}{-}\,\text{is outgoing at $r=r_{j}$.}\\[5pt]
  \end{aligned}
\end{equation}
In our case, as $r_{C} \geq r_{+} >0$, we have that
$\Omega_{+} \geq \Omega_{C} $. This allows for the possibility
$|m|\Omega_{+}> |\Re(\omega)| > |m|\Omega_{C}$, which for
$\omega \in \mathbb{R}$ corresponds to the superradiant regime, valid
only for bosons
\cite{tachizawa1993superradiance,Suzuki2000}. The physical definition
of superradiance is that the net flux of radiation with respect to
each horizon should have alternate signs. Given that the notion of
ingoing and outgoing is equivalent to the sign of the radiation flux, for
$\omega\in \mathbb{R}$, we have
\begin{equation}
  \label{eq:superradiance}
  \begin{aligned}
    \tilde{\omega}_{C}\,\tilde{\omega}_{+} &\geq 0 \entao
    \text{no superradiance},\\[5pt]
  \tilde{\omega}_{C}\,\tilde{\omega}_{+}  &< 0 \entao \text{superradiance}.
  \end{aligned}
\end{equation}
For further details, we refer the reader to \cite{Suzuki2000}.  Notice that definitions
\eqref{eq:in-out-conditions} and \eqref{eq:superradiance} are invariant under
$(\omega,m)\rightarrow (-\omega,-m)$.
\\

\noindent\textbf{Connection Coefficients for Radial Solutions.} We now wish to
obtain formulas relating the local solutions $R_{\pm}^{(j)}$, $j=+,C$.
This means writing a set of local solutions at a singular point as a linear combination
of a set of local solutions at another singular point in terms of certain \emph{connection coefficients}:
transmission coefficients $\mathcal{T}_{s}$ and reflection
coefficients $\mathcal{R}_{s}$. These coefficients
become scattering coefficients if the frequency is real, in which case
we can consistently define the radiation flux at each one of the
horizons. We classify the scattering problem between $r_{+}$ and
$r_{C}$ into two possibilities with respect to the boundary
conditions: purely ingoing or purely outgoing solutions at one of the
horizons.

We define the \emph{IN} mode  
   by imposing the boundary conditions
\cite{Harmark2010},
\begin{equation}
  \label{eq:classical_bdycond}
  \uin{+} \sim\left\{\begin{aligned}
     & \frac{1}{\mathcal{T}_{s}}\rlocal{C}{-}- \frac{\mathcal{R}_{s}}{\mathcal{T}_{s}}\, \rlocal{C}{+},&\quad
    r\rightarrow r_{C},\\[5pt]
     & \rlocal{+}{-} ,&\quad r\rightarrow
    r_{+},
  \end{aligned}\right.
\end{equation}
where $\mathcal{T}_{s}=\mathcal{T}_{s}(\omega,m)$ and
$\mathcal{R}_{s}=\mathcal{R}_{s}(\omega,m)$ are complex connection
coefficients. If $\Re(\omega)\Re(\tilde{\omega}_{j})>0$, this
corresponds to a purely ingoing solution at $r_{+}$. We can obtain
another linearly independent solution from
\eqref{eq:classical_bdycond} via the discrete transformation
\begin{equation}
  \label{eq:pt-reversal}
  \star: f
  _{s}(\omega,m) \mapsto f^{\star}_{s}(\omega,m) \equiv f_{-s}(-\omega,-m),
\end{equation}
and multiplying the solution by $\Delta_{r}^{-s}$, i.e.,
\begin{equation}
  \label{eq:classical_bdycond2}
  \uout[]{+} \defeq \Delta_{r}^{-s}
  (\uin{+})^{\star} \sim \left\{\begin{aligned}
      & \left(\Delta_{r}'(r_{C})\right)^{-s}
      \left(
        \frac{1}{\mathcal{T}^{\star}_{s}}\rlocal{C}{+}- \frac{\mathcal{R}^{\star}_{s}}{\mathcal{T}^{\star}_{s}}\,\rlocal{C}{-}
      \right),&\quad
      r\rightarrow r_{C},\\[5pt]
      & \left(\Delta_{r}'(r_{+})\right)^{-s}\,\rlocal{+}{+}
      ,&\quad r\rightarrow r_{+}.
  \end{aligned}\right.
\end{equation}
   If $\omega \in \mathbb{R}$, then
$(\omega,m) \rightarrow (-\omega,-m)$ is equivalent to taking the
complex conjugate solution, and so
\begin{equation}
  \label{eq:19}
\uout{+}= \Delta_{r}^{-s}
  (\uin[-]{+})^{*}.
\end{equation}
This implies that $\mathcal{R}^{\star}_{s} = \mathcal{R}_{-s}^{*}$ and
$\mathcal{T}^{\star}_{s} = \mathcal{T}_{-s}^{*}$ when $\omega \in \mathbb{R}$.

The ``Wronskian" between two solutions is
\begin{equation}
  \label{eq:radiation_flux}
  \mathcal{W}_{s}[u_{1}; u_{2}] \equiv -i\Delta^{s+1}_{r}(r)\left(u_{1}\Dfrac{u_{2}}{r} - u_{2}\Dfrac{u_{1}}{r}\right).
\end{equation}
It is a simple fact that this current is constant  in $r$, as long as 
$u_{1}$ and $u_{2}$ obey the same radial equation \eqref{eq:radialeq}.
Evaluating the quantities at $r=r_{+}$, we obtain the following
expression:
\begin{align}
  \label{eq:16}
\mathcal{W}_{\text{hor}}^{(+)} \equiv\mathcal{W}_{s}[\uin{+} ;\uout{+}]\Big\lvert_{r=r_{+}} =   -2 i\theta_{+} \Delta_{r}'(r_{+}),
\end{align}
whose real part represents the flux of radiation transmitted into the black hole
when $\omega \in \mathbb{R}$.  Evaluating the quantities at $r=r_{C}$,
yields for the total flux,
\begin{align}
  \label{eq:18}
  \mathcal{W}_{\text{cosm}}^{(+)} \equiv \mathcal{W}_{s}[\uin{+} ; \uout{+}]\Big\lvert_{r=r_{C}}=  2 i\theta_{C} \Delta_{r}'(r_{C})\frac{(\mathcal{R}_{s}\mathcal{R}^{\star}_{s}-1)}{\mathcal{T}_{s}\mathcal{T}^{\star}_{s}}=\mathcal{W}_{\text{ref}}^{(+)}+\mathcal{W}_{\text{in}}^{(+)},
\end{align}
where
\begin{equation}
  \mathcal{W}_{\text{in}}^{(+)}\equiv- 2 i\theta_{C} \Delta_{r}'(r_{C})\frac{1}{\mathcal{T}_{s}\mathcal{T}^{\star}_{s}}, \quad
  \mathcal{W}_{\text{ref}}^{(+)}\equiv 2 i\theta_{C}
  \Delta_{r}'(r_{C})\frac{\mathcal{R}_{s}\mathcal{R}^{\star}_{s}}{\mathcal{T}_{s}\mathcal{T}^{\star}_{s}}.
\end{equation}
Because the flux is conserved, 
$\mathcal{W}_{\text{cosm}}^{(+)}=\mathcal{W}_{\text{hor}}^{(+)}$ and
thus
\begin{equation}
  \label{eq:conservationcurrent1}
  \mathcal{R}_{s} \mathcal{R}^{\star}_{s} +\tau_{s}
  \mathcal{T}_{s}\mathcal{T}^{\star}_{s} = 1, \quad \tau_{s}(\omega,m)\equiv \frac{\theta_{+}\Delta_{r}'(r_{+})}{\theta_{C}\Delta_{r}'(r_{C})}=  
     \frac{2\chi^{2}(r_{+}^{2}+a^{2}) \tilde{\omega}_{+}-is \Delta'_{r}(r_{+})}{2\chi^{2}(r_{C}^{2}+a^{2}) \tilde{\omega}_{C}-is \Delta'_{r}(r_{C})}.
\end{equation}
The \emph{greybody factor} $G_{s}(\omega, m)$ is  defined to be the
ratio between the horizon flux and the incoming flux
\begin{equation}
  \label{eq:kerr-ds-greybody}
  G_{s}(\omega,m) = \frac{\mathcal{W}^{(+)}_{\text{hor}}}{\mathcal{W}^{(+)}_{\text{in}}} =
  \tau_{s}(\omega,m)\,\mathcal{T}_{s} (\omega,m)\mathcal{T}^{\star}_{s} (\omega,m).
\end{equation}
For $\omega\in \mathbb{R}$ and outside the superradiant regime, the
greybody factor must be real and positive. This has been proven in
\cite{Suzuki2000} using the Teukolsky--Starobinsky identities, which
relate $\mathcal{T}_{-s}$ with $\mathcal{T}_{s}$. This procedure also
makes the superradiance regime explicit. We refer the reader to \cite{Suzuki2000}
for details.

We can do the same exercise for the \emph{UP} mode, defined via the following boundary condition at the cosmological horizon:
\begin{equation}
  \label{eq:semiclassical_bdycond}
  \uout{C} \sim\left\{\begin{aligned}
      & \rlocal{C}{+} ,&\quad
      r\rightarrow r_{C},\\
      &\frac{1}{\mathcal{\tilde{T}}_{s} }\rlocal{+}{+} - \frac{\mathcal{\tilde{R}}_{s}}{\mathcal{\tilde{T}}_{s}}\, \rlocal{+}{-} ,&\quad r\rightarrow
      r_{+},
  \end{aligned}\right.
\end{equation}
and the reflected mode, which is defined via the boundary condition:
\begin{equation}
  \label{eq:semiclassical_bdycond2}
  \uin{C} \sim\left\{\begin{aligned}
      & \left(\Delta_{r}'(r_{C})\right)^{-s}\, \rlocal{C}{-},&\quad
      r\rightarrow r_{C},\\
      &\left(\Delta_{r}'(r_{+})\right)^{-s}\, \left(\frac{1}{\tilde{\mathcal{T}_{s}}^{\star}}\rlocal{+}{-}- \frac{\tilde{\mathcal{R}}_{s}^{\star}}{\tilde{\mathcal{T}_{s}}^{\star}}\, \rlocal{+}{+} \right) ,&\quad r\rightarrow
      r_{+},
  \end{aligned}\right.
\end{equation}
 where
$\tilde{\mathcal{T}_{s}}^{\star}\equiv\mathcal{\tilde{T}}_{-s}(-\omega,-m)$
and
$\tilde{\mathcal{R}}_{s}^{\star}\equiv\mathcal{\tilde{R}}_{-s}(-\omega,-m)$. We
then obtain similar results for the flux as above for the purely
in/outgoing solutions on the event horizon:
\begin{align}
  \label{eq:68}
  \mathcal{W}_{\text{cosm}}^{(C)} &\equiv\mathcal{W}_{s}[\uin{C} ;\uout{C}]\Big\lvert_{r=r_{C}} =   -2 i\theta_{C} \Delta_{r}'(r_{C}),\\[5pt]
  \mathcal{W}_{\text{hor}}^{(C)} &\equiv \mathcal{W}_{s}[\uin{C} ; \uout{C}]\Big\lvert_{r=r_{+}}=  2 i\theta_{+} \Delta_{r}'(r_{+})\frac{(\tilde{\mathcal{R}}_{s}\tilde{\mathcal{R}}_{s}^{\star}-1)}{\tilde{\mathcal{T}}_{s}\tilde{\mathcal{T}_{s}^{\star}}}=\mathcal{W}_{\text{ref}}^{(+)}+\mathcal{W}_{\text{in}}^{(+)}.
\end{align}
Then, $\mathcal{W}_{\text{cosm}}^{(C)}=\mathcal{W}_{\text{hor}}^{(C)}$ implies that
\begin{equation}
  \label{eq:conservationcurrent2}
  \tilde{\mathcal{R}}_{s} \tilde{\mathcal{R}}_{s}^{\star} +\tau_{s}^{-1}
  \tilde{\mathcal{T}}_{s}\tilde{\mathcal{T}_{s}^{\star}} = 1.
\end{equation}

Now, writing $\tau_{s}= |\tau_{s}|e^{i\phi}$,
$\phi = \arg (\tau_{s})$, we define the vectors
\begin{equation}
 \label{eq:70}
  \begin{aligned}
   \Phi_{+}(r) &\equiv |\tau_{s}|^{-1/2}e^{-i\phi/2}\,(\uin{+}(r)\;\;
  \uout{+}(r)),\\[5pt]
  \Phi_{C}(r) &\equiv \,(\uin{C}(r)\;\; \uout{C}(r))
\end{aligned}
\end{equation}
from the
above  linearly independent solutions.  Then it is straightforward to show
that \cite{Harmark2010,Castro2013b}
\begin{equation}
  \label{eq:connection}
   \Phi_{+}\mathcal{M}_{+C} =\Phi_{C}
\end{equation}
with the \emph{connection matrix} given by
\begin{equation}
  \label{eq:connectionmatrix}
\mathcal{M}_{+C} =  \begin{pmatrix}
    (\bar{\cal T}^{\star}_{s})^{-1} &  \bar{{\cal R}}_{s}(\bar{{\cal T}}_{s})^{-1}\\[5pt]
    \bar{{\cal R}}^{\star}_{s}(\bar{{\cal T}}^{\star}_{s})^{-1} & (\bar{{\cal T}}_{s})^{-1}
  \end{pmatrix}
\end{equation}
with
\begin{equation}
  \label{eq:20}
  \begin{aligned}
\bar{\mathcal{T}}_{s} &\equiv   |\tau_{s}|^{1/2}e^{i\phi/2}\left(\Delta_{r}'(r_{C})\right)^{-s} \mathcal{T}_{s}, &  
\bar{\mathcal{R}}_{s} &\equiv \left(\Delta_{r}'(r_{C})\right)^{-s}  \mathcal{R}_{s} ,\\[5pt]
\bar{\mathcal{T}}^{\star}_{s}  &\equiv |\tau_{s}|^{1/2}e^{i\phi/2}\left(\Delta_{r}'(r_{C})\right)^{s}\mathcal{T}^{\star}_{s} , & 
\bar{\mathcal{R}}^{\star}_{s}  &\equiv \left(\Delta_{r}'(r_{C})\right)^{s}  \mathcal{R}^{\star}_{s}.
\end{aligned}
\end{equation}
The condition of unit determinant for $\mathcal{M}_{+C}$ is equivalent
to the current conservation \eqref{eq:conservationcurrent1}
\begin{equation}
  \label{eq:ConservationRR+TT=1}
  \bar{\mathcal{R}}^{\phantom{'}}_{s}\bar{\mathcal{R}}^{\star}_{s}+\bar{\mathcal{T}}^{\phantom{'}}_{s}\bar{\mathcal{T}}^{\star}_{s} = 1.
\end{equation}


\subsection{Monodromy Approach to Angular Eigenvalues and
  Quasinormal Modes}  \label{sec:monodromy}

Let us start by reviewing
how we can write greybody factors in terms of monodromies
\cite{Castro2013b,Novaes2014c,daCunha:2015ana,daCunha:2015fna}. They
informed reader eager to skip this technical discussion can find the
main results for the Kerr-de Sitter greybody factor in
\eqref{eq:greybody-kerr-ds}, the quasinormal mode condition
\eqref{eq:qnmcondition-2} and the angular eigenvalue condition \eqref{eq:sigma-angular-eigenvalue-2}.

\begin{figure}
  \centering
  \includegraphics[width=.4\textwidth]{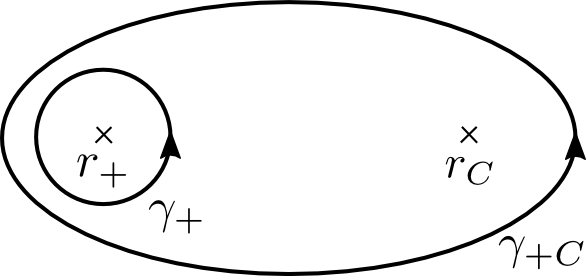}
  \caption{Example of local ($\gamma_+$) and composite  ($\gamma_{+C}$)  monodromy paths in the complex-$r$ plane around the singular points $r=r_+$ and $r_C$.}
  \label{fig:monodromy}
\end{figure}

A general solution of a second order linear ODE can be written as a
linear combination of local solutions, such as those in \eqref{eq:R+/-
  monodromies} for the radial equation. Thus, we can define a vector
of two general solutions by
\begin{equation}
  \label{eq:generalsol}
  \Phi(r) = \Phi_{k}(r)g_{k},
\end{equation}
with $g_{k}\in \mathrm{SL}(2,\mathbb{C})$ constant matrices and
$\Phi_{k}$ denotes a vector of local Frobenius solutions, with
$k=1,\ldots,n$ referring to any of the $n$ singular points $r=r_{k}$
of the ODE. If we go around a positively-oriented loop $\gamma_{k}$ in
the complex-$r$ plane enclosing only one of the singularities
$(r_{k})_{k=1,\ldots,n}$ (see figure~\ref{fig:monodromy}), then $\Phi$
transforms to
\begin{equation}
  \label{eq:46}
  \Phi_{\gamma_{k}}(r) = \Phi(r)M_{k},
\end{equation}
where the initial value $\Phi$  of the solution at the loop is multiplied by a
constant \emph{monodromy matrix} $M_{k}\in \mathrm{SL}(2,\mathbb{C})$.

The matrices $g_{k}$ diagonalize the monodromy matrix $M_{k}$ if
$2\theta_k \notin \mathbb{Z}$, i.e.
\begin{equation}
\label{eq:monodromymatrix}
  M_{k} = g_{k}^{-1}
  \begin{pmatrix}
    e^{2\pi i\theta_{k}} & 0\\
    0 &     e^{-2\pi i\theta_{k}}
  \end{pmatrix}
g_{k}. 
\end{equation}
In principle, there should be an overall phase $e^{i\pi s}$ in
\eqref{eq:monodromymatrix}, obtained from the common factor
$(r-r_k)^{-s/2}$ in the definition \eqref{eq:R+/- monodromies}, but we
absorb it, as a common overall factor, in the solution $\Phi$ to obtain
$\mathrm{SL}(2,\mathbb{C})$ representations of the monodromy
group. Notice that
\begin{equation}
    \Tr M_{k} = 2\cos(2\pi \theta_{k}).
\end{equation} 
Given two
singular points $r_{k}$ and $r_{j}$ with $j,k=1,\ldots,n$ and
$j\neq k$, we wish to study the \emph{connection matrix}
\begin{equation}
  \label{eq:connectiongi}
  \mathcal{M}_{jk} = g^{\phantom{1}}_{j}g_{k}^{-1},
\end{equation} 
where the last equality follows from Eq.~\eqref{eq:generalsol}. Thus,
the connection matrix is constant and determined by the $g_{k}$,
$k=1,\ldots,n$. Choosing a proper normalization for the local 
solutions, in terms of the $g_{k}$, such that $\det \mathcal{M}_{jk}=1$,
we can write this connection matrix in the form
\eqref{eq:connectionmatrix}.

Let us now choose a path $\gamma_{jk}$ which encloses two singular
points $r_{j}$ and $r_{k}$, as in figure~\ref{fig:monodromy}, in such
a way that the general solution changes to
$\Phi_{\gamma_{jk}}(z)=\Phi(z) M_{j}M_{k}$.
We define the \emph{composite monodromy coefficient} $\sigma_{jk}$ by
\begin{equation}\label{eq:compmono}
\Tr \left(M_{j}M_{k}\right) = -2\cos\left(2\pi\sigma_{jk}\right).
\end{equation}
The reason
for the extra minus sign here, in comparison with the definition of
the single traces, is explained in appendix~\ref{sec:altern-deriv}. 
Applying Eqs.~\eqref{eq:monodromymatrix}, \eqref{eq:connectiongi} and
\eqref{eq:compmono} to the case of \eqref{eq:connectionmatrix}, we find
\begin{align}
  \label{eq:greybody}
G_s=  \bar{\mathcal{T}}^{\phantom{'}}_{s}\bar{\mathcal{T}}^{\star}_{s} = 
  \frac{2\sin(2\pi\theta_j)\sin(2\pi\theta_{k})}{\cos (2\pi(\theta_{j}
  -\theta_{k}))+\cos(2\pi\sigma_{jk})}.
\end{align}
This is the greybody factor  for a connection matrix normalized as per
Eq.~\eqref{eq:connectionmatrix}.
\\

\noindent\textbf{Quasinormal Modes.}  For the radial equation
\eqref{eq:radialeq}, according to \eqref{eq:R+/- monodromies} and
\eqref{eq:70}, we can rewrite \eqref{eq:greybody} as
\begin{equation}
  \label{eq:greybody-kerr-ds}
G_s(\omega,m)=  \tau_{s} \mathcal{T}_{s}(\omega,m)\mathcal{T}_{-s}(-\omega,-m)=\frac{\sin(\pi(2\theta_+-s))\sin(\pi(2\theta_{C}-s))} 
  {\cos\left(\pi(\sigma+\theta_+-\theta_{C})\right)\cos\left(\pi(\sigma-\theta_++\theta_{C})\right)},
\end{equation}
where $\sigma\equiv \sigma_{+C}$. We wish to calculate QNM frequencies
in Kerr--de Sitter. These frequencies are defined by the condition that the
modes are purely ingoing at the black hole horizon and purely outgoing
at the cosmological horizon.\footnote{This is in contrast with Kerr,
  where the QNMs are purely outgoing at $r=\infty$.}
According to Eq.~\eqref{eq:classical_bdycond}, the QNM condition
corresponds to $(\mathcal{T}_{s})^{-1} = 0$,
i.e., poles of $\mathcal{T}_{s}$,
which appears in Eq.~\eqref{eq:greybody-kerr-ds}. The poles
$\omega=\omega_p$
on
the right hand side of Eq.~\eqref{eq:greybody-kerr-ds} are
\begin{equation}
  \label{eq:qnmcondition}
  \sigma^{(\pm)}(
  \omega_p
  ) =\pm( \theta_{C}-\theta_{+}) +N+\frac12,\quad
  N\in \mathbb{Z}.
\end{equation}
However, these poles 
may correspond to poles of  $\tau_{s}$ or
$\mathcal{T}^{\star}_{s}=\mathcal{T}_{-s}(-\omega,-m)$ as well.
Not being able to distinguish between poles of $\mathcal{T}_{s}$ and
those of $\mathcal{T}^{\star}_{s}$ or $\tau_{s}$ is a shortcoming of
the monodromy formula above, as it only gives information about the
product of these coefficients. However, below we recover the exact
greybody factor in the rotating Nariai limit $r_{C}\rightarrow r_{+}$
by using Eq.~\eqref{eq:qnmcondition}. This will also allow us to
identify the poles of $\mathcal{T}_{s}(\omega,m)$ with
$N\in \mathbb{Z}^{+}$, and the poles of $\mathcal{T}_{-s}(-\omega,-m)$
with $N \in \mathbb{Z}^{-}$. While the former corresponds to the
actual QNMs, the later corresponds to the reflected QNMs, and thus, in
principle, they are not QNMs. In order to assess the possible poles
coming from $\tau_{s}$, one needs to generalize the
Teukolsky--Starobinsky identities obtained in Ref.~\cite{Suzuki1999}
for complex $\omega$. We leave this for future work.

In section~\ref{sec:access-param-from}, we are going to apply the APE
derived in Ref.~\cite{Lencses:2017dgf} to find QNMs.  Because the APE has
the symmetry $\sigma \rightarrow - \sigma$, which is also compatible
with Eq.~\eqref{eq:greybody},
changing the sign in front of $( \theta_{C}-\theta_{+})$ in
Eq.~\eqref{eq:qnmcondition} is equivalent to letting $N\to
-N-1$. Therefore, from the APE point of view, there is no distinction
between the positive and negative $N$ possible poles in
\eqref{eq:qnmcondition} and, from here on, we choose the QNM condition
to be
\begin{equation}
  \label{eq:qnmcondition-2}
    \sigma^{(+)}(\QNMf) = \theta_{C}-\theta_{+} +N+\frac12,\quad
  N\in \mathbb{Z}^{+}.
\end{equation}
As mentioned above, the condition $N \geq 0$ will be justified in the
rotating Nariai limit in the next subsection.  Finally, it is worth
mentioning that, in this paper, we focus on the radial equation for
which $2\theta_{k} \notin \mathbb{Z}$, and so we do not cater for any
potential QNM frequencies that could correspond to
$2\theta_{k} \in \mathbb{Z}$. This can introduce logarithm
singularities in the solutions and thus change the QNM condition
above.
\\

\noindent\textbf{Accumulation points of Quasinormal Modes}.
As a function of complex frequency, the transmission coefficient may
have not only poles (QNMs) but also branch cuts.  Apart from spurious
(unphysical) branch cuts arising from the angular eigenvalue
\cite{casals2016high,BONGK:2004}, the transmission coefficient may
also have cuts arising from the behaviour of the radial potential
which may have physical consequences.  Let us first write the radial
equation \eqref{eq:radialeq} as
$ \left(\partial^2_{r_*}+\mathcal{V}(r)\right)\mathcal{R}(r)=0$, where
$\mathcal{R}(r)\equiv \Delta_r^{s/2}(r^2+a^2)^{1/2}R_{s,\omega,m}(r)$
and the radial potential $\mathcal{V}(r)$ is straightforwardly
read-off from Eq.~\eqref{eq:radialeq}.  It is easy to check that, if
$r_C\neq r_+$, then
$\mathcal{V}(r)\to \mathcal{V}_{+,C}+\mathcal{O}(e^{\alpha_{+,C}\
  r_*})$, as $r\to r_{+,C}$, for some constants $\mathcal{V}_{+,C}$
and $\alpha_{+}>0$ and $\alpha_{C}<0$.  Based on the arguments
of~\cite{Ching:1995tj} (see also~\cite{PhysRevD.64.104021}), this
asymptotic behaviour does not lead to branch points in the complex
$\omega$-plane for the radial solutions which obey purely-outgoing (or
purely-ingoing) boundary conditions at either $r_+$ or $r_C$.  On the
other hand, in the case $r_+=r_C$, this singular point becomes an
irregular singular point and it is easy to show that, then,
$\mathcal{V}(r)\to \mathcal{\bar V}_{C}+\mathcal{O}(1/r_*)$ as
$r\to r_+=r_C$, for some constant $\mathcal{\bar V}_C$. This is
essentially due to the fact that $r_*=\mathcal{O}(1/(r-r_+))$ as
$r\to r_{+}$ when $r_+=r_C$, as opposed to
$r_*=\mathcal{O}\left(\ln(r-r_+)\right)$ as $r\to r_{+,C}$ when
$r_+\neq r_C$.  Furthermore,
$\Re\left(\sqrt{\mathcal{\bar V}_C}\right)$ possesses a zero at the
superradiant bound frequency (namely, at the frequency such that
$\tilde\omega_C=0$; recall that, in this extremal case, it is
$\tilde\omega_+=\tilde\omega_C$).  According to
Ref.~\cite{Ching:1995tj}, this asymptotic behaviour means that radial
solutions obeying purely-outgoing (or purely-ingoing) boundary
conditions at $r\to r_{+}$ (see Eq.\eqref{eq:R+/- tortoise}) possess a
branch point at $\tilde\omega_C=(\tilde\omega_+=)0$ when $r_+= r_C$.\footnote{We note that the mentioned
  asymptotic behaviours of the radial potential are independent of the
  `centrifugal' barrier.}
Such branch point is carried over to the corresponding transmission
coefficient ($\mathcal{T}^{\phantom{'}}_{s}$, $\mathcal{T}^\star_{s}$,
$\tilde{\mathcal{T}}^{\phantom{'}}_{s}$,
$\tilde{\mathcal{T}}^\star_{s}$).

The above relationship between asymptotic behaviour of the radial
potential and branch points in the complex $\omega$-plane is already
well-known in Kerr.  Indeed, the transmission coefficient in Kerr
possesses a branch point at $\omega=0$, associated with the irregular
singular point at $r=\infty$.  This branch point gives rise to a
power-law tail decay at
late-times~\cite{barack1999late,PhysRevLett.84.10,casals2016high}.
More relevantly to the study here, in extremal ($r_+=r_-$) Kerr, the
transmission coefficient has an extra branch point at
$\tilde\omega_+=0$, which gives
rise~\cite{casals2016horizon,Gralla:2017lto} to the so-called Aretakis
phenomenon at the horizon~\cite{aretakis2012horizon}.  It has been
shown~\cite{detweiler1980black,yang2013quasinormal,hod2008slow,casals2016horizon}
that, in near-extremal Kerr, QNMs accumulate near $\tilde\omega_+=0$,
effectively leading to a branch cut originating at $\tilde\omega_+=0$
in (exactly) extremal Kerr spacetime.  We expect that a similar
accumulation of QNMs occurs in Kerr--de Sitter both near
$\tilde\omega_+=0$ and near $\tilde\omega_C=0$ as the extremal limit
$r_+\to r_C$ is approached.  In section~\ref{sec:quasinormal-modes}, we
show that this is indeed the case.
\\

\noindent\textbf{Renormalized Angular Momentum Parameter}. In Ref.~\cite{Suzuki1998}, the authors use a series of hypergeometric functions to express the solution of Heun equations. The hypergeometric functions depend on the so-called \textit{renormalized angular momentum parameter} $\nu$.
As discussed in \cite{daCunha:2015fna}, $\nu$ parametrizes the monodromy at infinity  of the hypergeometric functions. Because the hypergeometric differential equation possesses only three regular singular points, we see that its monodromy at infinity is equivalent to its composite monodromy. 
In its turn, this composite monodromy of the hypergeometric differential equation is equal to the composite monodromy 
$\sigma_{01}$ of the full Heun equation modulo $2n$, where $n\in \mathbb{Z}$. That results in the following relationship: $\sigma_{01}=
2(\nu-\theta_0+\theta_1) + 1\, (\text{mod}\, 2n)$, where $n\in \mathbb{Z}$.


\vspace{0.5cm}

\noindent\textbf{Angular Eigenvalues.} In order to find the angular eigenvalues
$\lamslm$, we need to impose a boundary condition on the solutions of
the spin-$s$ angular equation \eqref{eq:angulareq}. We are interested
in solutions which are regular in the domain $u\in [-1,1]$,
corresponding to $z\in [0,x]$ in the Heun equation
\eqref{eq:heuncanonicalangular}. It is well-known in the literature \cite{Suzuki1998}
that $(s,\ell,m)$ are either all integers or half-integers and that
\begin{equation}
  \label{eq:angular-constraint}
  \ell \geq \max(|s|,|m|).
\end{equation}
Therefore, $2\theta_{0}=m-s$ and $2\theta_{x}=-m-s$ are integers and so
the difference between Frobenius exponents for solutions at either
$z=0$ or $z=x$ is an integer. This opens up the possibility of a
logarithmic singularity in one of the local solutions at each singular
point. According to standard Frobenius analysis
\cite{Iwasaki:1991,slavyanov2000}, the local solutions in this case
are
\begin{equation}
  \label{eq:1}
  \left\{\begin{aligned}
    S_{\text{reg}}^{(k)} &\sim (u-u_{k})^{|\theta_{k}|},\\[5pt]
    S_{\text{irr}}^{(k)}&\sim (u-u_{k})^{-|\theta_{k}|}+
    A_{k}(u-1)^{|\theta_{k}|}\log(u-u_{k}),
\end{aligned}\right.\quad  u\rightarrow u_{k},\, (k=0,x),
\end{equation}
for some coefficients $A_{k}$.  If we define the vectors
$S^{(k)} \equiv (S_{\text{reg}}^{(k)}\,\, S_{\text{irr}}^{(k)})$, we
find the following form for their monodromy matrices (recall
Eq.~\eqref{eq:46}) at, respectively, $z=0$ and $x$:
\begin{equation}
  \label{eq:angular-monodromy}
  J_{0} \equiv e^{2\pi i |\theta_{0}|}\begin{pmatrix}
   1 & 1\\
    0 & 1
  \end{pmatrix},\quad
  J_{x} \equiv
e^{2\pi i |\theta_{x}|}\begin{pmatrix}
   1 & 1\\
    0 & 1
  \end{pmatrix}.
\end{equation}
We thus define a general solution of the master angular equation \eqref{eq:angulareq} as
\begin{equation}
  \label{eq:angular-solution-regular}
  S^{(\text{reg})}_{s,\ell,m} \sim
  \left\{\begin{aligned}     
      &A_{\text{reg}}S_{\text{reg}}^{(x)}- A_{\text{irr}}
      S_{\text{irr}}^{(x)}, &&  u\rightarrow 1,\\[5pt] & 
      S^{(0)}_{\text{reg}},& &u\rightarrow
      -1,
       \end{aligned}\right.
\end{equation}
and 
\begin{equation}
  \label{eq:angular-solution-irregular}
  S^{(\text{irr})}_{s,\ell,m} \sim
  \left\{\begin{aligned}     
      &\tilde{A}_{\text{irr}}S_{\text{irr}}^{(x)}- \tilde{A}_{\text{reg}}
      S_{\text{reg}}^{(x)}, &&  u\rightarrow 1,\\[5pt] & 
      S^{(0)}_{\text{irr}},& &u\rightarrow
      -1,
       \end{aligned}\right.
\end{equation}
for some coefficients $A_{\text{reg}}=A_{\text{reg}}(\omega,m)$ and
$A_{\text{irr}}=A_{\text{irr}}(\omega,m)$, and similarly for the tilde
connection coefficients.  Therefore, in this angular case, we have a
connection problem similar to the QNM one above.  Given that the
monodromy matrices of the general solution
$(S^{(\text{reg})}_{s,\ell,m}\,\,S^{(\text{irr})}_{s,\ell,m}) =
S^{(k)}g_k$ are given by $M_{k}=g_{k}^{-1} J_{k} g_{k}$, with $k=0,x$,
we calculate $\Tr \left(M_{0}M_{x}\right)=2\cos (2\pi\sigma)$ using
Eqs.~\eqref{eq:connectionmatrix}, \eqref{eq:connectiongi},
\eqref{eq:compmono} and \eqref{eq:angular-monodromy} to obtain
\begin{equation}
  \label{eq:angular-condition-eq}
  A_{\text{irr}}
  =\pm\sqrt{2(1+(-1)^{|m-s|+|m+s|}\cos(2\pi\sigma))}.
\end{equation}
The regularity condition corresponds to choosing
$ A_{\text{irr}}=0$ in
Eq.~\eqref{eq:angular-solution-regular}, which is equivalent to
\begin{align}
  \label{eq:sigma-angular-eigenvalue}
   \sigma&=\ell+\frac12,\quad\ell \in \mathbb{Z},\, \text{if $\max(|s|,|m|)
                \in \mathbb{Z}$};\quad \ell \in \mathbb{Z}+\frac12, \, \text{if $\max(|s|,|m|)
      \in \mathbb{Z}+\frac{1}{2}$}.
\end{align}
As we are going to see in section \ref{sec:access-param-from}, the parameter $\ell$ introduced
in Eq.~\eqref{eq:sigma-angular-eigenvalue}
is the same $\ell$ that labels  the angular eigenvalues $\lamslm$ and
so the same as in 
Eq.~\eqref{eq:angular-constraint}.
Taking this into account, the
condition Eq.~\eqref{eq:sigma-angular-eigenvalue} together with the
constraint \eqref{eq:angular-constraint} yield
\begin{equation}
  \label{eq:sigma-angular-eigenvalue-2}
  \sigma = \ell + \frac12,\quad  \ell \geq \max(|s|,|m|).
\end{equation}

In the next section, we shall use the constraints
\eqref{eq:qnmcondition-2} and \eqref{eq:sigma-angular-eigenvalue-2} on
the composite monodromy parameter $ \sigma_{0x}$ in order to find,
respectively QNM frequencies and angular eigenvalues.


\subsection{Quasinormal Modes in the Rotating Nariai Limit}
\label{sec:near-cosm-horiz}

In this subsection, we focus on the region between $r_{+}$ and $r_{C}$
(thus discarding the other horizons) and study the extremal limit
$r_+ \rightarrow r_C$, which is also called the rotating Nariai
limit~\cite{Booth1999,anninos_holography_2009,anninos2010sitter}.  The
rotating Nariai limit is applied directly on the metric and it yields
a new geometry which is topologically equivalent to
$dS_2 \ltimes S^2$.\footnote{The semi-direct product $\ltimes$ here
  means that the isometries of $dS_{2}$ must be accompanied by $S^{2}$
  isometries so that they are isometries of the full spacetime.} Here
we review this limit by finding the QNM frequencies of a scalar field
which possesses general coupling with the curvature of this new
extremal-limit geometry. This slightly extends the results in
Ref.~\cite{anninos2010sitter} for minimal coupling to general
coupling.  We also show how to obtain the scalar radial equation of
the rotating Nariai metric directly from the spin-$s$ master radial
equation \eqref{eq:radialeq} of Kerr--de Sitter. The result for QNMs
that we obtain in this limit here will be used as a check of the
leading-order term in the higher-order expansion in the extremality
parameter $\epsilon = (r_{C}-r_{+})/L$ that we obtain in the next
section.

The metric in the rotating Nariai limit can be obtained from the Kerr--de Sitter metric in
Eq.~\eqref{eq:kerradsmetric} via the following coordinate
transformation:\footnote{The definitions used in
  \cite{anninos2010sitter} can be recovered by letting
  $\epsilon \rightarrow \frac{r_{C}}{L}\epsilon$ and
  $\bar{\kappa} \rightarrow \frac{L}{\chi^{2}r_{C}}\bar{\kappa}$.}
\begin{equation}
  \label{eq:nhekcoords}
  \bt \equiv \bar{\kappa}\,\epsilon t, \quad \by \equiv
  \frac{r-\bar{r}_{+}}{\epsilon
   L 
  },\quad \bphi \equiv \phi - \Omega_{+}t,
\end{equation}
where
\begin{equation}
  \label{eq:44}
  \bar{\kappa}\equiv\frac{(\bar{r}_+-\bar{r}_-)(3\bar{r}_+ + \bar{r}_-)}{\chi^{2}L (a^2 + \bar{r}_+^2)},
\end{equation}
where $\bar{r}_{\pm}\equiv \left.r_{\pm}\right|_{\epsilon=0}$.
The location of the cosmological horizon is $\bar y=1$. 
If we now take $\epsilon \rightarrow 0$  in the metric
\eqref{eq:kerradsmetric} after the coordinate change
\eqref{eq:nhekcoords}, we find~\cite{anninos2010sitter}
\begin{equation}
  \label{eq:rotatingnariai}
  ds^2 = \Gamma(\theta)\left(-\by(1-\by)d\bt^2 +
    \frac{d\bar y^2}{\by(1-\by)} + \alpha(\theta)d\theta^2\right)
  + \gamma(\theta)(d\bphi + \by\, \bar{\Omega} d\bt)^2 ,
\end{equation}
where $\by \in (0,1)$, $\bphi \sim \bphi + 2\pi$ and
\begin{gather}
  \label{nariaiparam1}
  \Gamma(\theta) \equiv \frac{\rho_+^2 L}{\chi^{2}\bar{\kappa} (a^2 + \bar{r}_+^2)} ,
  \quad \alpha(\theta) \equiv \frac{\chi^{2}  \bar{\kappa} (a^2 + \bar{r}_+^2)}{L\Delta_\theta} , \quad \gamma(\theta)
  \equiv\frac{\Delta_\theta(\bar{r}_+^2 + a^2)^{2}\sin^2\theta}{\rho_+^2\chi^4} ,\\[5pt]
\label{nariaiparam2}
  \bar{\Omega} \equiv\frac{2a\bar{r}_+ L}{\bar{\kappa} (a^2 + \bar{r}_+^2)^2}, \quad \rho_+^2 \equiv \bar{r}_+^2 + a^2
  \cos^2\theta .
\end{gather}

Let us now consider a scalar field, with general coupling parameter
$\xi$, propagating on the rotating Nariai geometry of  Eq.~\eqref{eq:rotatingnariai}.
The corresponding Klein--Gordon equation separates by variables and, for a field mode  $e^{-i\bar{\omega}\bt+im\bphi}S_m(\theta)R_{\omega  m}(\by)$,
with $\bar{\omega}\in\mathbb{C}$ and $m\in\mathbb{Z}$,
the resulting angular and radial equations are, respectively,
\begin{align}
  \label{eq:rotating-nariai-angular-eq}
  \left[
  \Dfrac{}{\theta}(\Delta_{\theta} \sin{\theta})\Dfrac{}{\theta} +
  \frac{m^{2}\chi^{4}\rho_{+}^{4}}{(\bar{r}_{+}^{2}+a^{2})\Delta_{\theta} \sin\theta}+\left(
\frac{12\xi}{L^{2}}a^{2}\cos^{2}\theta -\lambda_{\ell m}
  \right)\sin\theta \right]S_{m}(\theta) &= 0,
  \\[10pt]
\label{eq:rotating-nariai-radial-eq}
  \left[ \frac{d}{d\by}\by(\by-1)\frac{d}{d\by} +
  \frac{\left( \bar{\omega} + m \bar{\Omega} y  \right)^2}{\by(\by-1)} + \bar{\lambda}_{\ell m}  \right] R_{\omega  m}(\by) &= 0,
\end{align}
where 
\begin{equation}
  \label{eq:48}
  \bar{\lambda}_{\ell m}
  \equiv \frac{
    L^2}{(3\bar{r}_+ + \bar{r}_-)(\bar{r}_+ - \bar{r}_-)}
  \left(
    \lambda_{\ell m} +\frac{12\xi }{L^{2}}\bar{r}_+^2
  \right),
\end{equation}
and $\lambda_{\ell m}$ is the eigenvalue of the angular equation
\eqref{eq:rotating-nariai-angular-eq}. Notice that
\eqref{eq:rotating-nariai-angular-eq} does not depend on the frequency
$\bar{\omega}$. Therefore, the angular eigenvalue $\lambda_{\ell m}$
does not depend on $\bar{\omega}$
either. 

The radial equation \eqref{eq:rotating-nariai-radial-eq} has three regular
singular points at $\by = 0,1$ and $\infty$ and can be transformed to
the hypergeometric equation~\cite{anninos2010sitter}
\begin{equation}
  \label{eq:hypergeometric-eq} \left[\frac{d^2}{d\by^2}+\left(\frac{C}{\by}+\frac{A+B+1-C}{\by-1}\right)\frac{d}{d\by}
  +\frac{A B}{\by(\by-1)}\right]\bar{f}_{\omega m}=0,
\end{equation}
where $\bar{f}_{\omega m}(\by)\equiv\by^{(1-C)/2}(\by-1)^{(C-A-B)/2}R_{\omega m}(\by)$, with parameters
\begin{gather}
  \label{eq:ABC-nariai}
  A \equiv \frac{1+i\beta}{2}+im \bar{\Omega},\quad B \equiv \frac{1-i\beta}{2}+im\bar{\Omega},\quad C \equiv
  1-2i\bar{\omega},
\end{gather}
and
\begin{equation}
  \label{eq:61}
  \beta \equiv
  \sqrt{4(m^{2}\bar{\Omega}^{2}+\bar{\lambda}_{\ell m})-1}.
\end{equation}
The Frobenius exponents of \eqref{eq:rotating-nariai-radial-eq} $\theta_{0,1,\infty}$  at, respectively, $\by = 0,1$ and $\infty$
can be found from these
parameters to be 
given via
\begin{gather}
  \label{eq:60}
  2\theta_{1} = C-A-B =-2 i\left(\bar{\omega}+m \bar{\Omega}\right),\quad 2\theta_{0} =1-C = 
  2i\bar{\omega},\quad  2\theta_{\infty} =A-B= i\beta.
\end{gather} 
Let $R_{\text{in}}$ be a solution of
\eqref{eq:rotating-nariai-radial-eq} defined via the boundary
condition
\begin{equation}
\label{eq:in-out-solutions}
 R_{\text{in}}
\sim \left\{
\begin{aligned}
  &(\by)^{-\theta_{0}} &\quad\text{as}\quad &\by \rightarrow 0, \\
  &\Aout(1-\by)^{-\theta_{1}}+ \Ain(1-\by)^{\theta_{1}} &\quad\text{as}\quad &\by \rightarrow 1,
\end{aligned}
\right.
\end{equation}
for some coefficients $\Aout$ and $\Ain$. 
Assuming that $\Re\bar{\omega} \geq 0$,
Eq.~\eqref{eq:in-out-solutions} represents scattering in the region $\by \in (0,1)$  with a purely ingoing condition at $\by = 0$,
 while the notion of ingoing
and outgoing at $\by=1$ depends on the sign of
$\Re \left(\bar{\omega}+m \bar{\Omega}\right)$. According to the asymptotic
behaviour of hypergeometric functions~\cite{NIST:DLMF}, we have
\begin{align}
\label{eq:in-out-amplitudes}  
  \Ain &= \frac{\Gamma(1-2i\bar{\omega})\Gamma\left(-2i\left(\bar{\omega}+m\bar{\Omega}\right)\right)}{\Gamma((1+i\beta)/2-im\bar{\Omega}-2i\bar{\omega})\Gamma((1-i\beta)/2-im\bar{\Omega}-2i\bar{\omega})}~,\\[10pt]
  \Aout &= \frac{\Gamma(1-2i\bar{\omega})\Gamma\left(2i\left(\bar{\omega}+m\bar{\Omega}\right)\right)}{\Gamma((1-i\beta)/2+im\bar{\Omega})\Gamma((1+i\beta)/2+im\bar{\Omega})}~.
\end{align}
We find the QNM frequencies by requiring that $\Ain =0$ in the case
$\Re\left(\bar{\omega}+m \bar{\Omega}\right) > 0$.\footnote{For
  $\Re\left(\bar{\omega}+m \bar{\Omega}\right) < 0$, we should instead
  look for zeros of $\Aout$, but it does not possess any non-trivial
  zeros.}  This condition straightforwardly yields that the QNM
frequencies $\bar{\omega}=\bar{\omega}_{N\ell m}^{\pm}$ are given by
\begin{equation}
\label{eq:ads2qnm}
2\bar{\omega}_{N\ell m}^{\pm} = -
i\left(N+\frac{1}{2}\right)-m \bar{\Omega}\pm
\sqrt{\bar{\lambda}_{\ell m}+m^{2}\bar{\Omega}^{2} -\frac{1}{4}},\quad
N=0,1,2,\dots\quad
\end{equation}

The greybody factor for real frequencies follows
readily from Eq.~\eqref{eq:in-out-amplitudes}:
\begin{align}
\label{eq:nariai-aa10}
  G(\bar{\omega},m) &= 1 -
                             \left|
                             \frac{\Aout}{\Ain}
                             \right|^2 = \frac{2\sinh(2\pi \bar{\omega})\sinh(2\pi(\bar{\omega}+m\bar{\Omega}))}{\cosh(2\pi(2\bar{\omega}+m\bar{\Omega}))+\cosh(\pi\beta)}.
\end{align}
Eqs.~\eqref{eq:ads2qnm} and \eqref{eq:nariai-aa10} agree with \cite{anninos2010sitter} for $\xi=0$ (Eq.~\eqref{eq:ads2qnm}
also agrees
with the QNM frequencies of the geometry resulting from the extremal Nariai limit $r_C\rightarrow r_+$ of Schwarzschild--de Sitter spacetime~\cite{Cardoso2003b} by setting $a=0$, $\xi =0$, $r_{-}=0$
and $\lambda_{\ell m} = \ell(\ell+1)$).\footnote{When comparing, one should taking into account the different coordinate times.}
Eqs.~\eqref{eq:ads2qnm} and \eqref{eq:nariai-aa10} are new results for the nonminimal coupling case ($\xi\neq 0$).

Alternatively, the result \eqref{eq:nariai-aa10} can be derived easily
if we use the monodromy technique \cite{Castro2013b,daCunha:2015fna}
reviewed in the previous section.  The greybody factor is given by
Eq.~\eqref{eq:greybody} with 
$\sigma$ replaced by
$\theta_{\infty}$ (as explained below):
\begin{align}
  \label{eq:nariai-transcoeff}
  G(\bar{\omega},m) &=
            \frac{2\sin(2\pi\theta_0)\sin(2\pi\theta_{1})}{\cos\left(2\pi(\theta_{0}-\theta_{1})\right)+\cos(2\pi\theta_{\infty})}\;\nonumber\\[10pt]
                &=\frac{2\sinh(2\pi\bar{\omega}) \sinh\left(2\pi(\bar{\omega}+m\bar{\Omega})\right)}{\cosh\left(2\pi(2\bar{\omega}+m\bar{\Omega})\right)+\cosh(\pi\beta)},
\end{align}  which is exactly the same result
as Eq.~\eqref{eq:nariai-aa10}. Therefore, we can avoid the calculation of
the exact connection coefficients using the general formula
\eqref{eq:greybody}. 

The poles of the greybody factor Eq.~\eqref{eq:nariai-transcoeff} are
given by \eqref{eq:ads2qnm} but with $N\in \mathbb{Z}$ instead of
$N=0,1,2,\dots$.  This is an intrinsic limitation of finding QNMs from
the greybody factor, as the two possibilities of $N\in\mathbb{Z}^+$
and $N\in\mathbb{Z}^-$ appear here because we take the ratio of
connection coefficients in \eqref{eq:nariai-aa10}.  From the exact
connection coefficients \eqref{eq:in-out-amplitudes}, we see that
positive $N$ corresponds to $\bar{\omega}, m >0$ and negative $N$
corresponds to $\bar{\omega}, m < 0$.  The transformation
$\{\bar\omega,m\}\to \{-\bar\omega,-m\}$ is a symmetry of the radial
equation \eqref{eq:rotating-nariai-radial-eq} but changes the notion
of ingoing and outgoing solutions. We can decide which branch to take by going
back to the solutions \eqref{eq:in-out-solutions}.
\\

\noindent\textbf{Spin-$s$ Quasinormal Modes from the Kerr--de Sitter Master Equation.} 
The formula \eqref{eq:nariai-transcoeff} deserves further explanation. Firstly, notice that \eqref{eq:greybody} is valid for a generic connection problem between two singular points of a Fuchsian equation which possesses \emph{any} number of singular points, as long as the monodromy matrices are diagonalizable. Secondly, as the hypergeometric
equation is a Fuchsian equation with three regular singular points, the composite monodromy  around $\by=0$ and $\by=1$ must be equal to the monodromy $\theta_\infty$ at $\by=\infty$. This can be made even more precise by
noticing that we can recover the rotating Nariai radial equation \eqref{eq:rotating-nariai-radial-eq} for $\xi= 1/6$ in canonical form directly from the full Heun equation in Kerr--de Sitter  \eqref{eq:heuncanonicalradial}.
Instead of the coordinate choice
\eqref{eq:radial-mobius}, we choose
\begin{equation}
\label{eq:radial-mobius-choice-2}
\begin{gathered}
 z(r) = \zeta_{\infty}\frac{r-r_+}{r-r_{--}},\quad \zeta_{\infty} = \frac{2r_{C}+r_{+}+r_{-}}{r_{C}-r_{+}},\quad x = \zeta_{\infty}\frac{r_{-}-r_{+}}{2r_{-}+r_{C}+r_{-}},\\[10pt]
  r=(r_{+},r_{C},r_{-},r_{--},\infty)\quad \mapsto\quad
 z= (0,1,x,\infty,\zeta_{\infty}).
\end{gathered}
\end{equation}
This means that, with respect to the monodromies in Eq.~\eqref{eq:heun_thetas}, we make the replacement
\begin{equation}
\label{eq:identification}
    (\theta_0,\theta_1,\theta_x,\theta_\infty) \mapsto  (\theta_1,\theta_x,\theta_0,\theta_\infty).
\end{equation}
Both $\zeta_{\infty}$ and $x$ diverge in the extremal limit
$r_{+}\rightarrow r_{C}$ as $\epsilon^{-1}$, and this implies that $\theta_{0}$ and
$\theta_{1}
$ diverge as $x\rightarrow \infty$. We can regularize  the
monodromies in Eq.~\eqref{eq:heun_thetas} by defining 
\begin{equation}
  \label{eq:omegabar}
\bar{\omega} \equiv \frac{\omega - m\Omega_{+}}{\bar{\kappa}\epsilon}, 
\end{equation}
which gives, for small $\epsilon$ and fixed $\bar{\omega}$,
\begin{equation}
\begin{aligned}
  \label{eq:nariai-monodromies}
&&  \theta_{0} &= 
\bar\theta_{0}
+ \mathcal{O}(\epsilon),& 
\theta_{1} &= 
\bar\theta_{1}
+ \mathcal{O}(\epsilon), && 
\\[5pt]
 && \theta_{x} &= im \bar{\Omega}\frac{(3\bar{r}_{+}+\bar{r}_{-})}{4\bar{r}_{+}}+\frac{s}{2}+ \mathcal{O}(\epsilon),&\quad \theta_{\infty} &= im \bar{\Omega}\frac{(\bar{r}_{+}-\bar{r}_{-})}{4\bar{r}_{+}}+\frac{s}{2}+ \mathcal{O}(\epsilon), &&
\end{aligned}
\end{equation}
with
\begin{equation}
    \bar\theta_{0}\equiv i\bar{\omega}+\frac{s}{2},\quad 
    \bar\theta_{1}\equiv 
    -i\left(\bar{\omega} + m\bar{\Omega}\right) +\frac{s}{2},
\end{equation}
where the leading order terms should be calculated at
$\epsilon=0$. The accessory parameter $K_{x}$ in
\eqref{eq:canonical-accessory-parameter-2}, 
with the appropriate identifications of
Eqs.~\eqref{eq:radial-mobius-choice-2} and \eqref{eq:identification},
goes to zero as $x\rightarrow \infty$, but $x(x-1)K_{x}/(z-x)$ is
finite. Applying this limit (i.e., $x\to \infty$ -- or, equivalently,
$\epsilon\to 0$ -- with $z$ finite) to the Kerr--de Sitter
Eq.~\eqref{eq:heuncanonicalradial} yields
\begin{equation}
  \label{eq:hypergeometric-canonical-radial} \left[\frac{d^2 }{dz^2}+\left(\frac{1-2\bar\theta_{0}}{z}+\frac{1-2\bar\theta_{1}}{z-1}\right)\frac{d}{dz}
  +\frac{\lamslmbar-2s+i m(1-2s) \bar{\Omega}}{z(z-1)}\right]f=0,
\end{equation}
and
\begin{equation}
  \label{eq:spin-s-angular-eigen-bar-rotating-nariai}
  \lamslmbar \equiv \frac{L^{2}}{(3\bar{r}_{+}+\bar{r}_{-})(\bar{r}_{+}-\bar{r}_{-})}
    \left(
      \lamslm\big\rvert_{\epsilon\to0}-s\left(1-\frac{a^2}{L^2}\right)+ \frac{2}{L^{2}}(2s+1)(s+1)\bar{r}_{+}^{2}
    \right),
\end{equation}
where we remind the reader that $\lamslm$ is the Kerr--de Sitter angular eigenvalue. 
One can check that by taking the rotating Nariai limit on the master angular equation 
\eqref{eq:rotating-nariai-angular-eq} in Kerr--de Sitter then, for $s=0$ and $\xi=1/6$, one recovers Eq.~\eqref{eq:rotating-nariai-angular-eq} with
${}_{0}\lambda_{\ell m}\big\rvert_{\epsilon\to0} $
  in the place of $\lambda_{\ell m}$  (and so they are equal when imposing the eigenvalue condition). Therefore, 
Eq.~\eqref{eq:spin-s-angular-eigen-bar-rotating-nariai} matches Eq.~\eqref{eq:48} for $s=0$ and $\xi=1/6$.  The reason for this relationship is that Eq.~\eqref{eq:hypergeometric-canonical-radial} is equivalent to Eq.~\eqref{eq:rotating-nariai-radial-eq} with $z=\by$ after an appropriate homotopic transformation (see appendix~\ref{sec:from-self-adjoint} for the general discussion). If we go to the
next order in $\epsilon$ in the full master radial equation, we can
start to calculate corrections to the extremal QNM result
in Eq.~\eqref{eq:ads2qnm}. That is what we achieve in the next section.


\section{Accessory Parameter Expansion}
\label{sec:access-param-from}

In the previous section, we considered an arbitrary second order Fuchsian
ODE and we expressed the greybody factor (written as a product of
connection coefficients) in terms of the monodromies of local
solutions -- see Eq.~\eqref{eq:greybody}.  By looking for poles of
this formula we found condition \eqref{eq:qnmcondition} for
$\sigma_{+C}$, which sets the boundary condition for QNMs
\eqref{eq:qnmcondition-2}of the Kerr-dS master radial equation for
$N\in \mathbb{Z}^+$. We also similarly discussed the monodromy
condition \eqref{eq:sigma-angular-eigenvalue-2} for angular
eigenvalues. We thus mapped special boundary conditions for a solution
of the radial or angular equation to constraints on its monodromy
exponents. This approach was proposed for the confluent Heun equation
in \cite{Castro2013b,daCunha:2015ana} and for the Heun equation in
\cite{Novaes2014c,daCunha:2015fna}.  We have in fact shown above that
this method works in the particular case of the rotating Nariai
limit. This limit yields a simpler case since the composite monodromy
then reduces to a local monodromy of the hypergeometric equation.  

In this section, we apply the composite monodromy constraints to
solutions of the Heun equation \eqref{eq:heuneq} obtained from the
Kerr--de Sitter master equation.  Below we present our results using
the APE and the numerical checks done using Leaver's method.

\subsection{How to obtain the APE}
\label{sec:how-obtain-ape}

Here we review the mathematical background behind the APE and its
derivation done in \cite{Lencses:2017dgf}.
The monodromy group of functions on the complex plane with four
regular singular points are labeled by six parameters
\cite{Lencses:2017dgf}. Apart from the four monodromies
$\bm{\theta} = (\theta_{0}, \theta_{1}, \theta_{x},\theta_{\infty})$
at the singular points, there are two extra \emph{composite
  monodromies}, $(\sigma_{0x},\sigma_{1x})$. We thus define the
\emph{monodromy data} of the four-punctured sphere as the six
parameters
$\mathcal{M}_{\mathrm{SL}(2,\mathbb{C})}\equiv
(\theta_{0},\theta_{1},\theta_{x},\theta_{\infty};
\sigma_{0x},\sigma_{1x})$. This means that the Heun equation is not
the most general Fuchsian equation with four regular singular points,
as it has only 5 parameters
$\mathcal{M}_{\text{Heun}}=
(\theta_{0},\theta_{1},\theta_{x},\theta_{\infty}; K_x)$ (see below
Eq.~\eqref{eq:37}).  Instead, we can say that
$\mathcal{M}_{\text{Heun}}$ corresponds to a reduced case of
$\mathcal{M}_{\mathrm{SL}(2,\mathbb{C})}$ with only one composite
monodromy, say,
$\mathcal{M}_{\text{red}}\equiv(\theta_{0},\theta_{1},\theta_{x},\theta_{\infty};
\sigma_{0x}, \sigma_{1x}=\sigma_{1x}(\sigma_{0x}))$.  Then,
$\mathcal{M}_{\text{red}}$ can be mapped to
$\mathcal{M}_{\text{Heun}}$ via the
relation $K_{x}=K_{x}(\bm\theta,\sigma_{0x},x)$. For fixed $x$, the
APE provides the Riemann--Hilbert map between
$\mathcal{M}_{\text{red}}$ and $\mathcal{M}_{\text{Heun}}$.  The
small-$x$ expansion of this relation is called the \emph{accessory
  parameter expansion} (APE). In Ref.~\cite{Lencses:2017dgf}, the APE
was obtained using isomonodromic deformations, as we review below.
For other approaches to obtaining the APE, we refer the reader to, for
example, \cite{Litvinov:2013sxa,Menotti2014,Hollands2017}. For recent
numerical results on finding 2D conformal maps from the APE using the
approach of \cite{Novaes2014c,Lencses:2017dgf}, see
\cite{Anselmo2018}.

The APE obtained in \cite{Lencses:2017dgf} depends on monodromy data
$\mathcal{M}_{\text{red}}$ which is independent of $x$. This is
contrary to what happens in the case of the ODEs satisfied by
perturbations of Kerr--de Sitter black holes.  In the case of the
radial ODE in Kerr--de Sitter,  $x\propto \epsilon$ and the
monodromies diverge for small $\epsilon$. However, as discussed in
section~\ref{sec:scattering-kerr-de}, if we fix the frequency
$\bar\omega$ defined in Eq.~\eqref{eq:omegabar}, the monodromies
converge as an expansion in $\epsilon$ and the accessory parameter is
analytic for small $\epsilon$. Therefore, by setting
$\sigma\equiv \sigma_{0x}$ and expanding all  parameters
$(x(\epsilon),\bm{\theta}(\epsilon),\sigma(\epsilon))$ in terms of
$\epsilon$ as
\begin{subequations}
 \label{eq:parameters-epsilon-expansion}
  \begin{gather}
  \label{eq:parameters-epsilon-expansion-1}
  x = \epsilon( x_{0}+\epsilon x_{1}+\ldots),\quad\sigma =
  \sigma_{0}+\epsilon \sigma_{1}+\ldots,\\[5pt]
  \theta_{k} = \theta_{k,0} + \epsilon \theta_{k,1}+\ldots, \quad
  k\in \mathcal{J},
   \label{eq:parameters-epsilon-expansion-2}
\end{gather}
\end{subequations}
we obtain the APE
\begin{equation}
  \label{eq:APE-generic}
  K_{x} =
  \frac{K_{x,-1}}{\epsilon}+\sum_{n=0}^{\infty}K_{x,n}\epsilon^{n},
\end{equation}
for some 
coefficients $x_{j}$ and $\theta_{k,j}$ and
functions $ K_{x,n}= K_{x,n}(x_{j}, \theta_{k,j},
\sigma_{j})$, $j \geq  0$, $n\geq -1$, $k=0,1,x,\infty$.

In the case of the angular ODE, we have $x\propto \alpha$ and we set
$\epsilon=\alpha$. The monodromies in Eq.~\eqref{eq:12} converge for
small $\alpha$ and fixed $L$ and we obtain the APE for the angular
eigenvalue from Eq.~\eqref{eq:APE-generic}. In the case of QNMs, we
instead set $\epsilon =(r_{C}-r_{+})/L$ and use the frequency
definition in Eq.~\eqref{eq:omegabar}, which gives a regular expansion
for the monodromies as $\epsilon \rightarrow 0$ for fixed
$\bar{\omega}$. Given the APE expansions in both cases, we can equate
it with the exact accessory parameters coming from the angular and
radial ODEs (see, respectively, Eqs.~\eqref{eq:angularaccessory} and
\eqref{eq:heun_K0}), using the exact monodromies $\bm\theta$ and the
QNM condition Eq.~\eqref{eq:qnmcondition-2} and the angular eigenvalue
condition Eq.~\eqref{eq:sigma-angular-eigenvalue-2}.  This procedure,
which we shall present in more detail below, will give the expansions
in Eqs.~\eqref{eq:omega-expansion} and
\eqref{eq:angular-eigenvalue-expansion}.  First, however, we shall
show how to obtain the APE from a technique called isomonodromic
deformation \cite{Jimbo1981b,Jimbo:1981-2,Jimbo:1981-3}.


The most general monodromy data
$\mathcal{M}_{\mathrm{SL}(2,\mathbb{C})}$ for an ODE with four
singular points includes two composite monodromies,
$(\sigma_{0x},\sigma_{1x})$. However, as we commented above, the Heun
equation \eqref{eq:heuneq} has only one extra parameter $K_x$. To
account for these two composite monodromies, we write the most general
Fuchsian equation as
\begin{multline}
  \label{eq:deformed-heun}
  y''+ \left( \frac{1-2\theta_0}{z}+\frac{1-2\theta_1}{z-1}+
    \frac{2-2\theta_x}{z-t}-\frac{1}{z-\lambda(t)}
  \right)y'+\\[10pt]
  +\left( \frac{b_1 b_2
    }{z(z-1)}-\frac{t(t-1)K(\theta_{0},\theta_{1},\theta_{x}-\frac12,\theta_{\infty};\lambda(t),\mu(t),t)}{z(z-1)(z-t)}+
    \frac{\lambda(t)(\lambda(t)-1)\mu(t)}{z(z-1)(z-\lambda(t))} \right)y=0,
\end{multline}
where $\theta_\infty = b_1-b_2$, $\lambda(t)$ and $\mu(t)$ are some $t$-dependent  functions with
$t\in\mathbb{C}$
and\footnote{The parameter $\lambda$ here is unrelated to the angular eigenvalue $\lamslm$;
these are the symbols typically used in the literature of isomonodromic deformations.}
\begin{multline}
  \label{eq:kamiltonian}
   K(\theta_{0},\theta_{1},\theta_{x}-\tfrac12,\theta_{\infty};\lambda,\mu,t)
\equiv\\[5pt]
  \frac{\lambda(\lambda-1)(\lambda-t)}{t(t-1)}
  \left[\mu^2-\left(\frac{2\theta_0}{\lambda}+\frac{2\theta_1}{\lambda-1}+
      \frac{2\theta_x-2}{\lambda-t}\right)\mu+\frac{b_1 b_2}{\lambda(\lambda-1)}\right].
\end{multline}
The form of \eqref{eq:kamiltonian} follows by the requirement that
$\lambda(t)$ is an \emph{apparent singularity}, i.e., the solution has
no branch cut at this point.  The ODE parameters can be mapped to
$(\sigma_{0x},\sigma_{1x})$ via initial conditions for the
isomonodromic flow $(\lambda(t),\mu(t))$, whose Hamiltonian is given
by Eq.~\eqref{eq:kamiltonian}. However, there is a special set of
initial conditions at $t=x$ \cite{Novaes2014c,daCunha:2015fna} that
reduce the general Eq.~\eqref{eq:deformed-heun} to the canonical Heun
equation \eqref{eq:heuneq} with the same monodromy data
$\mathcal{M}_{\text{Heun}}$.  These initial conditions are given by
\begin{equation}
  \label{eq:initialcondition}
  t=x,\quad\lambda(x) = x, \quad \mu(x) =
   -\frac{K_{x}}{2\theta_{x}-1}.
\end{equation}
In this case, we reduce the number of parameters of the ODE from two
$(\lambda,\mu)$ to only one ($K_{x}$), and similarly for the composite
monodromies, since we have
$\sigma_{1x}=\sigma_{1x}(\sigma_{0x},x)$. This is explained in full
detail in Ref.~\cite{Lencses:2017dgf}. Here we just report on the
solution for the accessory parameter.  

The main difference between the current work and
\cite{Lencses:2017dgf} is that here we want $K_{x}$ as an expansion in
$\epsilon$ using the expansion of the monodromy data in
\eqref{eq:parameters-epsilon-expansion}, as described above. This
changes the expansion obtained in \cite{Lencses:2017dgf}, which
becomes more complicated. It is more convenient to carry out the
expansion for the accessory parameter of the ODE in the normal form
(see appendix~\ref{sec:from-self-adjoint} for definitions)
\begin{equation}
  \label{eq:htok}
    \mathcal{H}_{x} \equiv K_{x} +\frac{(1-2\theta_{0})(1-2\theta_{x})}{2x}+\frac{(1-2\theta_{1})(1-2\theta_{x})}{2(x-1)}.
\end{equation}
This is the form most straightforwardly connected to 2D CFT. Using the
procedure outlined in appendix~\ref{sec:iso-definitions}, we find
\begin{multline}
  \label{eq:APE}
  \mathcal{H}_x =\frac{-4 \theta _{0,0}^2-4 \theta _{x,0}^2+4 \sigma _0^2+1}{4 x_0 \epsilon }+ \frac{\left(-4 \theta _{\infty ,0}^2+4 \theta _{1,0}^2+4 \sigma _0^2-1\right) \left(-4 \theta _{0,0}^2+4 \theta _{x,0}^2+4 \sigma _0^2-1\right)}{32 \sigma _0^2-8}\\[5pt]
  +\frac{x_1 \left(4 \theta _{0,0}^2+4 \theta _{x,0}^2-4 \sigma
      _0^2-1\right)+8 x_0 \left(-\theta _{0,0} \theta _{0,1}-\theta
      _{x,0} \theta _{x,1}+\sigma _0 \sigma _1\right)}{4
    x_0^2}+\mathcal{O}(\epsilon),
\end{multline}
where we made use of the expansions in
Eqs.~\eqref{eq:parameters-epsilon-expansion}. We have calculated this
series up to order $\epsilon^{3}$ in order to compare with the
numerical results, but the higher-order terms are particularly long so
we do not display them here.  \MC{Are we providing them in an online
  nb which also calculates the QNMs and eigenvalues?} \FN{Maybe I can
  put them out in the second arXiv version?}

The key point is that now we can equate
two small-$\epsilon$ expansions for $\mathcal{H}_x$.
One is the APE \eqref{eq:APE}. The other expansion is obtained from Eq.~\eqref{eq:htok} with the
 accessory parameter given by Eq.~\eqref{eq:heun_K0} or \eqref{eq:angularaccessory} in, respectively, the radial or angular cases,  and expanded for small $\epsilon$.
 We call the equality between these two small-$\epsilon$ expansions for $\mathcal{H}_x$ the ``APE equation". 
 We shall next proceed to do this, first for the angular case and then for the radial case.
 
\subsection{Angular Eigenvalues}
\label{sec:angular-eigenvalues}

Here we wish to apply the APE in Eq.~\eqref{eq:APE} to find the angular eigenvalues $\lamslm$ in
Eq.~\eqref{eq:angulareq}. The  expansion parameter  in APE here is
$\epsilon=\alpha= a/L$ and we expand in $\alpha$ only \emph{after} replacing $a$ by $\alpha L$ in the monodromies in Eq.~\eqref{eq:12}. 
Since $a$ in Eq.~\eqref{eq:12} only appears in the combination
$a\omega/\alpha$, the mentioned expansion procedure
is equivalent to expanding for both small $\alpha$ and small $a\omega$ (without the replacement of $a$ by $\alpha L$), both quantities being of the same order in smallness. From the APE point of view, we expand it up to order $\alpha^3$, which, from the previous observation, gives all powers $\alpha^n (a\omega)^m$ for $n+m=0,\ldots,3$ of $\lamslm(a\omega,\alpha)$, as in \eqref{eq:angular-eigenvalue-true-expansion}.

The monodromies \eqref{eq:12} are polynomial in $\alpha$ (after replacing $a$ by $\alpha L$)
and we expand $x$ to one order higher in $\alpha$
\begin{equation}
  \label{eq:x-expansion-angular}
  x = \frac{4 i \alpha}{(i+\alpha)^{2}}=
  -4 i \alpha+8 \alpha ^2+12 i \alpha ^3-16 \alpha ^4+ O(\alpha^5).
\end{equation}
 Using Eqs.~\eqref{eq:angularaccessory}, \eqref{eq:htok} and \eqref{eq:x-expansion-angular}
we obtain the following expansion for the accessory parameter:
\begin{multline}
  \label{eq:45}
  x(x-1)\mathcal{H}_{x}=
\left(4 i \alpha ^3+3 \alpha ^2-2 i \alpha -1\right) \lamslm +\alpha
^3 \left(4 L \omega  (m-s)-2 i m^2-10 i m s+6
  i\right)\\[5pt]
+2\alpha ^2 (-2 i L \omega  (m-s)- m
(m+3 s)+2)+2\alpha  ( L \omega  (s-m)+ i( m
(m+s)- 1))+\\[5pt] +\frac{m^2-s^2-1}{2} +O(\alpha^4).
\end{multline}
Equating this with the APE \eqref{eq:APE}, using the monodromies
\eqref{eq:12} and the angular eigenvalue condition
\eqref{eq:sigma-angular-eigenvalue-2} yields the APE equation in the
angular case. We then use this APE equation to isolate for $\lamslm$,
thus yielding an expression which we expand for small $\alpha$.  We
thus obtain Eq.~\eqref{eq:angular-eigenvalue-expansion}, where the
explicit form of the coefficients $\lambda_{\omega,k}$ expanded for
small $\alpha$ is given in appendix~\ref{sec:eigen coeffs}.  We have
checked that this result matches (in the orders contained in both
expansions) the angular eigenvalue expansion up to order
$(\alpha^{2},\omega^{2})$ calculated in \cite{Suzuki1998}.  We have
also checked that, in the Kerr limit $L\rightarrow \infty$, all the
orders that we give agree with the small-$a\omega$ expansion given
in~\cite{berti2006eigenvalues}.  Thus, the eigenvalue expansion that
we have obtained already exists in the literature (split
between~\cite{Suzuki1998} and~\cite{berti2006eigenvalues}) -- the
derivation here serves to validate the APE and provides a way of
extending the expansion to higher orders.

\subsection{Quasinormal Mode Frequencies}
\label{sec:quasinormal-modes}

Let us now calculate the QNM frequencies in Kerr--de Sitter around the
rotating Nariai limit using the APE. This calculation is more
complicated than in the angular eigenvalue case, as both sides of the
APE equation depend explicitly on powers of the frequency
$\omega$. Therefore, we cannot just isolate $\omega$ as we did for the
angular eigenvalue in the angular case. However, as mentioned above,
the QNM frequency can still be calculated order-by-order in $\epsilon$
as displayed in Eq.~\eqref{eq:omega-expansion}.  We show here the
explicit calculation near the rotating Nariai limit and just display
the result up to order $\epsilon$.  We, in fact, obtained the
expansion up to order $(\epsilon^{3},a^{3})$ and provide it in
appendix~\ref{sec:omegas}. We numerically-checked the higher order
expansion, which is discussed in the next section.

In this section, we will use the change of coordinates given by
Eq.~\eqref{eq:radial-mobius}, where, in particular, $r=r_{C}$ is
mapped to $z=0$ and $r=r_{+}$ is mapped to $z=x$. The expansion
parameter here is $\epsilon = (r_{C}-r_{+})/L$. Therefore,
\begin{align}
  \label{eq:62}
  x &= 
      \left(
      \frac{2
      r_{-}+r_{+}+r_{C}}{2r_{+}+r_{C}+r_{-}}
      \right)\frac{r_{C}-r_{+}}{r_{C}-r_{-}}\sim 
      \frac{2(\bar{r}_{+}+\bar{r}_{-})}{(3\bar{r}_{+}+\bar{r}_{-})(\bar{r}_{+}-\bar{r}_{-})}\,
      L\epsilon\quad\text{as}\quad r_{C}\rightarrow r_{+}.
\end{align}
So the extremal limit $r_{C}\rightarrow r_{+}$ corresponds to $x\rightarrow0$. We expand the 
monodromies \eqref{eq:heun_thetas} by expressing $\omega = m\Omega_{+} + \epsilon \bar{\kappa} \bar{\omega}$
and expanding for small $\epsilon$ with $\bar{\omega}$ fixed:
\begin{equation}
\begin{aligned}
  \label{eq:monodromies-APE}
&&  \theta_{0} &= -i
  \left(
  \bar{\omega} + m\bar{\Omega}
  \right)  +\frac{s}{2}+ \mathcal{O}(\epsilon),& \theta_{1} &= im \bar{\Omega}\frac{(3\bar{r}_{+}+\bar{r}_{-})}{4\bar{r}_{+}}+\frac{s}{2}+ \mathcal{O}(\epsilon),&& 
\\[5pt]
 && \theta_{x} &=i\bar{\omega}+\frac{s}{2}+
 \mathcal{O}(\epsilon),&\quad \theta_{\infty} &= im
 \bar{\Omega}\frac{(\bar{r}_{+}-\bar{r}_{-})}{4\bar{r}_{+}}+\frac{s}{2}+
 \mathcal{O}(\epsilon), &&
\end{aligned}\
\end{equation}
where $\bar{\Omega}$ is defined in Eq.~\eqref{nariaiparam2} and $\bar{\kappa}$ in Eq.~\eqref{eq:44}. 

We now set out to obtain the APE equation.
First, we
expand
\begin{equation}
  \label{eq:omegabar-expansion}
  \bar{\kappa}\bar{\omega} = \bar{\omega}_{0}+ \bar{\omega}_{1}\epsilon
  +\bar{\omega}_{2}\epsilon^2 
  +\mathcal{O}\left(\epsilon^3\right),
\end{equation}
for some coefficients $\bar{\omega}_{0,1,2}$ which we set out to
determine.  In order to obtain one of the sides of the APE equation,
we use Eq.~\eqref{eq:htok} for $\mathcal{H}_{x}$ with the accessory
parameter given by Eq.~\eqref{eq:heun_K0} consistently expanded for
small $\epsilon$.  For this expansion, we use
Eq.~\eqref{eq:monodromies-APE} for the monodromies and
Eq.~\eqref{eq:omegabar-expansion} for the frequency, which in its turn
is also used for expanding the angular eigenvalue in $\epsilon$.

The other side of the APE equation is the APE \eqref{eq:APE}.  This
expression depends on the coefficients of the expansion of the
monodromies as well as on $\sigma$, for which we use the QNM condition
given by Eq.~\eqref{eq:qnmcondition-2}.  In order to obtain the final
expansion for $\mathcal{H}_x$ from the APE, because of the replacement
in Eq.~\eqref{eq:qnmcondition-2}, the monodromies need to be expanded
as per Eq.~\eqref{eq:monodromies-APE} and the frequency needs to be
expanded as per Eq.~\eqref{eq:omegabar-expansion}.
 
We have now explained how to obtain the two sides of the radial APE
equation.  This APE equation gives polynomial equations for the
coefficients of $\bar{\omega}$ at each order in $\epsilon$. The first
coefficient $\bar{\omega}_{0}$ obeys a quadratic equation and the
other $\omega_{1,2,\dots}$ obey linear equations, depending on the
previously obtained ones in a recursive way.  We next proceed to give
the explicit expression for $\bar\omega$ to leading order in
$\epsilon$, which coincides with the rotating Nariai limit result.

The first side of the APE equation, which is obtained from
Eqs.~\eqref{eq:heun_K0}, \eqref{eq:htok}, \eqref{eq:monodromies-APE}
and \eqref{eq:omegabar-expansion}, reads, to leading order in
$\epsilon$,
\begin{align}
  \label{eq:Kx-heun-limit}  
  x(x-1)\mathcal{H}_{x} &=\lamslmbar -2 s^2+2 s \theta _x+2 \theta
                _0 \left(s-\theta
                _x\right)-s-\frac{1}{2}+\mathcal{O}(\epsilon)\nonumber\\[5pt]
              &=-2\left(\bar{\omega}_0\right)^2 -2 m \bar{\Omega}\bar{\omega}_0
          +\lamslmbar -s (1+i m \bar{\Omega})      -\frac{s^2+1}{2}+\mathcal{O}(\epsilon),
  \end{align}
  where $\lamslmbar$ is given in 
  Eq.~\eqref{eq:spin-s-angular-eigen-bar-rotating-nariai}.
  Equating the expansion for $\mathcal{H}_{x}$ coming from \eqref{eq:Kx-heun-limit}  with the
  one coming from the APE \eqref{eq:APE}
  as explained above,
yields, to leading order,
  the following quadratic equation for $\bar\omega_{0}$:
  \begin{equation}
    \label{eq:64}
          \left(\frac{\bar{\omega}_{0}}{\bar{\kappa}}\right)^2+ \left(m \bar\Omega+ i 
    \left(N+1/2\right)\right)\frac{\bar{\omega}_{0}}{\bar{\kappa}} +\frac{1}{4}\left(\lamslmbar-2 i m \bar\Omega(N+1/2)+N(N+1)-s(s+1)\right)=0,
  \end{equation}
  with $N\in \mathbb{Z}$ (which comes from the QNM condition in Eq.~\eqref{eq:qnmcondition-2}).
The two solutions of this equation are
  \begin{equation}
    \label{eq:spin-s-rotating-nariai-qnm}
\frac{2\bar{\omega}_{0}}{\bar{\kappa}}  =  - i\left(N+\frac12\right) - m \bar{\Omega}\pm\sqrt{
  \lamslmbar + m^2 \bar\Omega^2- s (s+1)-\frac14}.
  \end{equation}
This expression agrees with Eq.~\eqref{eq:ads2qnm} for $s=0$ and $\xi=1/6$.
As dictated by Eq.~\eqref{eq:ads2qnm}, from now on $N$ is restricted to be a non-negative integer -- we note that the APE method does not determine the sign of $N$, this
is just imposed via an independent calculation (namely, the one that led to Eq.~\eqref{eq:ads2qnm}).\footnote{Even though Eq.~\eqref{eq:ads2qnm} is only valid for $s=0$, we assume that the condition $N=0,1,2,\dots$ also holds for the other spins.}

Putting together Eqs.~\eqref{eq:omegabar}, \eqref{eq:omegabar-expansion} and \eqref{eq:spin-s-rotating-nariai-qnm}, we have that the near-extremality
expansion for the  QNM frequencies in Kerr--de Sitter is:
  \begin{equation}\label{eq:QMM exp}
  _{s}\omega_{\ell m} = m \Omega_{+}+
\frac{\bar{\kappa}\epsilon}{2}
\left[
 - i\left(N+\frac12\right) - m \bar{\Omega}\pm\sqrt{
  \lamslmbar + m^2 \bar\Omega^2- s (s+1)-\frac14}
   \right]+\mathcal{O}\left(\epsilon^2\right),
  \end{equation}
  where $N=0,1,2,\dots$.
 We remind the
  reader that $\Omega_{+}$, $\bar{\Omega}$, $\bar{\kappa}$ and
  $ \lamslmbar$ are given in,
  respectively, Eqs.~\eqref{eq:heun_temps}, \eqref{nariaiparam2},
  \eqref{eq:44} and
  \eqref{eq:spin-s-angular-eigen-bar-rotating-nariai}. To the best of our knowledge, this result
  is new in the literature.
We emphasize that, even though we write Eq.~\eqref{eq:QMM exp} as a small-$\epsilon$ expansion, $\Omega_+$ in it is exact and so \emph{not} expanded for small $\epsilon$  (the reason being that $\Omega_+$ became buried within
$\bar\omega$ in Eq.~\eqref{eq:omegabar}, which we kept fixed when carrying out the expansions).

We remind the reader that, up to this point, all of our expressions in this subsection were considering the limit $r\to r_+$ (together with the small-$\epsilon$ expansion). By doing this, we obtained the modes that approach $m\Omega_+$ in Eq.~\eqref{eq:QMM exp}. In order to obtain the modes that, instead, approach $m\Omega_C$, all we have to do is, in the quantities in Eq.~\eqref{eq:QMM exp}, swap $r_+\leftrightarrow r_C$ (which, in particular, implies $\epsilon\to-\epsilon$) and take the complex conjugate. Under this transformation, we obtain:
  \begin{equation}\label{eq:QMM exp-rc}
  _{s}\omega_{\ell m} = m \Omega_{C}+\frac{\bar{\kappa}\epsilon}{2}
\left[- i\left(N+\frac12\right) + m \bar{\Omega}\pm\sqrt{  \lamslmbar + m^2 \bar\Omega^2- s (s+1)-\frac14}
   \right]+\mathcal{O}\left(\epsilon^2\right),
  \end{equation}
  
Eqs.~\eqref{eq:QMM exp} and \eqref{eq:QMM exp-rc} show that, as advanced in section~\ref{sec:monodromy}, both the lower and the upper superradiant bound limits (i.e., $m \Omega_{C}$ and $m \Omega_{C}$, respectively) are accumulation points of QNMs as the extremal limit $r_C \to r_+$ is approached.
In order to obtain the next order correction $\bar\omega_1$, we need to expand all quantities in Eqs.~\eqref{eq:62}, ~\eqref{eq:monodromies-APE} and ~\eqref{eq:Kx-heun-limit} to one order higher in $\epsilon$. 
We then obtain:
\begin{eqnarray}\label{eq:bar omega_1}
  \bar\omega_1 &=& \frac{L\bar\kappa\left[(\bar{r}_+-\bar{r}_-)(\bar{r}_-+3\bar{r}_+)\right]^{-1}}{4\left(2\bar{\omega}_{0}/\bar{\kappa}+m\bar{\Omega}+i(N+1/2)\right)}\left[i m\bar{\Omega}(2N+1)\bar{r}_+ \left(2-L\chi^2\bar{\kappa}\right)+ L\frac{d}{d\epsilon}\lamslm\bigg|_{\epsilon=0}+\right. \nonumber\\
               & &\left. +\frac{m\bar{\Omega}\left(\bar{r}_+ +\bar{r}_-\right)^2\left(2s^2+s-\lamslmbar\right)\left(m\bar{\Omega}+2\bar{\omega}_0/\bar{\kappa}\right)}{2\bar{r}_+\left(\lamslmbar+m^2\bar{\Omega}^2-s^2-s\right)}\right].
 \end{eqnarray}
The $\epsilon$-derivative of $\lamslm$ at $\epsilon=0$ can be written
as
\begin{equation}
\frac{d}{d\epsilon}\lamslm\bigg|_{\epsilon=0}=\bar\kappa\left(m\bar{\Omega}+2\bar{\omega}_0/\bar{\kappa}\right)\frac{1}{2}\frac{d}{d\omega}\lamslm\left(\omega= \frac{ma}{a^2+\bar{r}_+^2}\right).
\end{equation}

To the best of our knowledge, no orders higher than leading-order in
Eq.\eqref{eq:omegabar-expansion} have been previously obtained in the
literature for any spin.  Above we have given explicit expressions for
general $a$ for the two leading-order terms in
Eq.\eqref{eq:omegabar-expansion}.  Although one could, in principle,
obtain higher order terms, we found it difficult to obtain them if
keeping $\alpha=a/L$ \emph{general}.  On the other hand, by expanding
the various quantities for small $\alpha$, we managed to reach up to
one higher order in $\epsilon$.  Specifically, what we did is the
following. We obtained both Eqs.~\eqref{eq:APE} and
\eqref{eq:Kx-heun-limit} up to including
$\mathcal{O}\left(\epsilon^3\right)$.  We then expanded the
coefficient of each $\epsilon$-order (including the $\bar\omega_0$,
$\bar\omega_1$, \ldots which appear within these coefficients, as well
as the eigenvalue as given by \eqref{eq:angular-eigenvalue-expansion}
with Eqs.~\eqref{eq:50}--\eqref{eq:52}) for small $\alpha$ up to
including $\mathcal{O}\left(\alpha^3\right)$.  We then equated the
left- and right-hand-sides at each order in this double series in
$\alpha$ and $\epsilon$, thus obtaining an expansion for the QNM
frequencies $_{s}\omega_{\ell m}$ valid up to order
$(\epsilon^3,\alpha^3$).  This expansion is the one in
Eq.~\eqref{eq:omega-expansion}, where the explicit form of the
coefficients is given in appendix~\ref{sec:omegas}.


\section{Comparison with Numerical Calculation}
\label{sec:comp-with-numer}
\FN{word 'method' appears many times}

We have verified our expansions of the QNM frequencies and the angular
eigenvalues with a numerical calculation.  In this section, we first
briefly sketch the approach used -- we note that the method will be
described in detail in~\cite{Casals:Marinho}.  We then present the
numerical results as a check of the previous analyical expansions:
Eq.~\eqref{eq:omega-expansion} (with the coefficients given in
appendix~\ref{sec:omegas}) for the QNM frequencies and
Eq.~\eqref{eq:angular-eigenvalue-expansion} for the angular
eigenvalues.

The method we used for the numerical calculation is a direct
implementation of the one described in~\cite{Yoshida2010}, which, in
its turn, is a generalization to Kerr--de Sitter of the technique
introduced by Leaver in Kerr
spacetime~\cite{Leaver:1985,Leaver:1986a}.  Essentially, the method
consists of expressing a solution of the radial ODE
\eqref{eq:radialeq} as a powers series in
``$(r_C-r_{--})(r-r_+)/((r-r_{--})(r_C-r_+))$".  By construction, this
series satisfies the QNM boundary condition at the event horizon
(which is the same as the ingoing boundary condition
\eqref{eq:classical_bdycond} there).  However, this infinite series
only satisfies the QNM boundary condition at the cosmological horizon
(which is the same as the upgoing boundary condition
\eqref{eq:semiclassical_bdycond} there) if it converges at that
horizon.  This convergence condition leads to an equation in terms of
a `radial' continued fraction which a frequency must satisfy for it to
be a QNM frequency.

The method for calculating the angular eigenvalue is similar.  One
essentially expands a solution of the angular ODE \eqref{eq:angulareq}
about $\theta=\pi$, thus obeying the regularity condition there by
construction.  The regularity condition at the other endpoint,
$\theta=0$, is only obeyed if $\lamslm$ satisfies a certain equation
in terms of an `angular' continued fraction.

We numerically solve \emph{simultaneously} for both the `radial'
continued fraction equation and the `angular' continued fraction
equation.  The solution of this system of equations is then a QNM
frequency $\omega={}_{s}\omega_{\ell m}$ and an angular eigenvalue
${}_{s}\lambda_{ \ell m}$.

\begin{figure}[th!]
\begin{center}
  \includegraphics[width=0.73\textwidth]{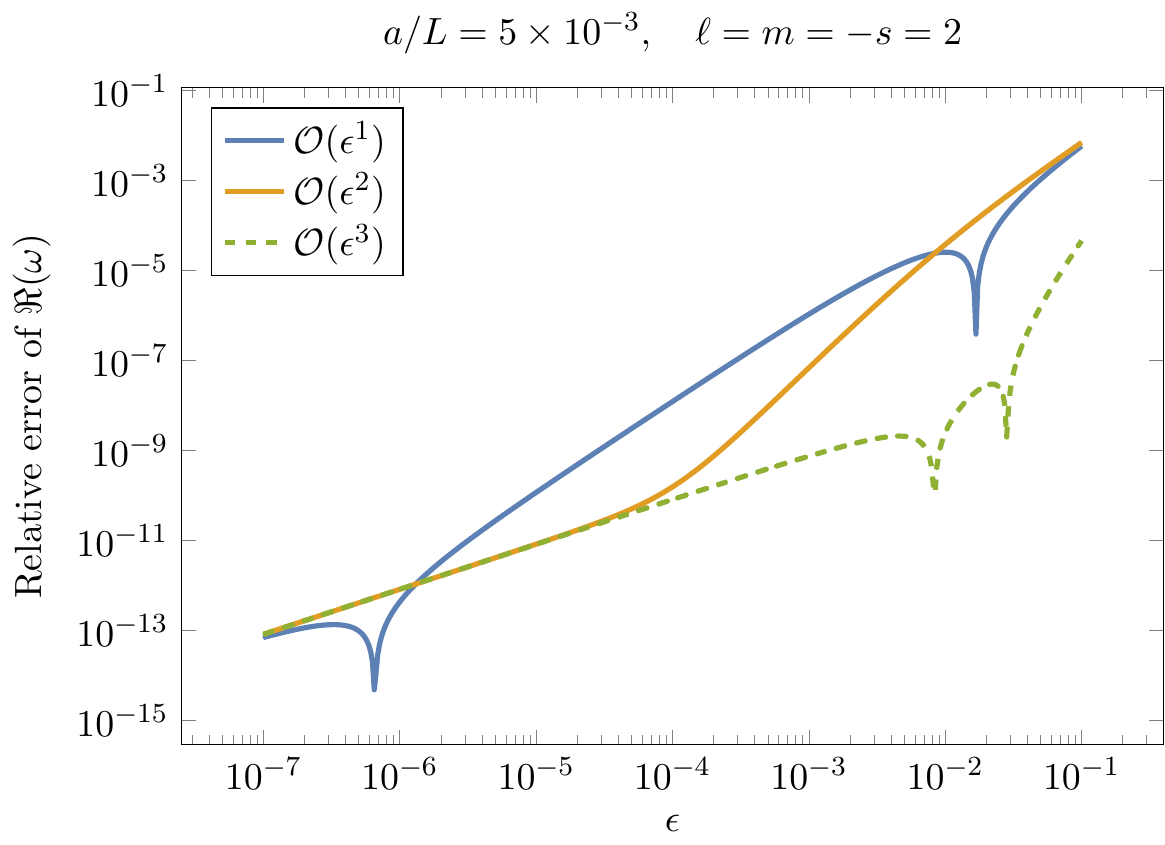}\vspace{.5cm}
   \includegraphics[width=0.73\textwidth]{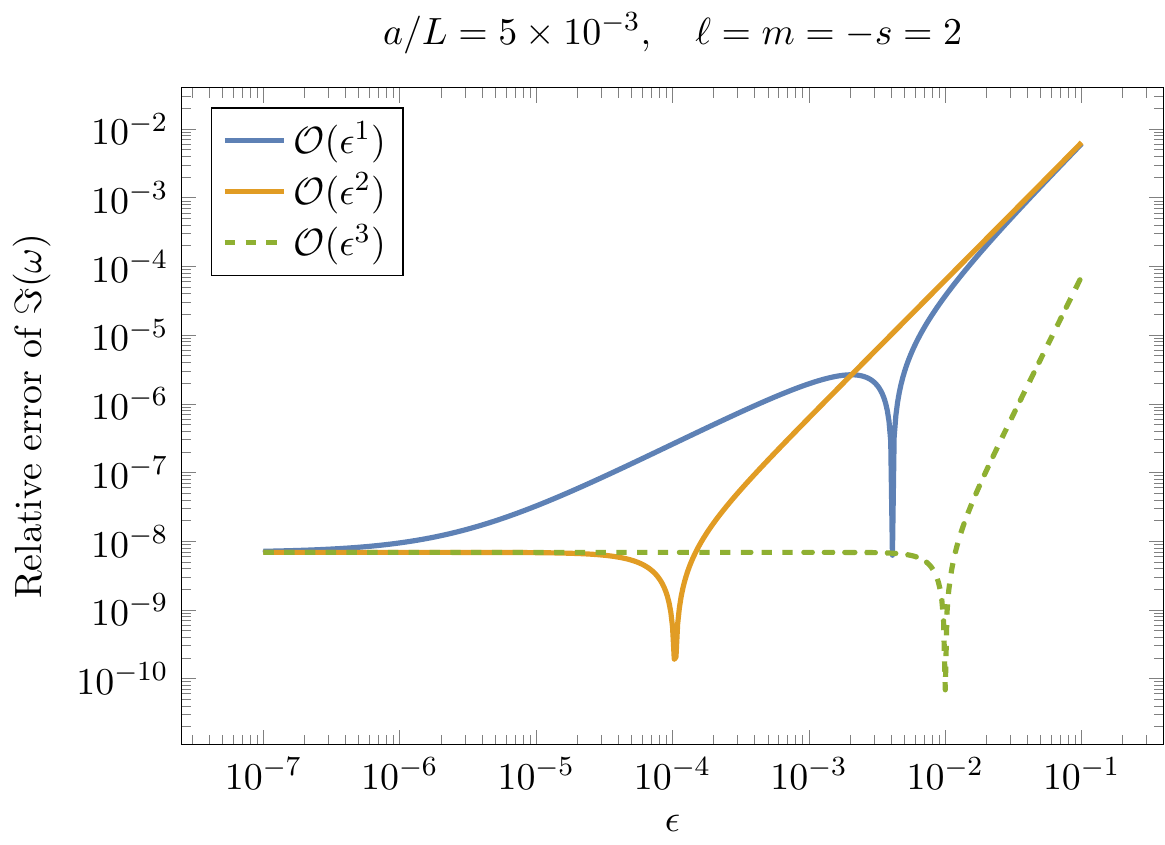}
   \end{center}
\caption{ Relative error in the QNM frequencies when comparing  our analytic expansion in Eq.~\eqref{eq:omega-expansion} with a numerical calculation.  It is plotted  as a function of $\epsilon$ and for the mode $s=-2$, $\ell=m=2$ and $a=5\cdot 10^{-3}L$.
The top graph is for the  real part of the QNM frequency and the bottom graph for the imaginary part.
Within each plot, the curve which is at the top corresponds to the relative error with the analytic expansion truncated to $O(\epsilon)$. The subsequent two lower curves correspond to truncation at $O(\epsilon^2)$ and $O(\epsilon^3)$, respectively.\FN{Fonts are too small}\CM{This depends on the journal. For a two-column paper yes, but not for A4-single-column journal. But I agree about the ticks.}\CM{When removed the comments, the figures will comeback to its normal position.}}
 \label{fig:Rel err}
\end{figure}

\begin{figure}
\begin{center}
  \includegraphics[width=0.77\textwidth]{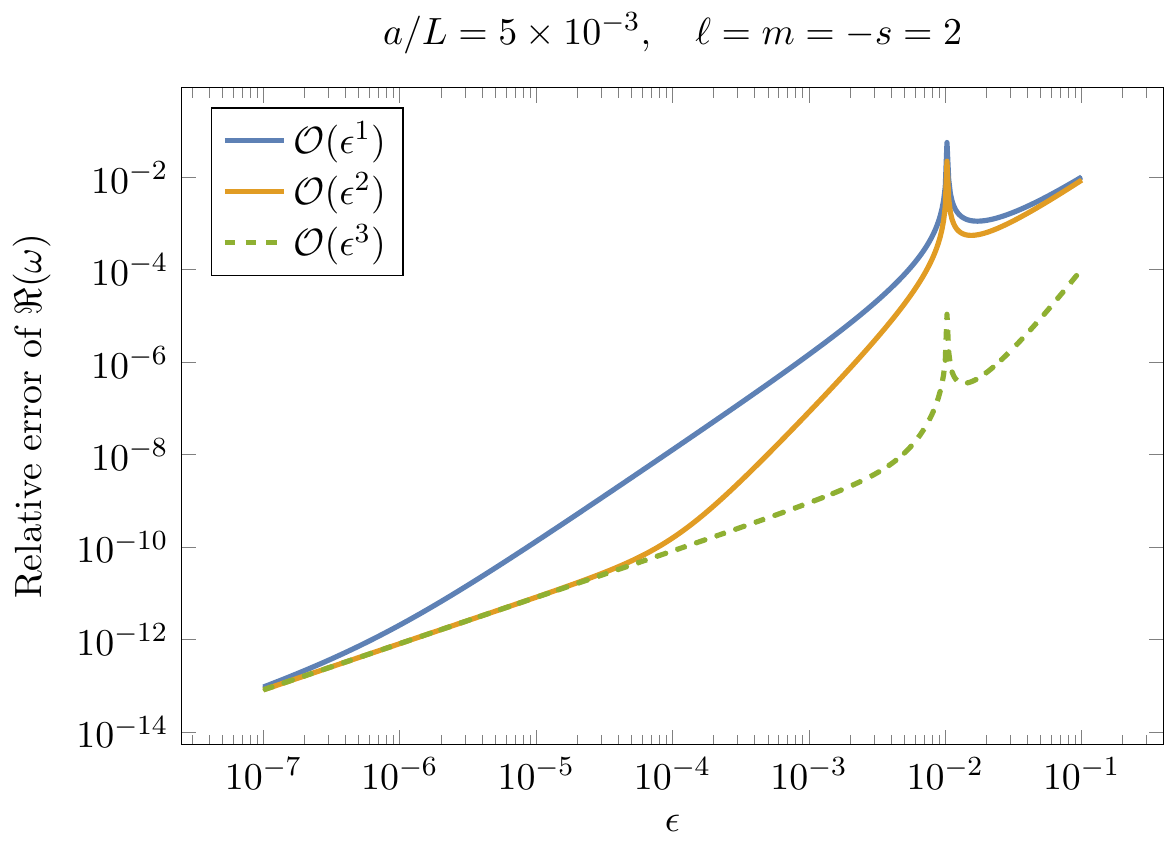}\vspace{.5cm}
   \includegraphics[width=0.77\textwidth]{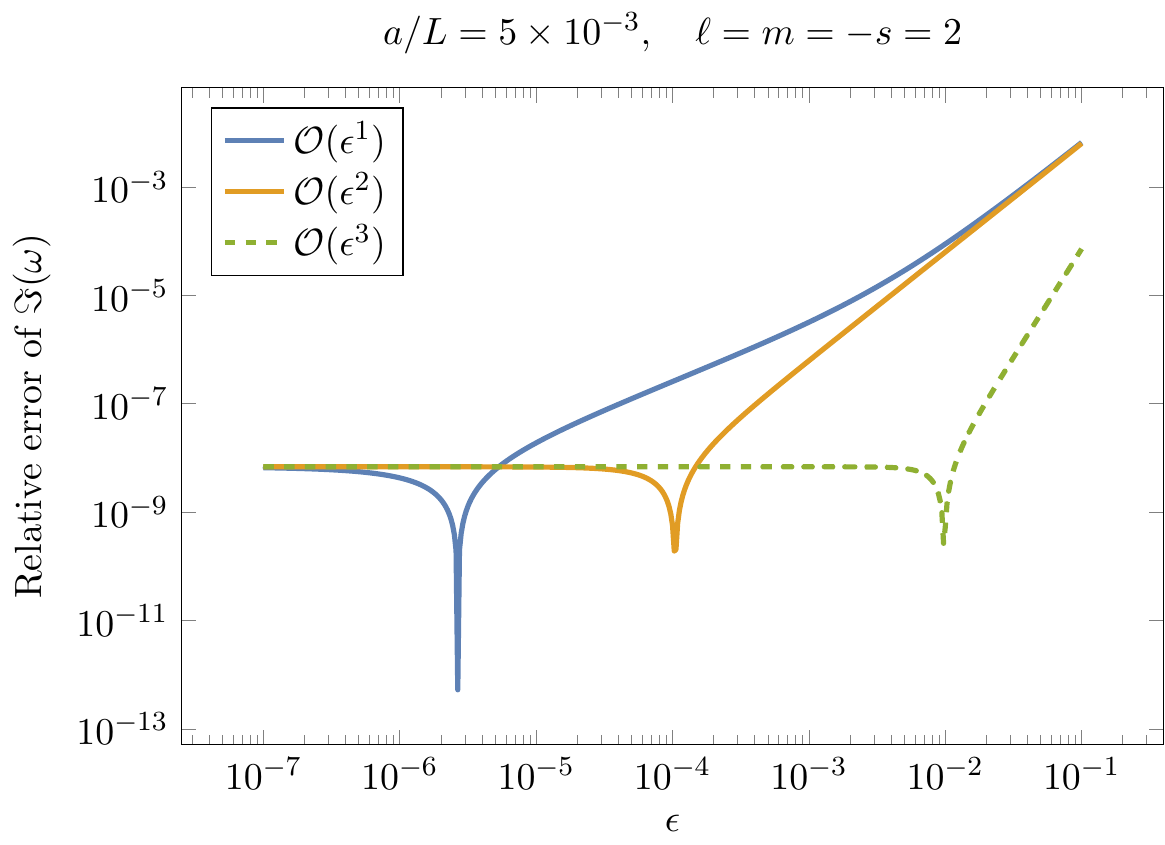}
   \end{center}
\caption{
Similar to figure~\ref{fig:Rel err} but here it refers to the QNM frequency expansion about $m\Omega_C$ instead of $m\Omega_+$ (i.e., the right hand side of Eq.~\eqref{eq:omega-expansion} after taking $\Omega_{+} \to \Omega_{C},\; \epsilon \rightarrow -\epsilon $ and complex conjugating).}
 \label{fig:Rel err rc}
\end{figure}

In Figs.~\ref{fig:Rel err} and \ref{fig:Rel err rc} we plot token
results of our numerical check.  The plots show that each order in our
small-$\epsilon$ expansion of the QNM frequencies consistently
improves the expansion relative to the numerical results.  The plots
are for $s=-2$, but we also carried out numerical checks for
$s=0, -1/2$ and $-1$ and found similarly good agreement with our
analytical expansions.

\section{Conclusions}
\label{sec:conclusions}

Quasinormal Modes are an interesting topic of research from
theoretical as well as observational points of view. In this work, we
provided a new method for calculating QNMs frequencies and angular
eigenvalues for Kerr--de Sitter black holes. This method consists of
using the accessory parameter expansion, given in terms of monodromies
of the associated Fuchsian ODE.  In particular, we have extended the
APE described in Ref.~\cite{Lencses:2017dgf} to the case where the
monodromies are expanded in a small parameter $\epsilon$. This allowed
us to obtain analytical expansions for both the angular eigenvalues
and the QNM frequencies.  Our expansion for the eigenvalues agrees
with results in the literature, whereas, to the best of our knowledge,
our high-order expansion for the frequencies is new.  Finally, we
provided numerical evidence that these analytical formulas are a very
good approximation for reasonably small $\epsilon$.

There are a couple of directions for improving or extending our
results. Firstly, as the $\tau$-function used here has a complete
expansion, there is a possibility that a deeper study of its
properties and symmetries can provide even more compact formulas or
with faster convergence than the ones used here. Our implementation of
the algorithm provided in \cite{Lencses:2017dgf} is limited as the
expressions of order higher than $\epsilon^{3}$ are too cumbersome to
be simplified with our available computing power.  We are confident
that simplification of these expressions is possible, but we leave
this optimization for future work.  Another possibility is to
implement the APE numerically, as it was done in
\cite{Anselmo2018}. There are also other approaches to obtain the APE,
like the ones in
\cite{Castro2013b,Menotti2014,Menotti2016,Hollands2017} or the direct
application of CFT operator product expansions, as a variation of what
was done in \cite{Litvinov:2013sxa,Lencses:2017dgf}. Secondly,
although our method works very well for a large region of
configuration space $(\epsilon,a)$, we were limited to study the APE
expansion close to the extremal limit $\epsilon\rightarrow 0$. We need
the other extremal limit $\epsilon\rightarrow 1$
($r_{+}\rightarrow r_{-}$) in order to try to cover the full
configuration space for Kerr--de Sitter black holes. This could in
principle be implemented with an expansion of the $\tau$-function as
$t\rightarrow 1$, but, in this case, the expansion would be in terms
of the other composite monodromy $\sigma_{1x}$. One would therefore
need to study the relationship between $\sigma_{1x}$ and $\sigma_{0x}$
\cite{Litvinov:2013sxa,Iorgov2013}, which sets the relevant boundary
conditions on the modes.

There are also other possible generalizations and applications of the
APE method itself. For example, in~\cite{Amado2017}, the scalar-field
perturbation equations of a $5$-dimensional Kerr--anti-de Sitter black
hole were written as Heun equations and the eigenvalue and QNM
conditions were written in terms of the monodromies (similarly to
section~\ref{sec:monodromy} here in the case of Kerr--de Sitter).  The
APE method applied to this black hole setting, verified numerically,
is available in \cite{amado2018}. Moreover, the connection between the
AdS/CFT approach to QNMs and the CFT structure behind isomonodromic
method still remains to be completely understood.

One could also try to extend the APE to the case of an ODE with either
less or more singular points.  In particular, the APE could be used to
study confluent cases of the Heun equation, corresponding to two
regular and one irregular singular points.  The linear perturbations
of Kerr and Schwarzschild black holes are described by confluent Heun
equations and the monodromy setup for these cases has been studied in
\cite{daCunha:2015ana}. However, this example still lacks a systematic
expansion using the Painlevé V $\tau$-function
\cite{Gamayun:2013auu,Lisovyy2018}, like in \cite{Lencses:2017dgf} for
the Painlevé VI case.  The same principle of finding the APE could
also be used for an ODE possessing more than four singular points,
using the respective $\tau$-function
\cite{Iorgov:2014vla,Gavrylenko2016}.  It is known that linear
perturbation equations for Kerr-NUT-(Anti-)de Sitter black holes are
also reducible to separable ODEs for dimensions higher than four
\cite{Frolov2017}. In particular, the number of singular points of
these equations increases as the dimension increases.

Another potentially interesting, but speculative, line of
investigation is to try to relate the isomonodromic approach with the
Kerr/CFT correspondence\cite{Guica2009}, which so far only deals with
the (near-)extremal ($r_+=r_-$) case \cite{Compere2012}.  The
near-extremality expansion used here might be helpful to understand
whether the CFT picture in the extremal case could in some sense be
deformed to cater for the non-extremal case. In this sense, it is
curious that the $\tau$-function can be understood as a $c=1$ CFT
chiral correlator (even in the non-extremal case). However, the
connection with the Kerr/CFT description, if any, still remains to be
understood. Discussions along these lines can be found in
\cite{Novaes2014c,daCunha:2016crm,Amado2017}. In particular, there is
also room to explore the Kerr-dS/CFT
correspondence~\cite{anninos_holography_2009}, which has been less
explored than Kerr/CFT. The approach of \cite{Piatek2017} to obtain
the full solutions of Heun's equation from CFT is also an interesting
related direction.

Finally, the APE can also be used outside the realm of black hole
physics. Particular examples are the Rabi model in quantum optics and
its extensions \cite{Braak2011,Cunha2016} and finding conformal
mappings in 2-dimensions \cite{Anselmo2018}, as well as potential
applications in condensed matter systems.

The APE method has opened new directions for studying spectral
problems of Fuchsian ODEs -- here, in particular, within the realm of
black hole physics and Heun equations. We hope that the study
presented here may be helpful for a better understanding of the
integrable structure behind black holes, their QNMs and their
connections to CFT.


\acknowledgments{
  We thank Alejandra Castro, Bruno Carneiro da Cunha and Amílcar
  Queiroz for stimulating discussions about this work. F.N. and M.L. acknowledge
  financial support from the Brazilian Ministry of Education
  (MEC). 
  M.C. acknowledges partial financial support by CNPq (Brazil), process number 310200/2017-2.
  }

\appendix
\section{Gauge Transformations of Fuchsian Equations}
\label{sec:from-self-adjoint}

Gauge transformations of Fuchsian equations are transformations that
preserve the monodromy coefficients and the character of its singular
points.\footnote{Except when these are apparent or removable
  singularities, which can be made explicit after a series of gauge
  transformations} There are two main gauge transformations in this
sense: homographic transformations of the independent variable (also
known as Möbius transformations) and s-homotopic transformations of
the solution (i.e., multiplication by a solution of certain first
order ODEs with polynomial coefficients) \cite{slavyanov2000}. Here we
apply these two transformations to reduce the master radial and
angular equations \eqref{eq:master-eqs-2} to its \emph{canonical} and
\emph{normal} forms.

A faster route for obtaining the canonical form of
Eqs.~\eqref{eq:angulareq} and~\eqref{eq:radialeq} is to work directly
with the corresponding linear system. Consider the following equation
in \emph{self-adjoint} form\footnote{We call the equation \eqref{eq:SAform}
  self-adjoint even if its coefficients are complex.}
\begin{equation}
  \label{eq:SAform}
  \partial_r(U(r)\partial_r\psi(r)) - V(r)\psi(r) = 0.
\end{equation}
We consider the case that $U$ and $V$ are 
rational functions of $r$ and \eqref{eq:SAform}
has $n$ regular singular points at $r=r_{k},\; k=1,\dots ,n$, with $r_{n}
= \infty$.  
This equation can be rewritten as
\begin{equation*}
  \label{eq:linear_system}
  (\partial_r -A(r))\Psi(r) = 0 \;, \quad A \equiv
  \begin{pmatrix}
    0 & U^{-1}\\
    V & 0
  \end{pmatrix}\;,\quad \Psi \equiv
  \begin{pmatrix}
    \psi \\ U\partial_{r}\psi
  \end{pmatrix}.
\end{equation*}
As in electromagnetism, we shall call $A(r)$ the \emph{gauge
  connection} of these Fuchsian equations. What happens to this system
after performing a homographic transformation and a s-homotopic
transformation?  First, let us apply the following homographic
transformation
 \begin{equation*}
   z \equiv \frac{d_1 r+d_2}{d_3 r+d_4} \entao \partial_{r} =F^{-1}(r)\partial_{z}
   \defeq \frac{d_1d_4-d_2 d_3}{(d_3 r+d_4)^{2}}\partial_{z} \;,
 \end{equation*}
 where $d_{1,2,3,4}$ are complex constants. This maps the points
 $r=r_{k}$ to new points $z=z_{k}$, $k=1,\ldots,n$,
 and leads to a new gauge connection
\begin{equation}
(\partial_z -\tilde A(z))\Psi(z) = 0,\qquad
  \label{eq:A2}
  \tilde A =
  \begin{pmatrix}
    0 & \tilde U^{-1}\\
    \tilde V & 0
  \end{pmatrix} \defeq 
   \begin{pmatrix}
    0 & FU^{-1}\\
    FV & 0
  \end{pmatrix}
\end{equation}
with $\Psi = (\psi(z), \;\tilde U(z)\partial_{z}\psi(z))^{T} $.  Now,
let us apply the following s-homotopic transformation to the solution
\begin{gather*}
  \label{eq:21}
  \psi = G(z)\tilde\psi
=  \prod_{i=1}^{n}(z-z_{i})^{-\rho_{i}/2}~\tpsi,\entao  \Dp\psi = G\Dp\tilde{\psi} + \left(\Dp G\right)\tilde{\psi} = G(\Dp + B)\tilde{\psi},
\end{gather*}
 for some $\rho_{i}\in \mathbb{C}$,
where
\begin{equation}
  \label{eq:B_expansion}
   B \defeq -\fosum[n]{\rho_{i}/2}.
\end{equation}
Thus,
\begin{align*}
  \Psi &= (G\tpsi,\; \tilde U G(\Dp + B)\tpsi)^{T} =
  \hat{U}(z)\tPsi, \\[5pt]
  \Dp\Psi &= \hat{U}\Dp\tPsi + \Dp \hat{U} \tPsi,
\end{align*}
where
\begin{equation*}
\hat{U}(z)\equiv  \begin{pmatrix}
    G & 0 \\ 
    \tilde U GB &  G
  \end{pmatrix},\qquad \tPsi\equiv (\tpsi,\; \tilde U \Dp\tpsi)^{T},
\end{equation*}
 and the new
gauge potential is
$\bar{\tilde A} = \hat{U}^{-1}\tilde A \hat{U} - \hat{U}^{-1}\Dp
\hat{U}$. Calculating all terms, we have that
\begin{equation}
  \bar{\tilde A} =
    \begin{pmatrix}
    0 & \tilde U^{-1}\\
    \tilde V - \tilde U (B^{2}+ \Dp B) - \Dp{\tilde U} B\quad & -2B
  \end{pmatrix}.
\end{equation}
This can be  simplified further by writing the linear system in terms
of $ (\tilde{\psi} , \Dp\tilde{\psi})^{T}$, removing $\tilde U$ by making
\begin{equation}
    \tilde\Psi = \begin{pmatrix} 
    1 & 0 \\
     0 & \tilde U
      \end{pmatrix}(\tilde{\psi} , \Dp\tilde{\psi})^{T}.
\end{equation}
Removing the tildes and primes in the new gauge
connection and linear system solution  (i.e., relabel $\bar{\tilde A}$ by $A$, $\tPsi$ by $\Psi$ and $\tilde\psi$ by $\psi$),
we are left with a system of the form
\begin{gather}
  \Dp \Psi = A(z)\Psi,\quad
   A =
  \begin{pmatrix}
    0 & 1\\
    \tilde V\tilde U^{-1}- (B^{2}+ \Dp B - B\Dp\log\tilde U)\quad  & -2B
    - \Dp\log\tilde U
  \end{pmatrix},
\end{gather}
where $\Psi = (\psi , \Dp\psi)^{T}$.  This results in the following
ODE for the transformed $\psi$
\begin{subequations}
  \label{eq:canonical_ODE}
  \begin{gather}
  \partial^{2}_{z}\psi + P(z)\Dp \psi + Q(z)\psi = 0,\\[5pt]
P(z) \equiv \Dp \log\tilde U + 2B, \quad Q(z) \equiv B^{2} + \Dp B +
B\Dp\log\tilde U  - \tilde V\tilde U^{-1}.
\end{gather}
\end{subequations}

\subsection{Radial Master Equation}
\label{sec:schw-de-sitt}

The master radial equation \eqref{eq:radialeq} is of the form
\begin{equation}
  \label{eq:SAform-radial}
  \partial_r(U(r)\partial_rR_{s,\omega,m}(r)) - V(r)R_{s,\omega,m}(r) = 0,
\end{equation}
with
\begin{align}
  \label{eq:A9}
  U&= \Delta_{r}^{s+1},\\[5pt]
  V&=
     U
     \left[
     -\left(
     \frac{W}{\Delta_{r}}
     \right)^{2}
     -is
     \left(
     2\frac{W'}{\Delta_{r}} -\frac{W\Delta'_{r}}{\Delta^{2}_{r}}
     \right)+\frac{Y_{s}}{\Delta_{r}}
     \right].
\end{align}
We now apply the Möbius transformation
\begin{equation}
  \label{eq:9}
  z 
  =
  \zeta_{\infty}\frac{(r-r_{1})}{(r-r_{4})}
  \defeq\frac{(r_{2}-r_{4})}{(r_{2}-r_{1})}\frac{(r-r_{1})}{(r-r_{4})}, \quad x\equiv \zeta_{\infty}\frac{r_{31}}{r_{34}},\quad r_{ij}\equiv r_i-r_j,
\end{equation}
with $i,j=1,2,3,4$,
and the inverse transformation
\begin{equation}
  \label{eq:inverse_homographic}
  r= \frac{r_{4}z - r_{1}\zeta_{\infty}}{z-\zeta_{\infty}}
  \quad\entao\quad r- r_{i} =\frac{r_{4i}z+ r_{i1}\zeta_{
      \infty}}{z-\zeta_{\infty}}, \quad i=1,\dots,4.
\end{equation}
Then
\begin{equation*}
  \partial_{r} = \frac{r_{14}\zeta_{\infty}}{(r-r_{4})^{2}}\partial_{z} =
  -\frac{(z-\zeta_{\infty})^{2}}{r_{41}\zeta_{\infty}}\Dp \entao F(z) = -\frac{r_{41}\zeta_{\infty} }{(z-\zeta_{\infty})^{2}},
\end{equation*}
and
\begin{equation}
  \label{eq:5}
  \Delta_{r} = -\Delta_{r}'(r_{4})F(z)\frac{f(z)}{(z-\zeta_{\infty})^{2}},
  \quad f(z)\equiv z(z-1)(z-x).
\end{equation}
We also apply the homotopic transformation
\begin{equation}
  \label{eq:4}
  R_{s,\omega,m}(r) =G(z) \psi(z),\quad G(z)= (z-\zeta_{\infty})^{\beta}\prod_{i=1}^{3}(z-z_{i})^{-\rho_{i}/2},
\end{equation}
for some constant $\beta$, which generates a new ODE of the form
\eqref{eq:canonical_ODE}. Notice that we omitted the indices
$s,\omega,m$ in the solution, as they do not play any role in this
appendix. To make \eqref{eq:canonical_ODE} more explicit, first, we
have
\begin{equation}
  \label{eq:6}
  \tilde{\Delta}_{r}\equiv F^{-1}\Delta_{r}  \entao \tilde{U} =
  F^{-1}\Delta_{r}^{s+1}= F^{s}\tilde{\Delta}_{r}^{s+1}
\end{equation}
and
\begin{align}
  \label{eq:7}
 \frac{\tilde{V}}{\tilde{U}} &= F^{2}\frac{V}{U}\nonumber \\[5pt]
&=  
     \left[
     -\left(
     \frac{W}{\tilde{\Delta}_{r}}
     \right)^{2}
     -is
     \left(
     2\frac{\del[z]W}{\tilde{\Delta}_{r}} -\frac{W\del[z](F\tilde{\Delta}_{r})}{F\tilde{\Delta}^{2}_{r}}
     \right)+F\frac{Y_{s}}{\tilde{\Delta}_{r}}
     \right]\nonumber \\[5pt]
&= \left[
     -\left(
      \frac{W}{\tilde{\Delta}_{r}} 
     \right)^{2}
     -is
     \left(
     2\del[z]
  \left(
  \frac{W}{\tilde{\Delta}_{r}}
  \right)
  +\frac{W}{\tilde{\Delta}_{r}}\del[z]\log
  \left(
  \frac{\tilde{\Delta}_{r}}{F}
  \right)
     \right)+F\frac{Y_{s}}{\tilde{\Delta}_{r}}
     \right].
\end{align}
Then, we can find that
\begin{equation}
  P(z) =
   \frac{2\left(\beta-(2s+1)\right)}{z-\zeta_{\infty}}+
  \sum_{i=1}^{3}\frac{1+s-\rho_{i}}{z-z_{i}}, 
\end{equation}
and
\begin{align}
  \label{eq:Q_function}
  Q(z) = 
         -\frac{\tilde{V}}{\tilde{U}}+\frac{\beta^{2}-\beta(4s+3)}{(z-\zeta_{\infty})^{2}} +
         \frac{q_{\infty}}{z-\zeta_{\infty}} 
       +  \sum_{i=1}^{3}
         \left(
         \frac{\rho_{i}(\rho_{i}-2s)/4}{(z-z_{i})^{2}}+\frac{q_{i}}{z-z_{i}}
         \right),
\end{align}
with
\begin{align}
  \label{eq:Q_function-2}
q_{\infty} &= -\sum_{i=1}^{3}
\frac{ (1+s)\beta+(1+2s-\beta)\rho_{i}}{z_{i}-\zeta_{\infty}},\\[5pt]
  q_{i} &= 
        \frac{(1+s)\beta+(1+2s-\beta)\rho_{i}}{z_{i}-\zeta_{\infty}}
     +   \sum_{j=1,j\neq
  i}^{3}\frac{\rho_{i}\rho_{j}-(1+s)(\rho_{i}+\rho_{j})}{2(z_{i}-z_{j})},\quad i=1,\ldots, 3.
\end{align}

It is helpful to expand Eq.~\eqref{eq:7} into partial
fractions. Essentially, we have the following two types of terms:
\begin{equation}
  \label{eq:14}
\frac{  A r^{2} + B}{\tilde{\Delta}_{r}}, \quad 
\left(
  \frac{  C r^{2} + D}{\tilde{\Delta}_{r}}\right)F(z),
\end{equation}
for some constants $A,B,C,D$.
The main difference between these two terms is that the first one has no pole at
$z=\zeta_{\infty}$, whereas the second one has a pole there due to the $F(z)$
factor. Thus, we have
\begin{equation}
  \label{eq:A22,W_as_residues}
  \frac{W}{\tilde{\Delta}_{r}} =
  \sum_{k=0,1,x}\frac{W_{k}}{z-z_{k}},\quad W_{k} \equiv \res_{r=r_{k}}\frac{W(r)}{\Delta_{r}(r)}= \frac{\chi^{2}(\omega(r_{k}^{2}+a^{2})-am)}{\Delta'_{r}(r_{k})},
\end{equation}
and
\begin{equation}
  \label{eq:A23}
  F\frac{Y_{s}}{\tilde{\Delta}_{r}} =
  \frac{2(s+1)(2s+1) (\sum_{i=1}^{3}r_{i}-r_{4})/(r_{14}\zeta_{\infty})}{z-\zeta_{\infty}}
  -\frac{2(s+1)(2s+1)}{(z-\zeta_{\infty})^{2}}+
  \sum_{i=0,1,x}F(z_{i})\frac{Y_{s,i}}{z-z_{i}},
\end{equation}
with
\begin{equation}
  \label{eq:17}
  Y_{s,i} \equiv \frac{2(s+1)(2s+1) r_{i}^{2}/L^{2}-s(1-a^2/L^2)+\lamslm}{\Delta'_{r}(r_{i})}.
\end{equation}
Then, going back to Eq.~\eqref{eq:7}, we get
\begin{align}
  -\frac{\tilde{V}}{\tilde{U}} =& 
  \frac{2(s+1)(2s+1)}{(z-\zeta_{\infty})^{2}}- \frac{2(s+1)(2s+1) (\sum_{i=1}^{3}r_{i}-r_{4})/(r_{14}\zeta_{\infty})}{z-\zeta_{\infty}}
  +\nonumber \\[5pt]
& + \sum_{k=1}^{3}
  \left(
  \frac{W_{k}(W_{k}-i s)}{(z-z_{k})^{2}}+ \frac{v_{k}}{z-z_{k}}
  \right) ,
\end{align}
with
\begin{equation}
  \label{eq:A26,vk_def}
  v_{k} \equiv 
  - F(z_{k})Y_{s,k}+
  \sum_{j=1, j\neq k}^{3}\frac{2W_{k}W_{j}+i s (W_{k}+W_{j})}{z_{k}-z_{j}} ,\qquad
  k=1,2,3.
\end{equation}
Let us first check that the singularity $z=\zeta_{\infty}$ is
removable. We start by demanding that
\begin{align}
  \label{eq:21-residus of P and Q}
\res_{z=z_{k}}P(z) &=2 \left(\beta-(2s+1)\right)=0,\\[5pt]
  \res_{z=z_{k}}(z-z_{k})Q(z) &= \beta^{2}-(4s+3)\beta+(2s+1)(2s+2)=0.
\end{align}
Both equations have a solution for $\beta = 2s+1$ and we will assume
this value for $\beta$ from now on in this subsection. Now, quite surprisingly, substituting $\beta$ in
Eq.~\eqref{eq:Q_function}, we obtain
\begin{align}
  \label{eq:22}
  \res_{z=z_{k}}Q(z) &=
  -\frac{(2s+2)(2s+1)
                       (\sum_{i=1}^{3}r_{i}-r_{4})}{r_{14}\zeta_{\infty}}
  -\sum_{i=1}^{3}
\frac{ (1+s)(2s+1)}{z_{i\infty}} 
\\[5pt]
 &= \frac{(2s+1)(2s+2)}{r_{14}\zeta_{\infty}}
   \left(
r_{4}+   \sum_{i=1}^{3}\frac{(r_{i}-r_{4})}{2}
   -\sum_{i=1}^{3}r_{i}
   \right)\\[5pt]
  &=-\frac{(2s+1)(2s+2)}{2 r_{14}\zeta_{\infty}}
    \left(
   r_{4}+ \sum_{i=1}^{3}r_{i}
    \right) =0,
\end{align}
where in the second line we used
$z_{i\infty} = -r_{14}\zeta_{\infty}/r_{i4}$, coming from
Eq.~\eqref{eq:inverse_homographic}, and in the last line we used the
fact that there is no third order term in the polynomial
$\Delta_{r}(r)$. Therefore, the singularity $z=\zeta_{\infty}$ is
completely removable from the ODE. This fact has been proved for any
type D vacuum solution with cosmological constant in
\cite{Batic2007a}. Our derivation is an improvement to that work, as
it gives the explicit dependence on the $\rho_{i}$ and a more compact
and suggestive final result in terms of the monodromies.

At this stage, we have
\begin{subequations}
  \label{eq:25}
  \begin{align}
  P(z)& = \sum_{k=1}^{3}\frac{1+s-\rho_{k}}{z-z_{k}},\\[5pt]
  Q(z) &=  \sum_{k=1}^{3}
         \left(
         \frac{\rho_{k}(\rho_{k}-2s)/4+W_{k}(W_{k}-i s)}{(z-z_{k})^{2}}+\frac{q_{k}+v_{k}}{z-z_{k}} 
         \right),\\[5pt]
  q_{k} &= 
  \frac{(1+s)(1+2s)}{z_{k}-\zeta_{\infty}}+
  \sum_{j=1, j\neq k}^{3}\frac{\rho_{k}\rho_{j}-(1+s)(\rho_{k}+\rho_{j})}{2(z_{k}-z_{j})},\\[5pt]
  v_{k} &= 
  \frac{r_{41}\zeta_{\infty}}{(z_{k}-\zeta_{\infty})^{2}}\,Y_{s,k}+
  \sum_{j=1, j\neq k}^3\frac{2W_{k}W_{j}+i s (W_{k}+W_{j})}{z_{k}-z_{j}} ,\quad k=1,2,3.
\end{align}
\end{subequations}
We now have two possibilities to simplify the Heun equation
\eqref{eq:canonical_ODE}:
the canonical form and the normal form.

\vspace{0.5cm}

\noindent\textbf{Canonical Form}. In order to find the canonical form of the Heun
equation, we  choose $\rho_{k}$ so as to cancel the $(z-z_{k})^{-2}$
term in $Q(z)$ in \eqref{eq:Q_function}.  In order to find the Frobenius exponents $\rho_{k}$, we
take
\begin{align}
  \label{eq:23}
  \res_{z=z_{k}}(z-z_{k})Q(z) &= \frac{1}{4}\rho_{k}
                                (\rho_{k}-2s)+W_{k}(W_{k}-is)=0.
\end{align}
Its roots are given by
\begin{align}
  \label{eq:A35-P,Q,Qk}
  \rho^{(-)}_{k} = -2iW_{k},\quad &\rho_{k}^{(+)}=2iW_{k}+2s
  \entao \rho_{k}^{(\pm)} = s\pm (2iW_{k}+s),\quad k=1,2,3.
\end{align}
Then, if we define $\theta_{k} = iW_{k}+s/2$ and choose $\rho_{k} = \rho_{k}^{(+)}=s+2\theta_{k}$ we obtain
\begin{align}
  \label{eq:26}
  P(z)& = \sum_{k=1}^{3}\frac{1-2\theta_{k}}{z-z_{k}},\quad Q(z) =  \sum_{k=1}^{3}
        \frac{Q_{k}}{z-z_{k}}, \\[10pt]
  Q_{k} &= 
   \frac{(1+s)(1+2s)}{z_{k}-\zeta_{\infty}}-F(z_{k})Y_{s,k}
   +\sum_{j=1,j\neq k}^{3}\frac{(2s-1)(\theta_{k}+\theta_{j}-s)-2s}{z_{k}-z_{j}}.
\end{align}
We can rewrite the last term in two convenient ways
\begin{align}
  \label{eq:27}
  -F(z_{k})Y_{s,k} &= -\frac{L^{2}}{z_{k}-\zeta_{\infty}}\, \frac{(2(2s+1)(s+1)r_{k}^{2}/L^2+\lamslm -s(1-a^2/L^2))}{\prod^{3}_{j=1,j\neq k}(r_{k}-r_{j})}\\[5pt]
                   &=\frac{(r_{1}-r_{4})\zeta_{\infty}}{\Delta'_{r}(r_{4})}\,\frac{2(2s+1)(s+1)r_{k}^{2}/L^{2}+\lamslm -s(1-a^2/L^2)}{f'(z_{k})}\nonumber\\[5pt]
  &=-\frac{L^2}{(r_{1}-r_{2})(r_{3}-r_{4})}\frac{(2(2s+1)(s+1)r_{k}^{2}/L^2+\lamslm -s(1-a^2/L^2))}{f'(z_{k})}.
\end{align}
Comparing with the radial Heun equation \eqref{eq:heuncanonicalangular}, we can extract Heun's accessory parameter in \eqref{eq:26} by $K_{x} =-\res_{z=x}Q(z)= -Q_{3}$ and this gives
\begin{multline}
  \label{eq:canonical-accessory-parameter}
  K_{x} =
  \frac{(1-2s)(\theta_{x}+\theta_{0}-s)+2s}{x}+\frac{(1-2s)(\theta_{x}+\theta_{1}-s)+2s}{x-1}-\\[5pt]
  -\frac{1}{x-\zeta_{\infty}}
  \left[(1+s)(1+2s)-\frac{(2(2s+1)(s+1)r_{k}^{2}+\lamslm L^{2}-sL^2(1-a^2/L^2))}{(r_{3}-r_{1})(r_{3}-r_{2})}
  \right]  
\end{multline}
or
\begin{multline}
  \label{eq:canonical-accessory-parameter-2}
  K_{x} =
  \frac{(1-2s)(\theta_{x}+\theta_{0}-s)+2s}{x}+\frac{(1-2s)(\theta_{x}+\theta_{1}-s)+2s}{x-1}-\\[5pt]
  -\frac{(1+s)(1+2s)}{x-\zeta_{\infty}}
  +\frac{1}{(r_{1}-r_{2})(r_{3}-r_{4})}\frac{(2(2s+1)(s+1)r_{k}^{2}+\lamslm L^{2}-sL^2(1-a^2/L^2))}{x(x-1)},
\end{multline}
which is the best form to multiply it by $x(x-1)$.

\vspace{0.5cm}

\noindent\textbf{Normal Form.} In the APE, we actually use the normal form
of the accessory parameter, since our formulas for the $\tau$-function  are
adapted to this form. The normal form is obtained by making
$\rho_{k}=1+s$, $k=1,2,3$  in Eq.~\eqref{eq:25}, which yields
\begin{align}
  \label{eq:28}
  P(z)&=0,\quad Q(z) = \sum_{k=1}^{3}
        \left(
        \frac{\frac14 - \theta_{k}^{2}}{(z-z_{k})^{2}}+ \frac{\tilde{Q}_{k}}{z-z_{k}}
        \right),\\[10pt]
  \tilde{Q}_{k} &= \frac{(1+s)(1+2s)}{z_{k}-\zeta_{\infty}}-F(z_{k})Y_{s,k}
                  +\sum_{j=1,j\neq k}^{3}\frac{-\frac{1}{2}(1+2s+4\theta_{j}\theta_{k})+2s(\theta_{j}+\theta_{k}-s)}{z_{k}-z_{j}}
                  .
\end{align}
Then, $\mathcal{H}_{x} = -\res_{z=x}Q(z) = -\tilde{Q}_{3}$, which is related
to \eqref{eq:canonical-accessory-parameter-2} by \eqref{eq:htok}.

\subsection{Angular Master Equation}
\label{sec:reduct-angul-mast}

Most of the expressions in this angular subsection readily follow from those in the previous radial case presented in Subsection~\ref{sec:schw-de-sitt}.
Since they are separate subsections, we shall use some of the same symbols to denote quantities here defined as the quantities with the same symbol in section~\ref{sec:schw-de-sitt} merely under the change $r\to u$, and so we will not give the definitions of those symbols explicitly here.

The angular master equation \eqref{eq:angulareq} is of the form
\begin{equation}
  \label{eq:SAform-angular}
  \partial_u(U(u)\partial_uS(u)) - V(u)S(u) = 0\;,
\end{equation}
with 
\begin{align}
  \label{eq:A44}
  U&= \Delta_{u},\\[5pt]
  V&=
     -U
     \left[
     -\left(
     \frac{H+\frac{s}{2}\Delta_{u}'}{\Delta_{u}}
     \right)^{2}
     +2s\frac{H'}{\Delta_{u}}
     -\frac{X_{s}}{\Delta_{u}}
     \right].
\end{align}
Applying similar transformations as in section~\ref{sec:schw-de-sitt}, we find
\begin{align}
  \label{eq:29}
  P(z) &=
    \frac{2(\beta-1)}{z-\zeta_{\infty}}+
  \sum_{k=1}^{3}\frac{1-\rho_{k}}{z-z_{k}},\\[5pt]
Q(z) &=-\frac{\tilde{V}}{\tilde{U}}+ \frac{\beta^{2}-3\beta}{(z-\zeta_{\infty})^{2}}+\frac{q_{\infty}}{z-\zeta_{\infty}}
+\sum_{k=1}^{3}
       \left(
       \frac{\rho_{k}^{2}/4}{(z-z_{k})^{2}}+\frac{q_{k}}{z-z_{k}}
       \right)
,\\[5pt]
q_{k} &=
 \frac{\beta+(1-\beta)\rho_{k}}{z_{k}-\zeta_{\infty}}
 +
\sum_{j=1,j\neq
  k}^{3}\frac{\rho_{k}\rho_{j}-(\rho_{k}+\rho_{j})}{2(z_{k}-z_{j})},\\[5pt]
q_{\infty} &= -\sum_{k=1}^{3}
\frac{ \beta+(1-\beta)\rho_{k}}{z_{k}-\zeta_{\infty}}.
\end{align}
We also have
\begin{subequations}
  \begin{align}
  \label{eq:30}
  -\frac{\tilde{V}}{\tilde{U}} &= -
  \left[
  \frac{H}{\tilde{\Delta}_{u}}+\frac{s}{2}\del[z]\log(F\tilde{\Delta}_{u})
  \right]^{2}+2s
  \left[
  \del[z]
  \left(
  \frac{H}{\tilde{\Delta}_{u}}
  \right)+\frac{H}{\tilde{\Delta}_{u}}\del[z]\log\tilde{\Delta}_{u}
  \right]-F\frac{X_{s}}{\tilde{\Delta}_{u}}\nonumber\\[5pt]
&=
                                                              \frac{2}{(z-\zeta_{\infty})^{2}}+\frac{v_{\infty}}{z-\zeta_{\infty}}+
\sum_{k=1}^{3}
  \left(
  \frac{-(H_{k}+\frac{s}{2})^{2}}{(z-z_{k})^{2}}+\frac{v_{k}}{z-z_{k}}
  \right),\\[10pt]
v_{k} &=
\frac{2s^{2}}{z_{k}-\zeta_{\infty}}-F(z_{k})X_{s,k}
-\sum_{j=1,j\neq k}^3\frac{2(H_{k}-s/2)(H_{j}-s/2)}{z_{k}-z_{j}},\\[5pt]
v_{\infty} &= -\sum_{k=1}^3\frac{2s^{2}}{z_{k}-\zeta_{\infty}}-\frac{2(2s^{2}+1)\left(-u_{4}+\sum_{j=1}^3u_{j}\right)}{u_{14}\zeta_{\infty}},\\[5pt]
X_{s,k} &= \frac{2(2s^{2}+1)\alpha^{2}u_{k}^{2}-\lamslm}{\Delta_{u}'(u_{k})},
\end{align}
\end{subequations}
with $k=1,2,3$,
where we used that
\begin{align}
  \label{eq:31}
  \frac{H}{\tilde{\Delta}_{u}} = \sum_{k=1}^3\frac{H_{k}}{z-z_{k}},\quad \tilde{\Delta}_{u} = -\Delta_{u}'(u_{4})\frac{f(z)}{(z-\zeta_{\infty})^{2}},\quad F(z_{k}) = \frac{u_{41}\zeta_{\infty}}{(z_{k}-\zeta_{\infty})^{2}}.
\end{align}
Using Eq.~\eqref{eq:30} in Eq.~\eqref{eq:29}, we find
\begin{align}
  \label{eq:32}
  Q(z) = \sum_{k=1}^{3}
  \left(
  \frac{\rho_{k}^{2}/4-(H_{k}+s/2)^{2}}{(z-z_{k})^{2}}+\frac{q_{k}+v_{k}}{z-z_{k}}
  \right)+\frac{\beta^{2}-3\beta+2}{(z-\zeta_{\infty})^{2}}+\frac{q_{\infty}+v_{\infty}}{z-\zeta_{\infty}}.
\end{align}
For $\beta=1$, the two terms with poles at $z=\zeta_{\infty}$ vanish identically, as in the radial case. 

\vspace{0.5cm}

\noindent\textbf{Canonical Form.} If we substitute
$\rho_{k}= 2\theta_{k} = - 2(H_{k}+s/2)$ in \eqref{eq:32}, we find
\begin{equation}
  \label{eq:canonical-form-angular-eq}
  \begin{aligned}
    P(z) &= \sum_{k=1}^{3}\frac{1-2\theta_{k}}{z-z_{k}},\quad    Q(z) = \sum_{k=1}^{3}\frac{Q_{k}}{z-z_{k}}, \\[10pt]
    Q_{k} &=\frac{2s^{2}+1}{z_k-\zeta_{\infty}}-\frac{1}{u_{12}u_{34}}\frac{2(2s^{2}+1)u_{k}^{2}-\lamslm\alpha^{-2}}{x(x-1)}+\sum_{j=1,j\neq
      k}^3\frac{-(2s+1)(\theta_{k}+\theta_{j}+s)+s}{z_{k}-z_{j}},
  \end{aligned}
\end{equation}
with $k=1,2,3$.

\vspace{0.5cm}

\noindent\textbf{Normal Form.} Substituting $\rho_{k}=1$ in \eqref{eq:32}, we get
\begin{equation}
  \label{eq:normal-form-angular-eq}
  \begin{aligned}
    P(z) &= 0,\quad Q(z) = \sum_{k=1}^{3} \left(
      \frac{\frac14-\theta_{k}^{2}}{(z-z_{k})^{2}}+\frac{\tilde{Q}_{k}}{z-z_{k}}
    \right), \\[10pt]
    \tilde{Q}_{k} &=
          \frac{2s^{2}+1}{z_k-\zeta_{\infty}}-\frac{1}{u_{12}u_{34}}\frac{2(2s^{2}+1)u_{k}^{2}-\lamslm\alpha^{-2}}{x(x-1)}
    -\sum_{j=1,j\neq
      k}^3\frac{1+4(\theta_{k}-s)(\theta_{j}-s)}{2(z_{k}-z_{j})}.
  \end{aligned}
\end{equation}


\section{Alternative Derivation of Quasinormal Mode Equation }
\label{sec:altern-deriv}

In this appendix, we show how to obtain
Eq.~\eqref{eq:spin-s-rotating-nariai-qnm} using a path-multiplicative
solution \cite{ronveaux1995heun} to leading order in $x\rightarrow 0$.
First, let us rewrite Eq.~\eqref{eq:heuneq} in Euler form
\begin{equation}
  \label{eq:53}
  \left\{
     (z-x)D_{\text{HG}}[b_1,b_2; 1-2\theta_{0}]+ (1-2\theta_{x})zD_{z} -  x(x-1)K_{x}z
  \right\}y(z) =0,
\end{equation}
where $D_{z} \equiv z\Dp$ is the Euler derivative and we define
the\emph{ hypergeometric differential operator} as
\begin{equation}
  \label{eq:54}
D_{HG}[A,B;C] \defeq  z (D_{z}+A)(D_{z}+B)- D_{z}(D_{z}+C-1),
\end{equation}
with $A,B, C$ arbitrary constants.  The operator
$D_{HG}$ is the operator in the Euler form of the hypergeometric
equation, i.e., $D_{HG}[A,B;C]\gauss(A,B;C;z)=0$. 

Now, let us assume that $y(z) = z^{\nu}h_{\nu}(z)+\mathcal{O}(x)$,
with $\nu$ a constant. For $x=0$, the monodromy of this solution
around $z=0$ must be equal to the composite monodromy of the full
solution around $z=0$ and $z=x$ to leading order in $x$
\cite{Lencses:2017dgf}. This type of solution is called a
\emph{path-multiplicative} solution by \cite{ronveaux1995heun}. If we
plug the small-$x$ expansion of $y(z)$ into Eq.~\eqref{eq:53}, we
obtain
\begin{equation}
  \label{eq:55}
  \left(
    D_{HG}[b_{1,0}+\nu,b_{2,0}+\nu;2(1-\theta_{0,0}-\theta_{x,0}+\nu)]-\nu(\nu+1-2\theta_{0,0}-2\theta_{x,0})+\tilde{K_{x}}
  \right)h_{\nu}(z)=0,
\end{equation}
where $\tilde{K_{x}} \equiv - \lim_{x\rightarrow0}x(x-1)K_{x}$.  Here,
$\theta_{0,0}$, $\theta_{x,0}$ and $b_{k,0}$ denote the Frobenius
exponents $\theta_{0}$, $\theta_{x}$ and $b_{k}$ of \eqref{eq:53},
respectively, calculated at $x=0$.  Therefore, if we make the choice
\begin{equation}
  \label{eq:56}
  \nu\left(\nu+1-2\theta_{0,0}-2\theta_{x,0}\right)-\tilde{K_{x}}=0,
\end{equation}
we have that $h_{\nu}(z)$ is a solution of the hypergeometric equation
\begin{equation}
  \label{eq:57}
  D_{HG}[b_{1,0}+\nu,b_{2,0}+\nu;2(1-\theta_{0,0}-\theta_{x,0}+\nu)]h_{\nu}(z)=0.
\end{equation}
Now, we use the two solutions of the quadratic equation \eqref{eq:56}
for $\nu$, which we denote by $\nu_{\pm}$, to obtain the composite
monodromy as $x\to 0$
\begin{equation}
  \label{eq:58}
  \sigma_0= \frac{1}{2}(\nu_{+}-\nu_{-}) = \frac{1}{2}\sqrt{(2\theta_{0,0}+2\theta_{x,0}-1)^{2}+4\tilde{K_{x}}},
\end{equation}
where $\sigma_0\equiv\lim_{x\rightarrow 0} \sigma_{0x}$ (see
Eq.~\eqref{eq:parameters-epsilon-expansion-1}).  Finally, by equating
the QNM condition in Eq.~\eqref{eq:qnmcondition}
\begin{equation}
  \label{eq:59}
  \sigma_0 = \pm(\theta_{0,0}-\theta_{x,0}) +N +\frac12
\end{equation}
and the appropriate monodromies \eqref{eq:monodromies-APE} onto
Eq.~\eqref{eq:58}, we obtain the correct Eq.~\eqref{eq:64} and the
corresponding QNMs. Finally, we see that, since
$\nu_{\pm} = \theta_{0,0}+\theta_{x,0}-\frac{1}{2}\pm \sigma_{0}$, the
monodromy matrix of the solution $z^{\nu}h_{\nu}(z)$ is given by
\begin{equation}
  \label{eq:11}
  M_{0x} =
  \begin{pmatrix}
    e^{2\pi i \nu_{+}} & 0\\
    0 & e^{2\pi i\nu_{-}}
  \end{pmatrix}\sim \begin{pmatrix}
   - e^{2\pi i\sigma_{0}} & 0\\
    0 & -e^{-2\pi i\sigma_{0}}
  \end{pmatrix}.
\end{equation}
This is the origin of the minus sign in the composite trace
definition in Eq.~\eqref{eq:compmono}.


\section{Accessory Parameter Expansion From Isomonodromic \texorpdfstring{$\tau$}{t}-function}
\label{sec:iso-definitions}

Here we adapt the approach of \cite{Lencses:2017dgf} for the APE to the case of monodromies that depend on $x$.
The accessory parameter $K_{x}$ of the Heun equation can be obtained via the isomonodromic $\tau$-function by
\begin{align}
    K_{x} &= \left.\frac{d}{dt}\log\left(t^{-4\theta_{0}\theta_{t}} (1-t)^{-4\theta_{1}\theta_{t}}
            \tau(\bm{\theta}_{0,+}
            ;t )\right)\right|_{t=x,\lambda(x)=x}
            \label{eq:litvinov_cond2},
  \end{align}
where
\begin{equation}
  \label{eq:69}
  \bm{\theta}_{s_{1},s_{2}} \equiv \left(\lmtheta{\tfrac{s_{1}1}{2} }{\tfrac{s_{2}1}{2} }\right),\quad s_{1},s_{2}=0,\pm .
\end{equation}
The general definition of the $\tau$-function can be found in
\cite{Gamayun2012,Lencses:2017dgf}, for example.  The $\tau$-function
satisfies a second-order non-linear ODE which is related to the
Painlev\'e VI equation.  This second-order ODE has two integration
constants $(\sigma,\mathsf{s})$.\footnote{Note that this integration
  constant is denoted by a sans-serif letter $\mathsf{s}$ in order to
  distinguish it from the spin $s$.  Similarly, we use $\varepsilon$
  in Eq.~\eqref{eq:structconsts} in order to distinguish it from the
  extremality parameter $\epsilon$ elsewhere in the paper.} The
integration constant $\sigma$ is in fact the composite monodromy of
Heun solutions as discussed in this paper and $\mathsf{s}$ is related
to the other composite monodromies in a precise way.
 
The usefulness of the approach via Eq.~\eqref{eq:litvinov_cond2}
relies on the complete expansion of the $\tau$-function in terms of
monodromy data, first obtained in \cite{Gamayun2012}.  Such expansion
is
\begin{equation}
  \tau(\bm{\theta};t)\equiv t^{2\theta_{0}\theta_{t}}(1-t)^{4\theta_{1}\theta_{t}}\,t^{\sigma^{2}-\theta_{0}^{2}-\theta_{t}^{2}}\sum_{n\in\mathbb{Z}}C(\bm{\theta},\sigma+n)
  \mathsf{s}^{n} t^{n(n+2 \sigma)}{\cal
    B}(\bm{\theta},\sigma+n;t),
\label{eq:taufunctionexpansion}
\end{equation} 
where the structure constants $C(\bm{\theta},\sigma)$ are given in
terms of Barnes functions $G$ as\footnote{Barnes functions are defined
  by the functional relation $G(z+1)=\Gamma(z)G(z)$, with $G(1)=1$ and
  $\Gamma(z)$ being the Euler gamma function.}
\begin{equation}
\label{eq:structconsts}
C(\bm{\theta},\sigma)=\frac{\prod_{\varepsilon,\tilde\varepsilon=\pm}G(1+\theta_t+
  \varepsilon\theta_0+\tilde\varepsilon\sigma)G(1+\theta_1+\varepsilon\theta_\infty+
  \tilde\varepsilon\sigma)}{\prod_{\varepsilon=\pm}G(1+\varepsilon2\sigma)}.
\end{equation}
The ${\cal B}$'s in Eq.~\eqref{eq:taufunctionexpansion} are the  conformal blocks for central charge $c=1$,
given by the
Alday-Gaiotto-Tachikawa (AGT)~\cite{Alday2010a} combinatorial series
\begin{equation}
\label{eq:cblockstau}
{\cal B}(\bm{\theta},\sigma;t)=
\sum_{
  n,m\in\mathbb{Y}}{\cal
  B}_{n,m}(\bm{\theta},\sigma)t^{|n|+|m|}, 
\end{equation}
summing over pairs of Young diagrams $n,m$ with
\begin{multline}
\label{eq:cblockstaucoeffs}
{\cal
  B}_{n,m}(\bm{\theta},\sigma)=\prod_{(i,j)\in n}
\frac{((\theta_t+\sigma+i-j)^2-\theta_0^2)((\theta_1+\sigma+i-j)^2 
-\theta_\infty^2)}{ h_n^2(i,j)(n'_j+m_i-i-j+1+2\sigma)^2}
\times \\
\prod_{(i,j)\in m}
\frac{((\theta_t-\sigma+i-j)^2-\theta_0^2)((\theta_1-\sigma+i-j)^2 -\theta_\infty^2)}{ h_m^2(i,j)(n_i+m'_j-i-j+1-2\sigma)^2},
\end{multline}
where $(i,j)$ denotes the respective box in a Young diagram $n$,
$n_i$ the number of boxes in row $i$, $n'_j$ the number of
boxes in column $j$ and $h_{n}(i,j)=n_i+n'_j-i-j+1$
its hook length. 

As mentioned, the two integration constants of the $\tau$-function are $(\sigma,\mathsf{s})$
and, as shown in \cite{Lencses:2017dgf}, the condition $\lambda(x)=x$
introduces a constraint
$\mathsf{s}=\mathsf{s}(\bm{\theta},\sigma,x) = \mathsf{s}_{0}+ \mathsf{s}_{1}x+\mathcal{O}(x^2)$, with
$\mathsf{s}_{i}=\mathsf{s}_{i}(\theta_{k},\sigma)$. In order to find the APE, we
just need to plug this constraint into Eq.~\eqref{eq:litvinov_cond2} and consistently expand the result for small $x$. 

The main difference between the APE used in the present paper and the one in \cite{Lencses:2017dgf} is that here the monodromies depend on the moduli $x=x(\epsilon)$ via the parameter $\epsilon$. As explained in the main text, we assume the monodromies and moduli have regular expansions in $\epsilon$ given by  \eqref{eq:parameters-epsilon-expansion}. If we solve the condition $\lambda(x)=x$ for $s$ as an expansion in $\epsilon$ using the expanded monodromies and moduli, we  get $\mathsf{s} = \tilde{\mathsf{s}}_{0}+ \tilde{\mathsf{s}}_{1}\epsilon+\mathcal{O}(\epsilon^2)$, with
$\tilde{\mathsf{s}}_{i}=\tilde{\mathsf{s}}_{i}(\theta_{k,j},\sigma_{j})$, and this result can be used in \eqref{eq:litvinov_cond2} to obtain our main result \eqref{eq:APE}.

\section{Coefficients of the Quasinormal Mode Expansion}
\label{sec:omegas}

In this appendix, we provide expressions below are the coefficients in
Eq.~\eqref{eq:omega-expansion} expanded up to $\mathcal{O}(a^3)$. For
simplicity, in this appendix we take $L=1$ and we define
$\delta_{\sigma}\equiv \ell(\ell+1)$ and
$\zeta\equiv \sqrt{12\delta _{\sigma}-8s^2+5}$.

General-spin expressions for the first two coefficients are the following:

	\begin{dseries}[breakdepth={5}]
		$\bar{\omega}_0\hiderel{=}\frac{1}{4} \sqrt{3} \left(1-14 a^2\right) \zeta +\frac{3}{4} i \left(7 a^2 (2 N+1)+4 i \sqrt{3} a m-2 N-1\right)+\frac{\sqrt{3} a^2}{4\zeta\delta_{\sigma}\left(4\delta_{\sigma}-3\right)}\times\left\{516 \delta_{\sigma }^3+4 \delta _{\sigma }^2 \left(45 m^2-74 s^2-46\right)+36 m^2 s^2 \left(s^2+2\right)-3\delta_{\sigma}\left[m^2(40s^2+44)+4s^4-76 s^2+51\right]\right\}+\mathcal{O}(a^4),$
	\end{dseries}
	\begin{dseries}[breakdepth={5}]
		$\bar{\omega}_1\hiderel{=}-3 a m-\frac{9 a m s^2 \left(\zeta -i \sqrt{3} (2N+1) \left(\delta _{\sigma }-s^2+1\right)\right)}{2 \zeta  \delta _{\sigma } \left(3 \delta _{\sigma }-2 s^2+2\right)}+\frac{9i\sqrt{3}a^3m\left(2N+1\right)} {2\zeta^3 (\delta_{\sigma}-2)\delta_{\sigma}^3(4\delta_{\sigma}-3)(3\delta_{\sigma}-2 s^2+2)^2}\left[3240 \delta _{\sigma }^8-18 \delta _{\sigma }^7 \left(60 m^2+518 s^2+177\right)+6 \delta _{\sigma }^6 \left(3 m^2 \left(178 s^2+83\right)+1862 s^4+2044 s^2-1251\right)-24 m^2 s^4(s^2-1)^2 \left(8 s^4+11 s^2-10\right)+\delta _{\sigma }^5 \left(-6 m^2 \left(922 s^4+574 s^2-263\right)-7564 s^6-16162 s^4+15359 s^2+60\right)+\delta _{\sigma }^4 \left(m^2 \left(6580 s^6+510 s^4-2472 s^2-352\right)+3460 s^8+8476 s^6-9133 s^4-4781 s^2+3112\right)+4s^2\delta_{\sigma}^2\left(m^2 (394 s^8-655 s^6-1101 s^4+1538 s^2-257)+(s^2-1)^2(40 s^6+75 s^4+126 s^2-133)\right)-4 m^2 s^4 \left(56 s^8-287 s^6-222 s^4+919 s^2-466\right)\delta_{\sigma}-2\delta_{\sigma}^3 \left(m^2 \left(2262 s^8-1412 s^6-2328 s^4+879 s^2+140\right)+540 s^{10}+706 s^8-577 s^6-2817 s^4+2588 s^2-440\right)\right]+\frac{3 a^3 m}{2(\delta _{\sigma }-2)\delta_{\sigma}^3(4 \delta _{\sigma }-3)(3 \delta _{\sigma }-2 s^2+2)^2}\left[126 \delta _{\sigma }^7+6 \delta _{\sigma }^6 \left(45 m^2+68 s^2-32\right)-72 m^2 s^4 \left(s^6-3 s^2+2\right)-\delta _{\sigma }^5 \left(54 m^2 \left(14 s^2+9\right)+994 s^4+301 s^2+208\right)+\delta _{\sigma }^4 \left(6 m^2 \left(199 s^4+166 s^2-32\right)+780 s^6+1127 s^4-901 s^2+128\right)+12 s^2 \delta _{\sigma }^2 \left(m^2(41 s^6-3 s^4-90 s^2+25)+(s^2-1)^2(5 s^4+14 s^2+17)\right)-12 m^2 s^4 \left(7 s^6-18 s^4-39 s^2+50\right) \delta_{\sigma}-6 \delta_{\sigma }^3 \left(m^2 \left(186 s^6+80 s^4-67 s^2-28\right)+54 s^8+89 s^6-107 s^4-20 s^2-16\right)\right]+\mathcal{O}(a^4).$
	\end{dseries}
        We give the coefficient $\bar{\omega}_2$ separately for each
        different value of the spin $s$. 

For $s=0$:
\begin{dseries}
	$\bar{\omega}_2\hiderel{=}\frac{9}{8} \sqrt{3} a \left(8 a^2-5\right) m+\frac{1}{32} i \left(465 a^2-19\right)(2 N+1)-3i(2 N+1) \left(-24 \left(63 a^2+2\right) \delta _{\sigma }^2+8 \left(21 a^2 \left(69 m^2-25\right)-4\right) \delta _{\sigma }+3 a^2 \left(5216 m^2-695\right)+51\right)/\left[32(4\delta_{\sigma}-3)(12\delta_{\sigma}+17)^2\right]+\frac{27 (2 N+1)^2}{16 \sqrt{3} \zeta ^3 \left(4 \delta _{\sigma }-3\right) \left(12 \delta _{\sigma }+17\right)^2}\left\{720 \left(75 a^2-4\right) \delta _{\sigma }^4+12 \left(3 a^2 \left(300 m^2+3767\right)-596\right) \delta _{\sigma }^3+2 \left(3 a^2 \left(1818 m^2+7267\right)-1054\right) \delta _{\sigma }^2+\left(4439-3 a^2 \left(7122 m^2+26665\right)\right) \delta _{\sigma }-a^2 \left(22228 m^2+31937\right)+1785\right\}+\left\{-10368 \left(765 a^2-76\right) \delta _{\sigma }^5+432 \left(3 a^2 \left(8760 m^2-18341\right)+5236\right) \delta _{\sigma }^4+108 \left(3 a^2 \left(91484 m^2-49331\right)+12268\right) \delta _{\sigma }^3+6 \left(9 a^2 \left(212862 m^2+163295\right)-166930\right) \delta _{\sigma }^2+\left(-3 a^2 \left(5249298 m^2-3509465\right)-951175\right) \delta _{\sigma }-15 \left(a^2 \left(459664 m^2-155035\right)+12325\right)\right\}/\left[16 \sqrt{3} \zeta ^3 \left(4 \delta _{\sigma }-3\right) \left(12 \delta _{\sigma }+17\right)^2\right]+\mathcal{O}(a^4).$
\end{dseries}
For $s=\pm1/2$:
\begin{dseries}
	$\bar{\omega}_2\hiderel{=}9 \sqrt{3} a^3 m-\frac{45}{8} \sqrt{3} a m+\frac{1}{32} i \left(465 a^2-19\right) (2 N+1)-\frac{i (2 N+1)}{256 \delta _{\sigma}^3\left(2\delta_{\sigma}+1\right)^3\left(4\delta_{\sigma}+5\right)}\left\{32 \left(135 a^2-14\right) \delta _{\sigma }^6+48 \left(a^2 \left(214 m^2+203\right)-21\right) \delta _{\sigma }^5+\left(a^2 \left(6336 m^2+6954\right)-672\right) \delta _{\sigma }^4+\left(a^2 \left(2121-8238 m^2\right)-140\right) \delta _{\sigma }^3+27 a^2 \left(10-317 m^2\right) \delta _{\sigma }^2-2646 a^2 m^2 \delta _{\sigma }-270 a^2 m^2\right\}+\frac{3 \sqrt{3}(2 N+1)^2}{256 \zeta^3 \delta_{\sigma}^3 \left(2\delta_{\sigma}+1\right)^3\left(4 \delta_{\sigma}+5\right)}\left\{1920 \left(75 a^2-4\right) \delta _{\sigma }^8+32 \left(75 a^2 \left(12 m^2+175\right)-692\right) \delta _{\sigma }^7+8 \left(a^2 \left(6372 m^2+55869\right)-2900\right) \delta _{\sigma }^6+6 \left(a^2 \left(6604 m^2+36965\right)-1892\right) \delta _{\sigma }^5+\left(a^2 \left(31158 m^2+50721\right)-2624\right) \delta _{\sigma }^4+\left(3 a^2 \left(7927 m^2+1254\right)-230\right) \delta _{\sigma }^3+135 a^2 \left(78 m^2-1\right) \delta _{\sigma }^2+2133 a^2 m^2 \delta _{\sigma }+135 a^2 m^2\right\}+\frac{\sqrt{3}}{256\zeta\delta_{\sigma}^3\left(2\delta_{\sigma}+1\right)^3\left(4\delta_{\sigma}+5\right)}\left\{-256 \left(765 a^2-76\right) \delta _{\sigma }^8+32 \left(15 a^2 \left(584 m^2-1213\right)+1796\right) \delta _{\sigma }^7+48 \left(a^2 \left(15494 m^2-13239\right)+1296\right) \delta _{\sigma }^6+\left(30 a^2 \left(21096 m^2-10825\right)+31592\right) \delta _{\sigma }^5+\left(9 a^2 \left(19754 m^2-8579\right)+7592\right) \delta _{\sigma }^4-3 \left(a^2 \left(5485 m^2+2104\right)-230\right) \delta _{\sigma }^3-9 a^2 \left(1552 m^2-15\right) \delta _{\sigma }^2-1593 a^2 m^2 \delta _{\sigma }-135 a^2 m^2\right\}+\mathcal{O}(a^4).$
\end{dseries}
For $s=\pm1$:
\begin{dseries}
	$\bar{\omega}_2\hiderel{=}9 \sqrt{3} a^3 m-\frac{45}{8} \sqrt{3} a m+\frac{1}{32} i \left(465 a^2-19\right) (2 N+1)+\frac{81 \sqrt{3}\left[(2 N+1)^2+1\right]\left(\left(1-41 a^2\right) \delta _{\sigma }^2+18 a^2 m^2\right)}{16 \zeta ^3 \delta _{\sigma }^3 \left(4 \delta _{\sigma }-3\right) \left(4 \delta _{\sigma }+3\right)^2}-\frac{i (2 N+1)}{32 \delta _{\sigma }^3 \left(4 \delta _{\sigma }-3\right) \left(4 \delta _{\sigma }+3\right){}^2}\left\{-8 \left(729 a^2-50\right) \delta _{\sigma }^5+24 \left(a^2 \left(55 m^2-41\right)+8\right) \delta _{\sigma }^4+9 \left(a^2 \left(464 m^2+799\right)-25\right) \delta _{\sigma }^3-108 \left(a^2 \left(112 m^2-41\right)+1\right) \delta _{\sigma }^2-11664 a^2 m^2 \delta _{\sigma }-1944 a^2 m^2\right\}+\frac{3 \sqrt{3} (2 N+1)^2}{16 \zeta ^3 \delta _{\sigma } \left(4 \delta _{\sigma }-3\right) \left(4 \delta _{\sigma }+3\right)^2}\left\{240 \left(75 a^2-4\right) \delta _{\sigma }^5+4 \left(a^2 \left(900 m^2+597\right)-20\right) \delta _{\sigma }^4+2 \left(9 a^2 \left(10 m^2-737\right)+406\right) \delta _{\sigma }^3+3 \left(a^2 \left(754 m^2+901\right)-1\right) \delta _{\sigma }^2-9 \left(a^2 \left(1160 m^2-509\right)+17\right) \delta _{\sigma }-8640 a^2 m^2\right\}+\frac{3 \sqrt{3}}{16 \zeta^3 \delta_{\sigma } \left(4 \delta _{\sigma }-3\right) \left(4 \delta _{\sigma }+3\right)^2}\left\{-128 \left(765 a^2-76\right) \delta _{\sigma }^6+16 \left(a^2 \left(8760 m^2+603\right)+124\right) \delta _{\sigma }^5-4 \left(3 a^2 \left(2572 m^2-6911\right)+2012\right) \delta _{\sigma }^4-2 \left(3 a^2 \left(23474 m^2+5205\right)+278\right) \delta _{\sigma }^3+3 \left(a^2 \left(31222 m^2-7631\right)+467\right) \delta _{\sigma }^2+9 \left(a^2 \left(5572 m^2+1265\right)-35\right) \delta _{\sigma }-25164 a^2 m^2\right\}+\mathcal{O}(a^4).$
\end{dseries}

For $s=\pm2$:
\begin{dseries}
	$\bar{\omega}_2\hiderel{=}9 \sqrt{3} a^3 m-\frac{45}{8} \sqrt{3} a m+\frac{1}{32} i \left(465 a^2-19\right) (2 N+1)+\frac{i (2 N+1)}{32 \left(5-4 \delta _{\sigma }\right){}^2 \left(\delta _{\sigma }-2\right){}^3 \delta _{\sigma }^3 \left(4 \delta _{\sigma }-3\right)}\left\{-8 \left(3105 a^2-194\right) \delta _{\sigma }^8+8 \left(3 a^2 \left(703 m^2+11631\right)-1744\right) \delta _{\sigma }^7-3 \left(a^2 \left(105136 m^2+432723\right)-15973\right) \delta _{\sigma }^6+2 \left(3 a^2 \left(292720 m^2+532233\right)-39133\right) \delta _{\sigma }^5+\left(60340-36 a^2 \left(117968 m^2+122227\right)\right) \delta _{\sigma }^4+24 \left(a^2 \left(192848 m^2+135795\right)-725\right) \delta _{\sigma }^3-1152 a^2 \left(1139 m^2+900\right) \delta _{\sigma }^2-1486080 a^2 m^2 \delta _{\sigma }+1036800 a^2 m^2\right\}+\frac{3 \sqrt{3} (2 N+1)^2}{16 \zeta ^3 \left(5-4 \delta _{\sigma }\right){}^2 \left(\delta _{\sigma }-2\right)^3 \delta _{\sigma }^3 \left(4 \delta _{\sigma }-3\right)}\left\{240 \left(75 a^2-4\right) \delta _{\sigma }^{10}+4 \left(15 a^2 \left(60 m^2-3901\right)+3148\right) \delta _{\sigma }^9-2 \left(3 a^2 \left(5298 m^2-226909\right)+35482\right) \delta _{\sigma }^8-3 \left(a^2 \left(12302 m^2+1554911\right)-73607\right) \delta _{\sigma }^7+\left(3 a^2 \left(395920 m^2+3455569\right)-405995\right) \delta _{\sigma }^6-2 \left(15 a^2 \left(161384 m^2+512431\right)-218393\right) \delta _{\sigma }^5+12 \left(a^2 \left(717768 m^2+1242091\right)-20995\right) \delta _{\sigma }^4-24 \left(a^2 \left(261416 m^2+361395\right)-2475\right) \delta _{\sigma }^3-288 a^2 \left(4489 m^2-8100\right) \delta _{\sigma }^2+4898880 a^2 m^2 \delta _{\sigma }-2332800 a^2 m^2\right\}+\frac{3 \sqrt{3}}{16 \zeta ^3 \left(5-4 \delta _{\sigma }\right){}^2 \left(\delta _{\sigma }-2\right)^3 \delta _{\sigma }^3 \left(4 \delta _{\sigma }-3\right)}\left\{-128 \left(765 a^2-76\right) \delta _{\sigma }^{11}+48 \left(5 a^2 \left(584 m^2+6277\right)-2796\right) \delta _{\sigma }^{10}-4 \left(3 a^2 \left(171852 m^2+857513\right)-201988\right) \delta _{\sigma }^9+2 \left(3 a^2 \left(2349982 m^2+6875683\right)-1385222\right) \delta _{\sigma }^8-3 \left(5 a^2 \left(3918450 m^2+7190207\right)-1969787\right) \delta _{\sigma }^7+\left(3 a^2 \left(54007292 m^2+64065961\right)-8007761\right) \delta _{\sigma }^6-6 \left(a^2 \left(49556268 m^2+39151043\right)-1117417\right) \delta _{\sigma }^5+12 \left(a^2 \left(28901884 m^2+15878979\right)-262665\right) \delta _{\sigma }^4-72 \left(a^2 \left(3095108 m^2+1294785\right)-8775\right) \delta _{\sigma }^3+69984 a^2 \left(493 m^2+300\right) \delta _{\sigma }^2+44089920 a^2 m^2 \delta _{\sigma }-20995200 a^2 m^2\right\}+\mathcal{O}(a^4).$	
\end{dseries}


\section{Coefficients of the Angular Eigenvalue Expansion}\label{sec:eigen coeffs}

In this appendix we give the explicit form of the coefficients in the
expansion for the eigenvalue in
Eq.~\eqref{eq:angular-eigenvalue-expansion}, expanded for small $\alpha$. As in the previous appendix, we define $\delta_\sigma\equiv \ell(\ell+1)$

\begin{equation}
  \label{eq:50}
  \begin{aligned}
    \lambda_{\omega,0}&\equiv \delta_\sigma-s^2+\frac{\alpha^2}{\delta_\sigma(4\delta_\sigma-3)}\left[-2 s^4 \left(\delta_\sigma-3 m^2\right)+s^2\left(\delta_\sigma\left(4m^2+1\right)-6 m^2\right)\right.\\[5pt]
    &\left.  +\delta_\sigma\left(\delta_\sigma\left(2\delta_\sigma+6 m^2+1\right)-4 m^2-2\right)\right]
    +\mathcal{O}\left(\alpha^4\right) ,
  \end{aligned}
\end{equation}

\begin{equation}
  \label{eq:42}
  \begin{aligned}
    \lambda_{\omega,1}&\equiv
    -2 m\left(1+\frac{s^2}{\delta_\sigma}\right) -\frac{2
      \alpha ^2 m }{\delta_\sigma^3 \left(\delta_\sigma-2\right) (4\delta_\sigma-3)}\left[-2 s^6 \left(5\delta_\sigma^2-(7\delta_\sigma+6)
        m^2\right)
      +\delta_\sigma)^2 s^2 \times \right.\\[5pt]
    &\left.\left(3\delta_\sigma\left(2\delta_\sigma-2
          m^2-7\right)+22 m^2+8\right)+\delta_\sigma^3
      \left(\delta_\sigma-2\right)\left(6\delta_\sigma+2 m^2-5\right)\right.\\[5pt]
    &\left.+2 s^4 \left(\delta_\sigma\left(\delta_\sigma\left(3\delta_\sigma-5
            m^2+4\right)-7 m^2\right)-6 m^2\right)\right]+\mathcal{O}\left(\alpha^4\right),
  \end{aligned}
\end{equation}

\begin{equation}
  \label{eq:51}
  \begin{aligned}
    \lambda_{\omega,2}&\equiv \frac{2}{\delta_\sigma^3(4 \delta_\sigma-3)} \left\{s^4 \left[(5\delta_\sigma+3) m^2-3\delta_\sigma^2\right]+2\delta_\sigma^2 s^2 \left(\delta_\sigma-3
        m^2\right)+\delta_\sigma^3
      \left(\delta_\sigma+m^2-1\right)\right\}
        \\[5pt]
    &+\mathcal{O}\left(\alpha^2\right),
  \end{aligned}
\end{equation}

\begin{equation}
  \label{eq:52}
  \begin{aligned}
    \lambda_{\omega,3}&\equiv -\frac{4 m s^2}{\delta_\sigma^5
      \left(\delta_\sigma-2\right) (4\delta_\sigma-3)}\left(-\delta_\sigma^4 \left(3\delta_\sigma-5m^2-1\right)+2\delta_\sigma^2 s^2 \left(5\delta_\sigma^2-(7\delta_\sigma+6) m^2\right)+\right.\\[5pt]
    &\left. 
s^4 \left(\delta_\sigma\left(19
          m^2-\delta_\sigma\left(7\delta_\sigma-9
            m^2+6\right)\right)+6 m^2\right)\right)  +\mathcal{O}\left(\alpha^2\right).
  \end{aligned}
\end{equation}

The expressions for $\lambda_{\omega,0}$ and $\lambda_{\omega,1}$
which we provide agree with~\cite{Suzuki1998} (which also provide the
$\mathcal{O}\left(\alpha^2\right)$ term in $\lambda_{\omega,2}$).  The
leading-order terms in the expansions for $\lambda_{\omega,2}$ and
$\lambda_{\omega,3}$ (which are not provided in~\cite{Suzuki1998})
are, in fact, just the same as in Kerr, and agree
with~\cite{berti2006eigenvalues}.

\bibliographystyle{JHEP} 
\bibliography{qnms}

\end{document}